\documentclass[11pt,a4paper]{article}

\usepackage{jheppub}
\usepackage[T1]{fontenc}
\usepackage{fix-cm}
\usepackage{lmodern}
\usepackage{mathtools}
\usepackage{mathrsfs}
\usepackage{empheq}
\usepackage{physics}
\usepackage{bbm}
\usepackage{bm}
\usepackage{hhline}
\usepackage{tensor}
\usepackage{tikz}
\usepackage{xcolor}
\usepackage{pifont}
\usepackage{latexsym}
\usepackage{slashed}
\usepackage{multirow}
\usepackage{float}
\usepackage{afterpage}
\usepackage{soul}

\makeatletter
\newcommand*{\rom}[1]{\expandafter\@slowromancap\romannumeral #1@}
\makeatother

\newcommand\dphi{\delta\phi}
\newcommand\mO{\mathcal{O}}
\newcommand\mD{\mathcal{D}}
\newcommand\mP{\mathcal{P}}
\newcommand\mN{\mathcal{N}}
\newcommand\fft[2]{\frac{#1}{#2}}

\newcommand\mQ{\mathcal{Q}}

\newcommand{\cN}{\mathcal{N}}
\newcommand{\cmark}{\ding{51}}
\newcommand{\xmark}{\ding{55}}

\title{A compendium of logarithmic corrections in AdS/CFT}
\author[a]{Nikolay Bobev,}
\author[a]{Marina David,}
\author[c]{Junho Hong,}
\author[b]{Valentin Reys,}
\author[a]{and Xuao Zhang}

\affiliation[a]{Instituut voor Theoretische Fysica, KU Leuven, \\
	Celestijnenlaan 200D, B-3001 Leuven, Belgium}

\affiliation[b]{Université Paris-Saclay, CNRS, CEA, \\
	Institut de physique théorique, 91191, Gif-sur-Yvette, France}
	
\affiliation[c]{Department of Physics \& Center for Quantum Spacetime, Sogang University\,,\\ 35 Baekbeom-ro, Mapo-gu, Seoul 04107, Republic of Korea}

\emailAdd{nikolay.bobev@kuleuven.be}
\emailAdd{junhohong@sogang.ac.kr}
\emailAdd{marina.david@kuleuven.be}
\emailAdd{xuao.zhang@kuleuven.be}
\emailAdd{valentin.reys@ipht.fr}

\abstract{We study the logarithmic corrections to various CFT partition functions in the context of the AdS$_4$/CFT$_3$ correspondence for theories arising on the worldvolume of M2-branes. We utilize four-dimensional gauged supergravity and heat kernel methods and present general expressions for the logarithmic corrections to the gravitational on-shell action and black hole entropy for a number of different supergravity backgrounds. We outline several subtle features of these calculations and contrast them with a similar analysis of logarithmic corrections performed directly in the eleven-dimensional uplift of a given four-dimensional supergravity background. We find results consistent with AdS/CFT provided that the infinite sum over KK modes on the internal space is regularized in a specific manner. This analysis leads to an explicit expression for the logarithmic correction to the Bekenstein-Hawking entropy of large Kerr-Newmann and Reissner-Nordström black holes in AdS$_4$. Our results also have important implications for effective field theory coupled to gravity in AdS$_4$ and for the existence of scale-separated AdS$_4$ vacua in string theory, which come in the form of new constraints on the field content and mass spectrum of matter fields.}

\begin{document}
	
\maketitle

\section{Introduction}
\label{sec:intro}

The AdS/CFT correspondence is a cornerstone of modern theoretical physics and it is of clear interest to test this duality as precisely as possible. In this work we focus on subleading corrections to the supergravity approximation used in holography, specifically in the context of AdS$_4$/CFT$_3$. The motivation for our analysis comes from the CFT side of the duality and the recent proliferation of techniques to compute QFT observables in the large $N$ limit of supersymmetric holographic CFTs. More concretely, the partition functions $Z_{\mathcal{M}_3}$ of~3d~$\mathcal{N}\geq 2$ holographic SCFTs arising from a stack of $N$ M2-branes on a compact Euclidean manifold $\mathcal{M}_3$ can be calculated in the large $N$ limit using supersymmetric localization. In these theories, the leading term in the free energy $\log Z_{\mathcal{M}_3}$ scales as $N^{3/2}$ and in many examples agrees with the regularized on-shell action of classical asymptotically locally AdS$_4$ Euclidean supergravity solutions which can be thought of as smooth fillings of $\mathcal{M}_3$.\footnote{See \cite{Drukker:2010nc,Jafferis:2011zi,Benini:2015eyy,Hosseini:2016tor,Azzurli:2017kxo} for an incomplete selection of references and \cite{Pestun:2016zxk,Zaffaroni:2019dhb} for a review} The first subleading term scales as $N^{1/2}$ and it can be accounted for in the bulk by studying the leading four-derivative correction to the classical four-dimensional supergravity action, see \cite{Bobev:2020egg,Bobev:2021oku}. The goal of this paper is to study how the $\log N$ term that arises at the next order in the large $N$ expansion of the free energy can be computed from the bulk supergravity theory.

An alternative point of view on logarithmic corrections to gravitational path integrals is found in black hole physics. It is expected that the Bekenstein-Hawking formula for the black hole entropy, $S_{\rm BH} = \frac{1}{4}A_{\rm H}$ with $A_{\rm H}$ the area of the horizon in Planck units, receives perturbative quantum corrections. These arise from higher-derivative terms in the gravitational effective action, as well as from quantum effects due to matter fields propagating on a fixed gravitational background. While a lot is known about these corrections to the black hole entropy, they are in general hard to calculate and strongly depend on the details of the UV completion of the effective gravitational theory. It was shown by Ashoke Sen that the logarithmic corrections to black hole entropy are very special in that regard. Notably, the coefficient of the $\log A_{\rm H}$ correction to $S_{\rm BH}$ is determined solely by the one-loop quantum contributions to the gravitational path integral of all fields below the cutoff scale of the effective gravitational theory. This fact represents a powerful ``IR window'' into the UV-complete theory of quantum gravity and can be employed to derive strong consistency condition on the microscopic description of black hole physics. Moreover, in special situations with enough symmetry, it has been shown that the $\log A_{\rm H}$ corrections to black hole entropy can be successfully matched to microscopic results from string theory, see \cite{Mandal:2010cj,Banerjee:2011jp,Sen:2012kpz,Sen:2012cj,Bhattacharyya:2012wz,Sen:2012dw} and references thereof.

These two vantage points on logarithmic corrections to gravitational path integrals lead us to study how such logarithmic terms arise in 4d gravitational theories in AdS,~see \cite{Bhattacharyya:2012ye,Liu:2017vbl,Liu:2017vll,Gang:2019uay,PandoZayas:2020iqr} and \cite{Hristov:2021zai,David:2021eoq,Karan:2022dfy} for previous studies of logarithmic corrections in AdS$_4$/CFT$_3$ in 11d and 4d supergravity, respectively. We follow the approach of Sen and study the quantum effects of matter fields propagating on a fixed background of a 4d gravitational theory with a negative cosmological constant. Focusing on fields with half-integer spin up to 2 and general masses, we employ heat kernel methods to calculate their contributions to the logarithmic term in the gravitational path integral, see \cite{Vassilevich:2003xt} for a review of the heat kernel expansion. In the absence of other scales in the problem, the log correction is of the form $\log L^2/G_N$ where $G_N$ is the 4d Newton constant and $L$ is the length scale set by the cosmological constant.\footnote{The methods we use can be adapted to study situations with a more general cutoff scale not related to the 4d Planck scale. We will comment on such situations further below.} In general the coefficient of $\log L^2/G_N$ can receive contributions from zero modes and non-zero modes of the differential operator that controls the dynamics of the field with a given spin, as well as from boundary terms. For the 4d gravitational backgrounds of interest in this work, we find under reasonable assumptions that the boundary terms have a vanishing contribution to the $\log L^2/G_N$ coefficient. The analysis of the contribution from zero modes is subtle and we can only make precise quantitative statements for very symmetric spaces like Euclidean AdS$_4$ and AdS$_2 \times \Sigma_{\mathfrak{g}}$, where $\Sigma_{\mathfrak{g}}$ is a smooth compact Riemann surface of genus $\mathfrak{g}$. Nevertheless, we argue that even for more general gravitational backgrounds, the zero modes contribute a pure number to the coefficient of $\log L^2/G_N$ which is independent on any continuous parameters that may be present in the 4d gravitational solution.

The results for the contribution of the non-zero modes is more intriguing. We find that general massive fields of spin up to 2 lead to a coefficient of $\log L^2/G_N$ that depends on the continuous parameter of the gravitational background, like squashing deformations of the boundary metric or the angular velocity of a black hole. This seemingly innocuous fact has important repercussions. Focusing on 4d Euclidean supergravity backgrounds, we can compare our results for the coefficients of the $\log L^2/G_N$ corrections to the $\log N$ terms in the path integral of large $N$ 3d holographic SCFTs arising from M2-branes. Using results from supersymmetric localization for a number of explicit examples of such SCFTs, we observe that the coefficient of $\log N$ does not depend on such continuous parameters. We use this to show that the apparent contradiction between the supergravity and SCFT results can only be resolved if the total contribution of the non-zero modes to the $\log L^2/G_N$ term in the gravitational path integral only depends on a specific contribution in the heat kernel expansion that we identify.

We then set out to check this strong ``bootstrap'' constraint on explicit top-down AdS$_4$/CFT$_3$ examples. We start with the familiar ABJM theory at level $k=1$ dual to 11d supergravity on an asymptotically locally AdS$_4 \times S^7$ background. When trying to employ our 4d gravitational approach we are faced with an immediate difficulty, namely that the 4d~$\mathcal{N}=8$ supergravity theory is not a standard EFT with finitely many fields but rather a consistent truncation to 4d $\mathcal{N}=8$ ${\rm SO}(8)$ gauged supergravity coupled to an infinite tower of Kaluza-Klein (KK) modes with masses below the 4d Planck scale. To calculate the non-zero mode contribution to the heat kernel, we thus have to take into account the full KK tower and sum the contributions of the infinitely many KK modes. This calculation leads to a divergent sum which we need to regularize. Using three different regularization methods proposed in the literature, we show that \textit{all} non-zero mode contributions in the heat kernel expansion relevant for the $\log N$ correction vanish.

This result presents a puzzle. Using supersymmetric localization on the round $S^3$, it can be shown that the ABJM free energy has a logarithmic term in its large $N$ expansion given by $\frac{1}{4}\log N$, see \cite{Fuji:2011km,Marino:2011eh}. The factor of $1/4$ was successfully reproduced in 11d supergravity in \cite{Bhattacharyya:2012ye} using a one-loop analysis of the 11d supergravity fields on the dual AdS$_4 \times S^7$ background. In short, one finds that since the heat kernel expansion is used in odd dimensions, only zero modes of the 11d differential operators contribute. The only possible such zero mode in Euclidean AdS$_4$ arises from a 2-form, and it would naively seem that there are no 2-forms among the fields of 11d supergravity. However, the quantization of the 3-form potential of the 11d theory necessitates the introduction of a 2-form ghost, which in turn gives the only non-vanishing contribution to the $\log N$ term. This contribution was calculated in \cite{Bhattacharyya:2012ye} and was shown to agree precisely with the $1/4$ calculated in the holographically dual SCFT. Our 4d analysis of the logarithmic correction yields a different result. Since there are no 2-forms in the KK spectrum of 11d supergravity on AdS$_4 \times S^7$ even after quantization, we find no contribution from any zero modes. As explained above, we also find that the non-zero modes do not contribute. Moreover, one can show that there are no contributions from boundary terms. We are thus led to two possible conclusions: A) The 4d supergravity calculation of the logarithmic correction yields a vanishing result for $\log N$ in clear contradiction with the 11d analysis and with holography; or B) The regularization of the infinite sum over KK modes that contributes to the heat kernel coefficients can be performed in a way that gives a finite result consistent with 11d supergravity and holography. 

While option A) above may appear puzzling, it is not hard to understand how and why it may be possible. While 11d supergravity on AdS$_4 \times S^7$ should be equivalent to 4d $\mathcal{N}=8$ ${\rm SO}(8)$ gauged supergravity coupled to an infinite tower of KK modes at the classical level, there is no guarantee that the two theories produce the same quantum effect. Indeed in this explicit example we can trace the discrepancy to the 2-form ghost needed for the consistent quantization of the 11d 3-form field, which is not present in the 4d supergravity theory or in the full KK spectrum. If indeed option A) is to be taken seriously then it should be viewed as a cautionary tale for any application of lower-dimensional consistent supergravity truncations to holography, especially when studying quantum corrections to the leading supergravity approximation. 

Option B) on the other hand leads to more interesting conclusions. Assuming that the infinite sum over KK modes is regularized in a specific way that yields a non-vanishing heat kernel coefficient, we find that the non-zero modes contribute a factor of $\frac{\chi}{6}$ to the logarithmic correction, where $\chi$ is the Euler number of the four-dimensional background. If we further assume that, apart from the 2-forms mentioned above, there are no zero modes in the asymptotically locally Euclidean AdS$_4$ spaces relevant for holography, we find the following general result for the saddle point approximation to the gravitational path integral:
\begin{equation}
\label{eq:Igenchi}
	I = I_0 + \frac{\chi}{6} \log(L^2/G_N) + \ldots \, .
\end{equation}
Here $I_0$ is the regularized gravitational on-shell action (possibly including higher-derivative corrections) evaluated on the equations of motion, and the $\dots$ include higher order corrections in the semi-classical gravitational expansion. While the result in \eqref{eq:Igenchi} is not derived very rigorously, we show that it is in perfect agreement with results for the large $N$ path integral of the 3d $\mathcal{N}=6$ ABJM SCFT on various compact Euclidean three-manifolds. Moreover, if we assume that \eqref{eq:Igenchi} is valid, we arrive at the following prediction for the logarithmic correction to the entropy of a large AdS-Kerr-Newmann black hole embedded in AdS$_4\times S^7$
\begin{equation}
	S_{\text{AdS-KN}} = \frac{A_{\rm H}}{4 G_N} - \frac{1}{3} \log(A_{\rm H}/G_N) + \ldots \, .
\end{equation}
Here we work in an ensemble with fixed temperature and chemical potentials, the leading term is the standard Bekenstein-Hawking entropy of the black hole, and the $\dots$ are a placeholder for possible further quantum corrections to the black hole entropy. In the supersymmetric limit, our bulk result agrees with the logarithmic term in the large $N$ limit of the superconformal index of the dual SCFT. We consider this agreement to be a strong precision test of the microstate counting for 4d AdS black holes. 

To understand further the interplay between logarithmic corrections in holography, 11d supergravity, and 4d KK supergravity theories, we study four other top-down examples of AdS$_4$ vacua of M-theory with different amounts of supersymmetry and different topology of the internal manifold, and underscore several subtleties in the calculation of the logarithmic terms and the regularization of the infinite sums over KK modes. The conclusions we draw from these other examples are similar to the ones for AdS$_4 \times S^7$, i.e. one either should not use the 4d supergravity theory with the infinite tower of KK modes for the calculation of logarithmic corrections, or one should find a suitable regularization scheme for the infinite sums over KK modes in order to obtain one-loop results consistent with holography.

In addition to being of interest for top-down holographic models with explicit string/M-theory embeddings, our results have important implications for a more agnostic bottom-up approach to holography. Consider an effective theory of gravity coupled to a \textit{finite} number of matter fields of spin less than 2 which is valid up to an energy cutoff scale $\Lambda$ and admits an AdS$_4$ vacuum of scale $L$. Our results imply that the coefficient of the $\log L\Lambda$ term in the semi-classical expansion of the gravitational path integral depends in general on the continuous parameters of an asymptotically locally AdS$_4$ background. If we assume that the gravitational theory has a holographic description, this in turn implies that there exist $\log N$ or $\log \lambda$ terms in various physical observables, such as local correlation functions or the thermal effective action which captures the partition function of the theory on $S^1\times S^2$ in a saddle-point approximation. Note that here we are somewhat abstract and use the integer $N$ to denote the large number of degrees of freedom in the holographic CFTs, and $\lambda$ denotes some notion of a continuous parameter like an exactly marginal coupling. To the best of our knowledge such logarithmic terms do not appear in local correlation functions in any sequence of CFTs and they do not depend on continuous parameters in thermal effective actions. In this work we assume this to be true in general. Equipped with this assumption we are then led to conclude that general effective gravitational theories with a finite number of fields and an AdS$_4$ vacuum cannot be holographic. As we show in detail the only possible exception to this constraint, which somewhat poetically may be referred to as ``the unbearable lightness of the KK scale'' \cite{Kundera:1984}, arises from theories with a very finely tuned spectrum of quadratic fluctuations around the AdS$_4$ vacuum. In particular, we show that the pure 4d $\mathcal{N}=2$ and $\mathcal{N}=4$ gauged supergravity theories do not evade this constraint and are therefore not consistent UV-complete holographic gravitational theories. 

As a final application of our results we can also consider scale-separated vacua of string or M-theory, see \cite{Coudarchet:2023mfs} for a recent review and an extended bibliography. These are putative consistent AdS$_4$ vacua for which the length scale associated with the internal space is much smaller than the scale $L$ that sets the 4d cosmological constant.\footnote{A precise definition of this internal length scale is not crucial for our argument. The scale can be defined through the volume, diameter, or the spectrum of some differential operator on the internal space.} Clearly such a scale-separated AdS$_4$ vacuum defines an effective gravitational theory with finitely many fields valid up to the KK scale set by the internal manifold. If we assume that the scale-separated vacuum is associated with a consistent dual 3d CFT that does not have $\log N$ or $\log \lambda$ terms in local correlations functions, we arrive at very strong consistency constraints. In particular, we find that the mass spectrum of quadratic excitations around the AdS$_4$ vacuum should be such that three of the four heat kernel coefficients that contribute to the bulk logarithmic terms should vanish. These constraints are particularly strong for AdS$_4$ vacua preserving $\mathcal{N}=1$ or more supersymmetry, for which we show that most mass spectra with finitely many fields are inconsistent.

We continue in the next section with a summary of known QFT results for the $\log N$ terms in the large $N$ expansion of partition functions of 3d holographic SCFTs on compact Euclidean manifolds. In Section~\ref{sec:sugra} we review the heat kernel method and apply it to the calculation of logarithmic corrections in the saddle-point approximation of gravitational path integrals in asymptotically AdS$_4$ backgrounds. In Section~\ref{sec:bootstrap} we show how one can use the results for the logarithmic corrections of SCFT partition functions from Section~\ref{sec:SCFT} in conjunction with holography to deduce strong constraints on the Seeley-de Witt coefficients in the heat kernel expansion. Section~\ref{sec:KK} is devoted to a discussion of some top-down examples in AdS$_4$/CFT$_3$ arising from M-theory for which we can compute the $\log N$ corrections by summing over the infinite tower of supergravity KK modes. In Section~\ref{sec:Kundera} we change gears and discuss the important constraints implied by our results for EFTs with finitely many fields coupled to gravity in AdS$_4$, as well as the implications of these constraints for scale-separated AdS$_4$ vacua of string and M-theory. We conclude in Section~\ref{sec:ccl} with a discussion of some open problems and directions for further study. In the five appendices we collect many of the technical details needed for our analysis.

\section{Large $N$ partition functions of 3d holographic SCFTs}
\label{sec:SCFT}

Let us begin by briefly reviewing known results about the logarithmic terms in the large $N$ partition functions of 3d holographic SCFTs with $\mathcal{N}\geq2$ supersymmetry placed on compact Euclidean manifolds. Using supersymmetric localization, partition functions of 3d $\mathcal{N}=2$ SCFTs on various compact manifolds can be reduced to matrix models. Of particular interest to us are the squashed\footnote{We will be interested in a particular squashing of $S^3$ \cite{Hama:2011ea} that preserves a U(1)$\times$U(1) isometry and is parametrized by a real positive number $b$, with $b=1$ corresponding to the round sphere. We denote this manifold by $S^3_b$.}  3-sphere partition function \cite{Kapustin:2009kz,Hama:2011ea,Imamura:2011wg}, the Topologically Twisted Index (TTI) obtained by putting the theory on $S^1 \times \Sigma_\mathfrak{g}$ with a supersymmetry-preserving twist along the Riemann surface \cite{Benini:2015noa,Benini:2016hjo,Closset:2016arn,Closset:2017zgf}, and the Superconformal Index (SCI) where the 3d SCFT is put on $S^1 \times_\omega S^2$ after turning on a chemical potential $\omega$ for the angular momentum on the $S^2$ \cite{Bhattacharya:2008zy,Bhattacharya:2008bja,Kim:2009wb,Imamura:2011su}. Under favorable circumstances related to special values of the 3d background parameters, the matrix model for the $S^3_b$ partition function can be computed in closed form to all orders in the large $N$ expansion \cite{Fuji:2011km,Marino:2011eh,Mezei:2013gqa,Nosaka:2015bhf,Hatsuda:2016uqa}. The matrix models for the other supersymmetric partition functions have not yet been solved analytically, but closed-form expressions were conjectured for various $\mathcal{N}\geq2$ SCFTs based on non-trivial consistency checks with the bulk theory and numerical studies \cite{Bobev:2022jte,Bobev:2022eus,Bobev:2022wem,Bobev:2023lkx}. In all known cases, we have the large $N$ behavior
\begin{equation}
\label{logZ}
	\log Z_{\text{CFT}} = F_0 + \mathcal{C} \log N + \mathcal{O}(N^0) \, ,
\end{equation}
where $F_0$ denotes all terms that dominate over $\log N$ in the large $N$ expansion. In general, $F_0$ is a polynomial in some (possibly fractional) power of $N$ depending on the SCFT of interest, whose coefficients are functions of the parameters of the theory such as the discrete Chern-Simons (CS) level or continuous parameters like squashings, fugacities for flavor symmetries or real masses. In stark contrast, the logarithmic correction in \eqref{logZ} is \emph{universal} and independent of such parameters. We summarize a number of known results for the logarithmic coefficient $\mathcal{C}$ in Table~\ref{tab:CFT-log}, after giving some general explanations of the 3d holographic SCFTs and their observables.\\
\afterpage{
\begin{table}[H]
\centering
\renewcommand*{\arraystretch}{1.3}
\begin{tabular}{|c|c|c|c|c|}
\hline
Theory & $\mathcal{M}_3$ & log coefficient $\mathcal{C}$ & Ref. & 10/11d bulk \\
\hline\hline
\multicolumn{5}{|c|}{M2-brane theories (class I)} \\
\hline
%
%
\multirow{3}{*}{$(S^7/\mathbb{Z}_k)_{\text{free}}$ $(\dagger)$} & $S^3_b$ & $-\frac14$ & \cite{Fuji:2011km,Marino:2011eh,Nosaka:2015iiw,Bobev:2022eus} & \cmark\,\cite{Bhattacharyya:2012ye} (s.c.) \\
\cline{2-5}
& $S^1\times \Sigma_\mathfrak{g}$ & $-\frac12(1 - \mathfrak{g})$ & \cite{Liu:2017vll,Bobev:2022eus} & \cmark\,\cite{Liu:2017vbl} (s.c.) \\
\cline{2-5}
& $S^1\times_\omega S^2$ & $-\frac12$ & \cite{Bobev:2022wem} & \xmark \\
\hline
%
%
\multirow{3}{*}{$(S^7/\mathbb{Z}_{N_f})_{\text{f.p.}}$ $(\dagger)$} & $S^3_b$ & $-\frac14$ & \cite{Mezei:2013gqa,Hatsuda:2016uqa,Minahan:2021pfv,Bobev:2023lkx} & \multirow{3}{*}{\xmark} \\
\cline{2-4}
& $S^1\times \Sigma_\mathfrak{g}$ & $-\frac12(1 - \mathfrak{g})$ & \cite{Bobev:2023lkx} & \\
\cline{2-4}
& $S^1\times_\omega S^2$ & $-\frac12$ & \cite{Bobev:2022wem} & \\
\hline
%
%
\multirow{3}{*}{$N^{010}/\mathbb{Z}_k$} & $S^3_{b=1}$ & $-\frac14$ & \cite{Marino:2011eh,Bobev:2023lkx} & \xmark \\
\cline{2-5}
& $S^1\times \Sigma_\mathfrak{g}$ & $-\frac12(1 - \mathfrak{g})$ & \cite{Bobev:2023lkx} & \cmark\,\cite{PandoZayas:2020iqr} \\
\cline{2-5}
& $S^1\times_\omega S^2$ & $-\frac12$ & \cite{SCI-toappear} & \xmark \\
\hline
%
%
\multirow{2}{*}{$V^{52}/\mathbb{Z}_k$} & $S^1\times \Sigma_\mathfrak{g}$ & $-\frac12(1 - \mathfrak{g})$ & \cite{Bobev:2023lkx} & \cmark\,\cite{PandoZayas:2020iqr} \\
\cline{2-5}
& $S^1\times_\omega S^2$ & $-\frac12$ & \cite{SCI-toappear} & \xmark \\
\hline
%
%
\multirow{2}{*}{$Q^{111}/\mathbb{Z}_k$} & $S^1\times \Sigma_\mathfrak{g}$ & $-\frac12(1 - \mathfrak{g})$ & \cite{Bobev:2023lkx} & \cmark\,\cite{PandoZayas:2020iqr} \\
\cline{2-5}
& $S^1\times_\omega S^2$ & $-\frac12$ & \cite{SCI-toappear} & \xmark \\
\hline\hline
\multicolumn{5}{|c|}{M5-brane theories (class II)} \\
\hline
%
%
\multirow{3}{*}{$A_{N-1}$} & $S^3_b$ & $-\frac12$ & \cite{Gang:2014ema,Bobev:2020zov} & \xmark \\
\cline{2-5}
& $S^1 \times \Sigma_{\mathfrak{g}>1}$ & $(b_1(\mathcal{H}_3)-1)(1-\mathfrak{g})$ & \cite{Gang:2019uay,Benini:2019dyp} & \multirow{2}{*}{\cmark\,\cite{Benini:2019dyp}} \\
\cline{2-4}
& $S^1 \times_\omega S^2$ & $b_1(\mathcal{H}_3) - 1$ & \cite{Benini:2019dyp} & \\
\hline
%
%
\multirow{3}{*}{$D_{N}$} & $S^3_b$ & \multirow{3}{*}{$0$} & \multirow{3}{*}{\cite{Bobev:2020zov}} & \multirow{3}{*}{\xmark} \\
\cline{2-2}
& $S^1 \times \Sigma_{\mathfrak{g}>1}$ & & & \\
\cline{2-2}
& $S^1 \times_\omega S^2$ & & & \\
\hline\hline
\multicolumn{5}{|c|}{IIA theories (class III)} \\
\hline
%
%
\multirow{2}{*}{$\mathbb{CP}^3$ ($\dagger$)} & $S^3_b$ & $-\frac16$ & \cite{Marino:2009dp,Bobev:2022eus} & \multirow{2}{*}{\xmark} \\
\cline{2-4}
& $S^1\times \Sigma_{\mathfrak{g}}$ & $\frac23(1 - \mathfrak{g})$ & \cite{PandoZayas:2019hdb,Bobev:2022eus} & \\
\hline
%
%
$\mathbb{CP}_{\text{def}}^3$ & $S^3_{b=1}$ & $-\frac16$ & \cite{Hong:2021bsb} & \xmark \\
\hline
%
%
\multirow{3}{*}{$S^6_{\text{def}}$} & \multirow{2}{*}{$S^3_{b=1}$} & $-\frac29$ (fixed $k$) & \multirow{2}{*}{\cite{Liu:2019tuk}} & \multirow{3}{*}{\xmark} \\
\cline{3-3}
& & $-\frac16$ (`t Hooft) & & \\
\cline{2-4}
& $S^1\times \Sigma_{\mathfrak{g}}$ & $-\frac7{18}(1 - \mathfrak{g})$ & \cite{Liu:2018bac} & \\
\hline
\end{tabular}
\caption{The logarithmic coefficient in \eqref{logZ} for various 3d $\mathcal{N}\geq 2$ SCFTs. We indicate the compact manifold $\mathcal{M}_3$ on which the theory is placed and the possibility of mass deformations (${}\dagger$), along with relevant references where analytic and/or numerical computations are presented. We refer to the main text for detailed explanations and implicit restrictions of each entry. The last column indicates whether the result has been matched from a one-loop computation in 10d or 11d supergravity with the abbreviation s.c. used to indicate that the supergravity analysis is performed at the superconformal vacuum, i.e. for vanishing squashing and mass deformations.\label{tab:CFT-log}}
\end{table}
\clearpage}

The three broad classes of 3d $\mathcal{N}\geq 2$ SCFTs we consider arise from the worldvolume of M2- and M5-branes in M-theory, and from D2-branes in massive Type IIA string theory. SCFTs in the first class will be distinguished by the Sasaki-Einstein base SE$_7$ of the cone probed by a stack of $N$ M2-branes. Familiar examples in this class are the $\mathcal{N}=6$ ABJM theory~\cite{Aharony:2008ug}, for which SE$_7$ is a freely acting orbifold of the 7-sphere $(S^7/\mathbb{Z}_k)_{\text{free}}$, or the $\mathcal{N}=4$ ADHM theory or ``$N_f$ model''~\cite{Porrati:1996xi,Mezei:2013gqa,Grassi:2014vwa} where the base of the cone is a 7-sphere orbifold $(S^7/\mathbb{Z}_{N_f})_{\text{f.p.}}$ with fixed points. 

SCFTs in the second class are constructed by starting with the $\mathcal{N}=(2,0)$ theory on the worldvolume of $N$ M5-branes and placing it on a compact hyperbolic 3-fold $\mathcal{H}_3$, including a partial twist to preserve $\mathcal{N}=2$ supersymmetry in the three non-compact directions. Taking the limit vol$(\mathcal{H}_3) \rightarrow 0$ produces 3d holographic theories of class $\mathcal{R}$~\cite{Dimofte:2011ju}. In Table~\ref{tab:CFT-log}, we distinguish them by the ADE-type of the ``gauge group'' $G$ in the parent 6d theory. 

The third class of theories consists of 3d holographic theories whose gravity duals uplift to solutions of (massive) Type IIA string theory. They will be distinguished by the internal six-dimensional space. This class includes a particular limit of the ABJM theory where one takes both $N$ and $k$ large while keeping $N/k$ fixed. This implements a dimensional reduction along the Hopf fiber over $\mathbb{CP}^3$ inside $S^7/\mathbb{Z}_k$ and gives access to a regime of the theory dual to massless Type IIA string theory on an asymptotically AdS$_4 \times \mathbb{CP}^3$ background~\cite{Aharony:2008ug}. In addition, this class of theories includes the Gaiotto-Tomasiello (GT) theory~\cite{Gaiotto:2009mv} and the SCFT dual to 4d ISO(7)-gauged maximal supergravity~\cite{Guarino:2015jca} also belong in this class of theories, which uplift to massive Type IIA (mIIA) on a deformed $\mathbb{CP}^3$ and a deformed $S^6$, respectively. In these mIIA SCFTs, the CS level $k$ is controlled by the Romans mass and one can study either the regime of large $N$ and fixed $k$, or the limit of large $N$ and large $k$ with $N/k$ fixed. We call the former the fixed $k$ limit and the latter the `t Hooft limit in Table~\ref{tab:CFT-log}.\\

Once the theory is specified, it can be placed on a compact Euclidean manifold $\mathcal{M}_3$ and the corresponding supersymmetric partition functions can be computed using localization. The resulting matrix models can be studied at large $N$ and, in some cases, the logarithmic term in \eqref{logZ} can be extracted. For M2-brane theories, the squashed 3-sphere partition function ($\mathcal{M}_3 = S^3_b$), the TTI ($\mathcal{M}_3 = S^1 \times \Sigma_\mathfrak{g}$), and the SCI ($\mathcal{M}_3 = S^1 \times_\omega S^2$) have all been studied along these lines. In Table~\ref{tab:CFT-log} we collect the resulting logarithmic coefficients and give the relevant references. We note here that the coefficient of the $\log N$ term in the SCI has only been obtained in the Cardy-like limit where $\omega \rightarrow 0$, so we restrict our summary to this regime.\footnote{Since we expect the logarithmic correction to always be universal, it is likely that higher-order terms in $\omega$ will not change the results. Making this argument precise is however beyond the scope of this work.} Class $\mathcal{R}$ theories can be placed on the same three-manifolds and the logarithmic term in the large $N$ limit can be extracted using the 3d-3d correspondence~\cite{Dimofte:2014ija}. In the squashed sphere case, M5-brane results are available only in the limit $b \rightarrow 0$ and for hyperbolic three-folds with trivial $H^1(\mathcal{H}_3,\mathbb{Z})$. For the SCI, the results have again only been obtained in the Cardy-like limit. These restrictions are implicit in the relevant entries of Table~\ref{tab:CFT-log}. Finally, the logarithmic term in the round 3-sphere partition function of the GT theory has only been obtained for the two-node case, and we again implicitly restrict to this case in our summary.

SCFTs arising from M2-branes can also be deformed away from the superconformal point. For example, the ABJM theory has an SO(4)$\times$U(1) flavor symmetry and one can turn on three $\mathcal{N}=2$ supersymmetry-preserving real masses in the Cartan subgroup of this global symmetry group. The observables introduced above can be studied using localization in the deformed theory, and the logarithmic terms can again be extracted. Theories in which such deformations have been studied are denoted with a $(\dagger)$ in Table~\ref{tab:CFT-log}. We stress that the logarithmic coefficient turns out to be independent of the deformations in all known cases, in line with its universal character. To the best of our knowledge, such mass deformations have not been studied for SCFTs in the second and third class.\\

In Table~\ref{tab:CFT-log}, we also contrast the deluge of SCFT results with the drought of dual supergravity computations. As far as we are aware, there are only few bulk results available in the literature, and they have all been obtained in the eleven-dimensional low-energy effective description of M-theory. There, as mentioned in the Introduction, a one-loop calculation can be performed using the heat kernel method wherein the problem of computing the logarithmic correction reduces to a zero mode counting problem, see e.g.~\cite{Bhattacharyya:2012ye}. While this technique can be used to match with CFT results, it is limited to some simple 11d backgrounds where the spectrum of kinetic operators can be worked out in full. Typically, this requires being at the superconformal point where possible mass or squashing deformations are turned off. To emphasize this point, we include the abbreviation ``s.c.'' in the relevant entries of Table~\ref{tab:CFT-log}. A complementary, although broader, approach to logarithmic corrections in the bulk was put forward in~\cite{Hristov:2021zai} where it was shown how supergravity localization can efficiently compute the coefficient $\mathcal{C}$ in the 4d effective supergravity theory using index theorems. We will discuss this method in more details below.

The rest of the paper will be devoted to studying logarithmic corrections in the lower-dimensional 4d supergravity theories, with an eye towards replacing some of the ``\xmark'' entries in Table~\ref{tab:CFT-log} with bulk results matching the SCFT predictions. To this end, we set up the 4d Euclidean supergravity path integral and the appropriate heat kernel expansion in the next section.

\section{Logarithms in the 4d Euclidean path integral}
\label{sec:sugra}

The SCFTs we consider in Section \ref{sec:SCFT} are holographically dual in the large $N$ limit to ten- or eleven-dimensional supergravity theories on backgrounds that are asymptotically of the form AdS$_4 \times X$, where $X$ is a 6d or 7d internal manifold. We denote the (common) length scale of these spaces by $L$.\footnote{Here we focus on situations arising from standard top-down AdS/CFT examples in string and M-theory. We will discuss scale-separated AdS$_4$ vacua and more general 4d supergravity theories in Section~\ref{sec:Kundera}.} The KK reduction of the 10d/11d supergravity theory on $X$ produces an infinite tower of massive 4d fields in addition to the massless ones. The Euclidean path integral for the 4d KK supergravity theory involving all these 4d fields is then holographically dual to the 3d SCFT partition function \eqref{logZ}, 
\begin{equation}
	Z_\text{sugra}\approx Z_\text{CFT} \, ,
\end{equation}
where the approximate equality reminds us that we work in the large $N$ limit and are ignoring heavy string and brane states. In the holographic context, specifying a 3d SCFT amounts to choosing a particular KK supergravity theory and the corresponding field content in the bulk. Once the theory is specified, one can study various supersymmetric observables and their bulk incarnations. In particular, the partition function of a given 3d SCFT on a compact Euclidean manifold $\mathcal{M}_3$ is captured by the corresponding KK supergravity Euclidean path integral around a 4d supergravity background whose conformal boundary is $\mathcal{M}_3$. 

In the semi-classical limit where the Newton constant is small in units of the AdS radius, i.e.~$L^2/G_N \gg 1$, the 4d Euclidean path integral can be evaluated in the saddle-point approximation. This yields
\begin{equation}
\label{eq:sugra-saddle}
	\log Z_{\text{sugra}} = - \frac{1}{16\pi G_N}\,S_{\text{cl}}[\mathring{\phi}] + C\log(L/\sqrt{G_N}) + \mO(1)\, ,
\end{equation}
where $\mathring{\phi}$ generically denotes the on-shell values of all fields $\phi$ in the gravity theory for a given background. To leading order (LO) at large $L^2/G_N$, the saddle-point approximation of the supergravity path integral is controlled by the two-derivative on-shell action of the given 4d background. The next-to-leading order (NLO) term in the semi-classical expansion is a four-derivative correction to the action evaluated on the two-derivative solution. Both LO and NLO terms are grouped in the $S_{\text{cl}}[\mathring{\phi}]/G_N$ term in \eqref{eq:sugra-saddle}. This quantity has been computed and shown to match with $F_0$ in \eqref{logZ} in numerous examples, see~\cite{Bobev:2021oku,Bobev:2020zov,Bobev:2022eus,Bobev:2023lkx}, in line with expectations from AdS/CFT away from the strict semi-classical limit.

We now would like to focus on the logarithmic term and ask if the values of the logarithmic coefficient $\mathcal{C}$ in the large $N$ expansion of a given 3d SCFT partition function match the logarithmic coefficient $C$ in the saddle-point approximation \eqref{eq:sugra-saddle}. In this comparison, we will ultimately have to use the AdS$_4$/CFT$_3$ dictionary appropriate for each class of SCFTs introduced in Section~\ref{sec:SCFT} (as explained there, we split class III according to whether the bulk solution admits a IIA or massive IIA uplift):
\begin{equation}
	\fft{L^2}{G_N} \sim N^{\fft32} \;\; (\text{class I}) \, , \qquad \fft{L^2}{G_N} \sim N^{3} \;\; (\text{class II}) \, , \qquad \fft{L^2}{G_N} \sim \begin{cases} N^2 \;\; &(\text{IIA}) \\ N^{\fft53} \;\; &(\text{mIIA}) \end{cases} \, . \label{eq:holo-dict-gen}
\end{equation}
Since we are concerned with a logarithmic term we do not have to worry about the precise numerical factors or possible finite $N$ corrections to the above relations, as these would only contribute $\mathcal{O}(1)$ terms in \eqref{eq:sugra-saddle}. We will use heat kernel techniques to compute $C$. In Section \ref{sec:sugra:log} we first review how the logarithmic coefficient splits into a local and a non-local contribution based on the heat kernel expansion. In Section~\ref{sec:sugra:local} and Section~\ref{sec:sugra:non-local}, we study these contributions in detail.

\subsection{Logarithmic contributions to the Euclidean path integral}
\label{sec:sugra:log}

In this subsection, we review the general structure of the logarithmic coefficient $C$ in the Euclidean path integral for any 4d KK supergravity theory. Note that all supergravity fields with masses below the UV cutoff can run in loops in the Euclidean path integral and thereby contribute a logarithmic term in the bulk partition function. Up to $L$-independent terms, the resulting logarithmic correction generically takes the form \cite{Sen:2012kpz,Sen:2012cj,Sen:2012dw,Bhattacharyya:2012ye}
\begin{equation}
\label{eq:log-sugra}
	C\log(L/\sqrt{G_N}) = \sum_{\phi}\Bigg[-\frac{(-1)^F}{2}\log\text{det}'\mathcal{Q}_\phi + \log\int\mathcal{D}\dphi_0\Bigg] \, ,
\end{equation}
where $\mathcal{Q}_\phi$ is a second-order differential operator that captures the dynamics of a quantum fluctuation $\dphi = \phi - \mathring{\phi}$ around a given background, and the sum is taken over all 4d fields $\phi$ in the KK theory weighted by their fermion number $F$, including ghosts.\footnote{Quantization of a $p$-form requires a tower of $(p-j)$-form ghost fields with $j=1,\ldots, p$ labeling the ghost level~\cite{Siegel:1980jj}. For the purpose of computing the determinant factor, the kinetic terms of these ghost fields can be thought of as differential operators of order $2(j+1)$ and therefore the first term in \eqref{eq:log-sugra} has to be multiplied by a factor of $j+1$, see e.g. \cite{Bhattacharyya:2012ye}. This factor can be understood as arising from ghost number conservation, and the same remark applies for any field with gauge invariance. To  declutter formulas, we will mostly keep the ghost level factor implicit when summing over the spectrum of the theory.\label{foot:ghosts}} The prime on the determinant indicates that it should be computed after removal of the zero modes, which are modes that satisfy
\begin{equation}
\label{eq:ZM-def}
	\mathcal{Q}_\phi\dphi_0 = 0 \, .
\end{equation}
The second term in \eqref{eq:log-sugra} encodes the contribution to the path integral from these zero modes separately. In practice, we expect the spectrum of the $\mathcal{Q}_\phi$ differential operators to be well-behaved, and in particular that the eigenvalues satisfy certain ordering and positivity properties with only a finite number of them being zero. On non-compact spaces with a boundary, such properties depend on the boundary conditions imposed on the fluctuations~$\delta\phi$ and we will discuss some of these aspects in due course. One immediate consequence of this ``good'' spectral behavior is that only massless fluctuations can satisfy~\eqref{eq:ZM-def}, since any positive mass-squared term in~$\mathcal{Q}_\phi$ would lift the $\delta\phi_0$ zero-mode.

To evaluate the non-zero mode contribution to the logarithmic correction \eqref{eq:log-sugra}, it is convenient to introduce the heat kernel associated to the operator $\mathcal{Q}_\phi$ as \cite{Vassilevich:2003xt}
\begin{equation}
\label{eq:K}
	K(x,y;t;\mathcal{Q}_\phi) = \langle x| e^{-t\mathcal{Q}_\phi}|y \rangle \, .
\end{equation}
At coincident space-time points, this can be expanded in the small $t$ limit as
\begin{equation}
	K(x,x;t;\mathcal{Q}_\phi) = \sum_{k=0}^{\infty} a_{2k}(x;\mathcal{Q}_\phi)\,t^{k-2}\,,
\end{equation}
in terms of the Seeley-de Witt (SdW) coefficients $a_{2k}$. Using \eqref{eq:K}, we write the determinant factor in \eqref{eq:log-sugra} as an integral \cite{Sen:2012kpz,Sen:2012cj,Sen:2012dw,Bhattacharyya:2012ye}
\begin{equation}
\label{eq:one-loop-det}
	\log\text{det}'\mathcal{Q}_\phi = -\lim_{\epsilon \to 0}\int_{\epsilon}^\infty \frac{dt}{t}\,\left[\int d^4x \sqrt{g}\,K(x,x;t;\mathcal{Q}_\phi) - n_{\phi_0}\right] \, ,
\end{equation}
where we have explictly removed the $n_{\phi_0}$ zero modes of $\mathcal{Q}_\phi$ and $\epsilon$ is a UV cutoff. By a scaling argument, one can show that the one-loop determinant~\eqref{eq:one-loop-det} produces a logarithmic correction to $\log Z_{\text{sugra}}$ in the large $L^2/G_N$ limit that is entirely controlled by the $k=2$ term in the SdW expansion~\cite{Sen:2012dw,Bhattacharyya:2012ye}:
\begin{equation}
\label{eq:log-sugra:1}
	\log\text{det}'\mathcal{Q}_\phi=-2\Bigg[\int d^4x \sqrt{g}\,a_4(x;\mathcal{Q}_\phi)-n_{\phi_0}\Bigg]\log(L/\sqrt{G_N}) + \mO(1) \, .
\end{equation}

On the other hand, the path integral over the zero modes of the $\mathcal{Q}_\phi$ operator produces a logarithmic correction to \eqref{eq:log-sugra} that can be written as \cite{Sen:2012kpz,Sen:2012cj,Sen:2012dw,Bhattacharyya:2012ye}
\begin{equation}
\label{eq:log-sugra:2}
	\log\int\mathcal{D}\dphi_0 = (-1)^F\beta_{\phi_0}n_{\phi_0}\log(L/\sqrt{G_N}) + \mO(1) \, ,
\end{equation}
where $\beta_{\phi_0}$ is a pure number fixed by demanding locality of the path integral measure $\mathcal{D}\delta\phi_0$. For future reference, we record here the results for fields of spin $3/2$, spin $2$, and for $p$-forms in four dimensions, see \cite{Sen:2012cj,Bhattacharyya:2012ye}:
\begin{equation}
\label{eq:beta}
	\qquad \beta_{3/2} = 3 \, , \qquad \beta_2 = 2 \, , \qquad \beta_{A_p} = \frac{4-2p}{2} \, .
\end{equation}

Together, the non-zero mode contribution \eqref{eq:log-sugra:1} and the zero mode contribution \eqref{eq:log-sugra:2} yield the full logarithmic correction to the 4d Euclidean path integral. Accordingly, we split the resulting coefficient $C$ in \eqref{eq:log-sugra} into two parts,
\begin{equation}
\label{eq:C}
	C = C_{\text{local}} + C_{\text{non-local}} \, ,
\end{equation}
where the local contribution is controlled by the fourth SdW coefficient, 
\begin{equation}
\label{eq:C-local}
	C_{\text{local}} = \sum_{\phi} (-1)^F\int d^4x \sqrt{g}\,a_4(x,\mathcal{Q}_\phi) \, ,
\end{equation}
and the non-local contribution captures the effect of zero modes,
\begin{equation}
\label{eq:C-non-local}
	C_{\text{non-local}} = \sum_{\text{massless }\phi}(-1)^F n_{\phi_0} (\beta_{\phi_0} - j - 1) \, .
\end{equation}
Here the sum is over the massless spectrum, and we have explicitly displayed the ghost level factor discussed in Footnote \ref{foot:ghosts} since it will be important later. In the following, we study these contributions in turn.

\subsection{Local contributions}
\label{sec:sugra:local}

We now review some important features of the SdW coefficient $a_4$ that controls the local contribution (\ref{eq:C-local}). For this discussion, we assume that the second-order differential operator $\mathcal{Q}$ is of Laplace type.\footnote{We temporarily drop the subscript $\phi$ on $\mathcal{Q}_\phi$ since the analysis is valid for generic fields.} This means that it can be represented locally as
\begin{equation}
	\mathcal{Q} = \mD^\mu\mD_\mu + 2\,\omega^\mu\mD_\mu + P \, ,
\end{equation}
where $\mathcal{D}_\mu$ is the covariant derivative with respect to all bosonic local gauge transformations of the theory under consideration\footnote{In particular, $\mathcal{D}_\mu$ includes background gauge fields. We use the symbol $\nabla_\mu$ to denote the spacetime covariant derivative.} and $(\omega_\mu,P)$ are a set of matrices acting in field space. Introducing the differential operator $D_\mu = \mathcal{D}_\mu + \omega_\mu$, we complete the square and write
\begin{equation}
\label{eq:Laplace-op}
	\mathcal{Q} = D^\mu D_\mu + E \, ,
\end{equation}
where $E = P - \omega^\mu\omega_\mu - \mathcal{D}^\mu\omega_\mu$. We stress that the matrix multiplication in field space is implicit in our notation. The curvature associated to $D_\mu$ is denoted by $\Omega_{\mu\nu} = [D_\mu,D_\nu]$.

In general, the SdW coefficients associated to $\mathcal{Q}$ can be written in terms of traces of the matrix-valued quantities $E$ and $\Omega_{\mu\nu}$, together with curvature tensors of the background geometry. If we consider a general 4d Riemannian manifold $(\mathcal{M},g)$ with a non-empty boundary $\partial\mathcal{M}$, the formula for the fourth SdW coefficient of the Laplace operator~\eqref{eq:Laplace-op} integrated over $\mathcal{M}$ reads \cite{Vassilevich:2003xt}\footnote{Note that this reference uses the inward unit vector normal to the boundary. We will use the outward one, and this difference is reflected in various signs in the last term of~\eqref{eq:a4}.} 
\begin{equation}
\label{eq:a4}
	\begin{split}
		\int_{\mathcal{M}} d^4x \sqrt{g}\,a_4(\mathcal{Q}) =&\; \int_{\mathcal{M}} d^4x \sqrt{g}\,a_4^{\text{bulk}}(\mathcal{Q}) + \int_{\mathcal{M}} d^4x \sqrt{g}\,a_4^{\text{tot.der.}}(\mathcal{Q}) \\
		&+ \int_{\partial\mathcal{M}}d^3y\sqrt{\gamma}\,a_4^{\text{bdry}}(\mathcal{Q}) \, ,
	\end{split}
\end{equation}
where $\gamma_{ab}$ is an induced metric on the boundary $\partial\mathcal{M}$ and we define
\begin{align}
	(4\pi)^2a_4^{\text{bulk}}(\mathcal{Q}) =&\; \mathrm{Tr}\bigg[\frac12 E^2 + \frac16 R E + \frac1{12}\Omega^{\mu\nu}\Omega_{\mu\nu} + \frac{1}{360}\bigl(3W^2 - E_4 + 5R^2\bigr)\bigg] \, , \label{eq:a4-details:bulk} \\[1mm]
	(4\pi)^2a_4^{\text{tot.der.}}(\mathcal{Q}) =&\; \mathrm{Tr}\bigg[\frac16\square E + \frac1{30}\square R\bigg] \, , \label{eq:a4-details:tot}
\end{align}
\begin{align}
\label{eq:a4-details:bdry}
	(4\pi)^2a_4^{\text{bdry}}(\mathcal{Q}) =&\; \frac{1}{360}\mathrm{Tr}\bigg[(120\Pi_- - 240\Pi_+)\nabla_n E + (18\Pi_- - 42\Pi_+)\nabla_n R + 24 \widetilde{\nabla}^a\widetilde{\nabla}_aK \nonumber \\
	&\qquad\quad + 120 E K + 20 R K + 4 R_{an}{}^{an} K - 12 R_{anb}{}^n K^{ab} + 4 R_{abc}{}^b K^{ac} \nonumber \\
	&\qquad\quad + \frac1{21}\Big\{(280\Pi_+ + 40\Pi_-)K^3 + (168\Pi_+ - 264\Pi_-)K^{ab}K_{ab}K \nonumber \\
	&\qquad\quad + (224\Pi_+ + 320\Pi_-)K^{ab}K_{bc}K_a{}^c\Big\} \\
	&\qquad\quad - 720 S E - 120 S R - 144 S K^2 - 48 S K^{ab} K_{ab} + 480 S^2 K \nonumber \\[1.5mm]
	&\qquad\quad - 480 S^3 - 120\widetilde{\nabla}^a\widetilde{\nabla}_aS - 60\Omega_{an}\mP\widetilde{\nabla}^a\mP \nonumber \\
	&\qquad\quad + 12(10S - K)\widetilde{\nabla}^a\mP\widetilde{\nabla}_a\mP - 24K^{ab}\widetilde{\nabla}_a\mP\widetilde{\nabla}_b\mP\bigg] \, . \nonumber
\end{align}
Here we have used $i,j,k\in\{1,2,3,4\}$ and $a,b,c\in\{1,2,3\}$ for the local orthonormal frame indices of the tangent bundles on $\mathcal{M}$ and $\partial\mathcal{M}$, respectively. From here on we choose a bulk local orthonormal frame appropriately so that the 4d $i,j,k$ indices become identical to the union of 3d boundary $a,b,c$ indices and the normal vector index $n=4$ on the boundary $\partial\mathcal{M}$. Then the extrinsic curvature of the boundary can be written explicitly as $K_{ab} = -\Gamma^n_{ab}$. The trace of the extrinsic curvature is $K=K^a{}_a$. The boundary covariant derivative $\widetilde{\nabla}$ is distinguished from the bulk one $\nabla$, and the box symbol is defined with respect to the latter as $\square=\nabla^\mu\nabla_\mu$. The quantities $\Pi_\pm$ and $S$ specify the boundary conditions used for the fields on which the differential operator $\mathcal{Q}$ acts, and $\mP=\Pi_+-\Pi_-$. These boundary conditions will be discussed in more detail below. Finally, the trace Tr[\ldots] is taken in field space and over all free indices carried by the $E$ and $\Omega$ matrices, and we have introduced the usual curvature combinations
\begin{equation}
\label{eq:W2}
	\begin{split}
		W^2 =&\; R^{\mu\nu\rho\sigma}R_{\mu\nu\rho\sigma} - 2\,R^{\mu\nu}R_{\mu\nu} + \frac13\,R^2 \, , \\
		E_4 =&\;  R^{\mu\nu\rho\sigma}R_{\mu\nu\rho\sigma} - 4\,R^{\mu\nu}R_{\mu\nu} + R^2 \, .
	\end{split}
\end{equation}
The general structure of~\eqref{eq:a4} is that of a bulk contribution involving four-derivative terms, a total derivative contribution, and an intrinsic boundary contribution. We discuss each of them below.

\subsubsection{Bulk contributions}
\label{sec:sugra:local:bulk}

First we study the bulk contribution \eqref{eq:a4-details:bulk}. For simplicity, from here on we focus on the case where the background $\mathring{\phi}$ in the saddle-point approximation \eqref{eq:sugra-saddle} is a solution of Euclidean $\mN=2$ minimal gauged supergravity with bosonic action
\begin{equation}
\label{eq:EMaction}
	S_\text{minimal}=-\frac{1}{16\pi G_N}\int_{\mathcal{M}} d^4x\,\sqrt{g}\,\bigg[R-2\Lambda-F_{\mu\nu}F^{\mu\nu}\bigg]\,,\qquad \Lambda=-\fft{3}{L^2}\,,
\end{equation}
where the corresponding equations of motion read
\begin{equation}
\label{eom}
	\begin{split}
	R_{\mu\nu}-\fft12g_{\mu\nu}(R-2\Lambda)&=2F_{\mu\rho}F_\nu{}^\rho-\fft12g_{\mu\nu}F_{\rho\sigma}F^{\rho\sigma}\,,\\
	\nabla_\mu F^{\mu\nu}&=0\,.
	\end{split}
\end{equation}
In this case, the bulk contribution \eqref{eq:a4-details:bulk} can be conveniently rewritten using the background equations of motion \eqref{eom} and field redefinitions as \cite{David:2021eoq}
\begin{equation}
\label{eq:a4-bulk}
	(4\pi)^2 a_4^{\text{bulk}}(\mathcal{Q}) = -a_E\,E_4 + c\,W^2 + b_1\,R^2 + b_2\,R\,F^{\mu\nu}F_{\mu\nu} \, ,
\end{equation}
where $(a_E,c,b_1,b_2)$ are a set of coefficients that can be obtained from trace computations. These heat kernel coefficients will be central to our analysis since they govern the bulk part of the SdW contribution to the $C_{\text{local}}$ term. \\

For illustration, consider the quantum fluctuations of a massive neutral scalar (MNS) field around a given background. The relevant second-order operator $\mathcal{Q}_{\text{MNS}} = \square - m^2$ capturing the dynamics of these fluctuations is of Laplace type \eqref{eq:Laplace-op} with
\begin{equation}
	E=-m^2\,,\qquad \Omega_{\mu\nu}=0\,.
\end{equation}
Using this in \eqref{eq:a4-details:bulk} and taking the trace, we obtain the bulk contribution
\begin{equation}
	\text{MNS:} \qquad (4\pi)^2 a_4^{\text{bulk}} = \frac{m^4}{2} - \frac16\,R\,m^2 + \frac{1}{360}\bigl(3W^2 - E_4 + 5R^2\bigr) \, .
\end{equation}
This can be brought into the form~\eqref{eq:a4-bulk} by making use of the trace of the background equation of motion $R=-12/L^2$, which yields the coefficients
\begin{equation}
	\text{MNS:} \qquad a_E = \frac{1}{360} \, , \quad c = \frac{1}{120} \, , \quad b_1 = \frac{1}{288}\bigl((mL)^2 + 2\bigr)^2 \, , \quad b_2 = 0 \, .
\end{equation}
This simple example illustrates how obtaining the bulk contribution to the fourth SdW coefficient for a field $\phi$ amounts to identifying the appropriate second-order $\mathcal{Q}$ operator governing its fluctuations, extracting the $E$ and $\Omega$ matrices, and collecting the $(a_E,c,b_1,b_2)$ coefficients by computing traces and making use of the background equations of motion.

In Appendix~\ref{App:coeffs} we implement this method for both massless and massive fields of spin $0 \leq s \leq 2$. For scalars and fermions, we consider quadratic fluctuations of minimally coupled fields around a generic background satisfying the equations of motion \eqref{eom}. For fields of spin $1 \leq s \leq 2$ we opted to turn off the background Maxwell field for simplicity. In this way, we have obtained the coefficients $(a_E,c,b_1)$ in~\eqref{eq:a4-bulk} for all fields minimally coupled to an Einstein-Maxwell background, while we only have access to the $b_2$ coefficient for scalars and fermions. The results of these lengthy computations are summarized in Table~\ref{tab:coeffs}.
\begin{table}[h]
	\centering
	\renewcommand*{\arraystretch}{1.4}
	\begin{tabular}{|c|c|c|c|c|}
		\hline
		spin & mass & $a_E$ & $c$ & $b_1$ \\
		\hline\hline
		0 & $(mL)^2 = -2$ & $\frac{1}{360}$ & $\frac{1}{120}$ & 0 \\
		\hline
		0 & $m$ & $\frac{1}{360}$ & $\frac{1}{120}$ & $\frac{1}{288}\bigl((mL)^2 + 2\bigr)^2$ \\
		\hline\hline
		1/2 & 0 & $-\frac{11}{720}$ & $-\frac{1}{40}$ & 0 \\
		\hline
		1/2 & $m$ & $-\frac{11}{720}$ & $-\frac{1}{40}$ & $\frac{1}{144}(mL)^2\bigl((mL)^2 - 2\bigr)$ \\
		\hline\hline
		1 & 0 & $\frac{31}{180}$ & $\frac{1}{10}$ & 0 \\
		\hline
		1 & $m$ & $\frac{31}{180} + \frac{1}{360}$ & $\frac{1}{10} + \frac{1}{120}$ & $\frac{1}{288}\bigl(3(mL)^4 - 12(mL)^2 + 4\bigr)$ \\
		\hline\hline
		3/2 & $mL=1$ & $\frac{589}{720}$ & $\frac{137}{120}$ & 0 \\
		\hline
		3/2 & $m$ & $\frac{589}{720} - \frac{11}{720}$ & $\frac{137}{120} - \frac{1}{40}$ & $\frac{1}{72}\bigl((mL)^4 - 8(mL)^2 + 11\bigr)$ \\
		\hline\hline
		2 & 0 & $\frac{571}{180}$ & $\frac{87}{20}$ & 0 \\
		\hline
		2 & $m$ & $\frac{571}{180} + \frac{31}{180} + \frac{1}{360}$ & $\frac{87}{20} + \frac{1}{10} + \frac{1}{120}$ & $\frac{5}{288}\bigl((mL)^4 - 8(mL)^2 + 8\bigr)$ \\
		\hline
	\end{tabular}
	\caption{The quantities controlling the bulk contribution to the SdW coefficient~\eqref{eq:a4-bulk}. For massive fluctuations with gauge invariance, we indicate the effect of adding the appropriate Stückelberg fields as a separate contribution to $a_E$ and $c$. \label{tab:coeffs}}
\end{table}

Let us collect some important remarks on the results. For fields of spins $1\leq s \leq 2$, the operator $\mathcal{Q}$ can be brought to Laplace form \eqref{eq:Laplace-op} only after imposing gauge-fixing conditions inside the path integral. For massless fields, this requires the addition of appropriate ghost fields, as we review in Appendix~\ref{App:coeffs}. Our results for the corresponding $(a_E,c,b_1,b_2)$ coefficients are compatible with previous derivations in the literature, see e.g. \cite{David:2021eoq,Karan:2022dfy}. For massive fields, the mass term typically breaks gauge invariance and to remedy this we must introduce appropriate Stückelberg fields \cite{Stueckelberg:1938zz,Stueckelberg:1938hvi}, see \cite{Ruegg:2003ps} for a modern review. It is important to note that these fields are physical, and while they do not modify the classical equations of motion, they are allowed to run in loops. Therefore, they generically give non-trivial contributions to the heat kernel coefficients. To draw attention to this point, we explicitly separate the contribution from the Stückelberg fields to the $(a_E,c)$ coefficients in Table~\ref{tab:coeffs}. This also makes it clear that the massless case is not obtained as a limit of the massive one, which is a manifestation of the fact that the Stückelberg fields do not decouple at the one-loop level. Introducing the appropriate Stückelberg fields is therefore crucial to obtain the correct coefficients in~\eqref{eq:a4-bulk} for massive fields.

It is also worth mentioning that, because we study fields on an asymptotically AdS$_4$ space, by massless we mean $m^2 = -2/L^2$ for a scalar field and $m = 1/L$ for a gravitino field. This nomenclature takes into account the conformal coupling in the Lagrangian which, in the case of a (pseudo) scalar field for example, reads $\frac16 R \phi^2 = -2L^{-2}\phi^2$ on an asymptotically AdS background. Indeed, such a conformally coupled scalar field sits in the massless supermultiplet of a conserved current.

%
\begin{table}[t]
\centering
\renewcommand*{\arraystretch}{1.4}
\begin{tabular}{|c|c|}
\hline
fields & mass \\
\hline
$s=0,2$ & $(mL)^2 = \Delta(\Delta - 3)$ \\
\hline
$p$-form & $(mL)^2 = (\Delta - p)(\Delta + p - 3)$ \\
\hline
$s=\frac12,\frac32$ & $|mL| = \Delta - \frac32$ \\
\hline
\end{tabular}
\caption{Relation between 4d mass and 3d conformal dimension for various bulk fields.\label{tab:conf-m}}
\end{table}

Finally, the four-derivative quantities $(E_4,W^2,R^2)$ are not all linearly independent on an Einstein background for which $R_{\mu\nu} - \frac12 g_{\mu\nu}(R - 2\Lambda) = 0$. After turning off the background Maxwell field, the linear constraint implied by the Einstein equations \eqref{eom} is
\begin{equation}
	\text{Einstein background:} \qquad E_4-W^2-\fft16R^2 = 0 \, ,\label{Einstein:constraint}
\end{equation}
which translates to an ambiguity in the coefficients $(a_E,c,b_1)$ entering \eqref{eq:a4-bulk}. In Table~\ref{tab:coeffs}, this freedom has been fixed by demanding that $b_1=0$ for massless fields.

\subsubsection{Boundary contributions}
\label{sec:sugra:local:bdry}

For the total derivative contribution \eqref{eq:a4-details:tot}, the fact that $\mathrm{Tr}[E]$ is a two-derivative quantity shows that we can use the background equations of motion~\eqref{eom} together with Stoke's theorem to bring it to the form~\cite{David:2021eoq}
\begin{equation}
\label{eq:a4-tot-der}
	(4\pi)^2 \int_{\mathcal{M}} d^4x\sqrt{g}\,a_4^{\text{tot.der.}}(\mathcal{Q}) = \int_{\partial \mathcal{M}} d^3y\sqrt{\gamma}\,n^\mu \nabla_\mu\Bigl[\alpha_1R + \alpha_2F_{\mu\nu}F^{\mu\nu}\Bigr] \, ,
\end{equation}
where $n^\mu$ denotes the outward unit vector normal to the boundary and $\alpha_{1,2}$ are constants whose values depend on the field under consideration. We can therefore view the total derivative contribution to the integrated SdW coefficient as a surface term.\\

The other surface term in \eqref{eq:a4} depends explicitly on the choice of boundary conditions for each field fluctuation $\delta\phi$. In writing \eqref{eq:a4-details:bdry}, we have implicitly assumed that we can choose so-called ``mixed'' boundary conditions~\cite{Vassilevich:2003xt} for all fields. Such boundary conditions are parametrized as
\begin{equation}
\label{eq:bc-param}
	\Pi_-\delta\phi|_{\partial\mathcal{M}} = 0 \, , \qquad (\nabla_n + S)\Pi_+\delta\phi|_{\partial\mathcal{M}} = 0 \, .
\end{equation}
Within this class, various choices are avalaible for fields in Euclidean gravity. We will further restrict ourselves to boundary conditions that ensure that the differential operator $\mathcal{Q}_\phi$ is elliptic. Roughly speaking, this guarantees that the operator has only finitely-many zero modes so that we can define its determinant in the usual way, as we have indeed assumed we could do in writing~\eqref{eq:log-sugra}. For a nice review of such boundary conditions we refer the reader to~\cite{Witten:2018lgb}, which we follow.

Dirichlet and Neumann boundary conditions on the scalar Laplacian are elliptic. The former amount to choosing $\Pi_- = 1$ and $\Pi_+ = S = 0$ in~\eqref{eq:bc-param}, while the latter are achieved for $\Pi_- = S = 0$ and $\Pi_+ = 1$. To compute the one-loop determinant for a spinor field $\psi$, it turns out to be convenient to use the square of the Dirac operator for the $\mathcal{Q}_\psi$ operator, as we explain in Appendix~\ref{App:coeffs}. A choice of elliptic (and hermitian) boundary conditions for this operator is given by $\Pi_- = \frac12(1 + \mathrm{i}\gamma^n\gamma^5)$, $\Pi_+ = 1 - \Pi_-$ and $S = \frac12 K \Pi_+$, see~\cite{Vassilevich:2003xt}. For Yang-Mills fields, we will use so-called relative boundary conditions where the fluctuations $\delta A_\mu = A_\mu - \mathring{A}_\mu$ satisfy \cite{Vassilevich:2003xt}
\begin{equation}
\label{eq:A-bc-rel}
	(\delta A_a)|_{\partial\mathcal{M}} = 0 \, ,
\end{equation}
where the Latin index $a$ denotes the local orthonormal frame indices of the tangent bundle on the boundary introduced below (\ref{eq:a4-details:bdry}). By BRST invariance, this condition implies that the ghost field $c$ required for quantization vanishes at the boundary, $c|_{\partial\mathcal{M}} = 0$. The same must be true for the anti-ghost field $b$ in order to have a well-defined propagator for the $b$-$c$ system. Invoking BRST again,~\eqref{eq:A-bc-rel} then implies that $\mathcal{G}(\delta A)|_{\partial\mathcal{M}} = 0$ where $\mathcal{G}$ is the gauge-fixing function for the fluctuations. This set of boundary conditions is elliptic, see~\cite{Witten:2018lgb}. Working in the Lorenz gauge as in Appendix~\ref{App:coeffs}, it corresponds to
\begin{equation}
	(\Pi_-)_{ij} = \delta_{ij} - \delta_{in}\delta_{jn}\, , \quad (\Pi_+)_{ij} = \delta_{in}\delta_{jn} \, , \quad S_{ij} = K\delta_{in}\delta_{jn} \, , 
\end{equation} 
in the parameterization~\eqref{eq:bc-param}. Elliptic boundary conditions for spin-3/2 fluctuations can be obtained by using an appropriate combination of elliptic Yang-Mills and spinor boundary projectors $\Pi_\pm$ and $S$, and are also of mixed type.
 
The boundary problem for metric fluctuations is notoriously more involved. While it was shown in~\cite{Anderson:2006lqb} that there exists a set of elliptic boundary conditions, they involve tangential derivatives and therefore cannot be cast in the form~\eqref{eq:bc-param}. Instead, they fall into the more general class of ``oblique'' boundary conditions for which much less is known regarding the general form of $a_4^{\text{bdry}}$ in~\eqref{eq:a4}, see~\cite{Vassilevich:2003xt} for a discussion. To show this explicitly, let us recall that the conformal boundary conditions of~\cite{Anderson:2006lqb} are obtained by fixing the conformal structure of the boundary and requiring that the metric fluctuations preserve the trace of the extrinsic curvature. Writing the perturbed metric as $g_{\mu\nu} = \mathring{g}_{\mu\nu} + h_{\mu\nu}$, the first condition amounts to demanding that the traceless part of $h_{\alpha\beta}|_{\partial\mathcal{M}}$ vanishes, where $\alpha,\beta,\ldots$ denote spacetime indices tangent to the boundary. Just as in the Yang-Mills case, BRST invariance then implies that $\mathcal{G}_\mu(h)|_{\partial\mathcal{M}} = 0$, which gives two additional boundary conditions when split along $\mu =\,\perp$ and $\mu = \alpha$, see also~\cite{Anninos:2023epi}. The extrinsic trace condition gives us one last equation that the fluctuations should satisfy. For concreteness, we will work in the harmonic (de Donder) gauge for the metric perturbations, and use Gaussian normal coordinates for the background so that $\mathring{g}_{\perp\perp} = 1$ and $\mathring{g}_{\perp\alpha} = 0$. Then, the full set of elliptic metric boundary conditions can be summarized as
\begin{equation}
\label{eq:conf-bc}
	\begin{split}
	&\Bigl[h_{\alpha\beta} - \frac13\,\mathring{g}_{\alpha\beta}h^\gamma{}_\gamma\Bigr]\Big\vert_{\partial\mathcal{M}} = 0 \, , \\
	&\Bigl[\partial_\perp h^\gamma{}_\gamma - 2\widetilde{\nabla}^\alpha h_{\alpha\perp} - K h_{\perp\perp}\Bigr]\Big\vert_{\partial\mathcal{M}} = 0 \, , \\
	&\Bigl[\partial_\perp h_{\perp\perp} + K h_{\perp\perp} - 2 K^{\alpha\beta}h_{\alpha\beta}\Bigr]\Big\vert_{\partial\mathcal{M}} = 0 \, , \\
	&\Bigl[\partial_\perp h_{\alpha\perp} + K h_{\alpha\perp} + \widetilde{\nabla}^\beta h_{\alpha\beta} - \frac12\widetilde{\nabla}_\alpha(h_{\perp\perp} + h^{\gamma}{}_\gamma)\Bigr]\Big\vert_{\partial\mathcal{M}} = 0 \, .
	\end{split}
\end{equation}
Since the tangential covariant derivatives $\widetilde{\nabla}$ cannot be eliminated from these equations, we are indeed dealing with oblique boundary conditions. As far as we are aware, the problem of finding good boundary conditions for metric fluctuations in 4d and their explicit contribution to the fourth SdW coefficient has not been solved on a general non-compact manifold with boundary. We believe this is a hard mathematical problem that falls outside the scope of the present work. In order to make progress, we will formally impose the above conformal boundary conditions in order to ensure ellipticity of the $\mathcal{Q}_h$ operator. This will induce a modification of the boundary term~\eqref{eq:a4-details:bdry},
\begin{equation}
\label{eq:a4-oblique}
	a_4^{\text{bdry}} \longrightarrow a_4^{\text{bdry}} + a_4^{\text{oblique}} \, .
\end{equation}
As we will see in Section~\ref{sec:bootstrap}, the contribution from mixed boundary conditions $a_4^{\text{bdry}}$ can always be holographically renormalized away after integrating over a cutoff hypersurface and sending the cutoff to infinity. We will assume that the same is true for the unknown quantity $a_4^{\text{oblique}}$ in the rest of this paper. This assumption ensures that the oblique nature of the boundary conditions~\eqref{eq:conf-bc} does not affect the local contribution to the logarithmic term in the Euclidean path integral.

\subsection{Non-local contributions}
\label{sec:sugra:non-local}

We now discuss the non-local contribution \eqref{eq:C-non-local}. The first step is to identify whether the Laplace-type operator $\mathcal{Q}_\phi$ admits zero modes. In general, this spectral problem for generic Einstein-Maxwell backgrounds that solve the equations of motion \eqref{eom} is quite involved. We will therefore restrict our attention to two minimal supergravity backgrounds for which we can make explicit statements. These backgrounds are pure Euclidean AdS$_4$ (EAdS$_4$) and Euclidean AdS$_2 \times \Sigma_\mathfrak{g}$ with a non-trivial background gauge field along the Riemann surface. They will serve to illustrate expected general features of the non-local contribution to the log term in the supergravity path integral.

\subsubsection{Zero-modes on EAdS$_4$}
\label{sec:sugra:non-local-AdS4}

For pure EAdS$_4$, an analysis of the zero modes of Laplace-type differential operators acting on $p$-forms, spin-1/2 and metric tensors was conducted in \cite{Camporesi:1994ga,camporesi1994plancherel,Camporesi:1995fb}. There it was shown that there are no square-integrable zero modes of the Laplacian for symmetric tensors on EAdS$_4$. Fermions also do not allow for square-integrable zero modes of the Dirac operator. Lastly for $p$-forms, there are square-integrable zero modes only in even dimension $d$ and for $p = d/2$. Hence, zero modes can only come from 2-forms in a pure EAdS$_4$ background. The number of such zero modes is given by~\cite{camporesi1994plancherel,Bhattacharyya:2012ye} 
\begin{equation}
\label{n:A2}
	n_{A_2} = \fft{3}{4\pi^2L^4}\,\int_{\mathcal M}d^4x \sqrt{g} = 1 \, ,
\end{equation}
for each 2-form field present in the spectrum. In \eqref{n:A2} the divergent volume factor of EAdS$_4$ is renormalized as
\begin{equation}
	\text{vol}(\text{EAdS}_4) = \int_{\mathcal M}d^4x \sqrt{g} = \frac{4\pi^2L^4}{3} \, ,
\end{equation}
by adding appropriate boundary counter-terms. For spin-3/2 fields, one can construct zero modes of the Rarita-Schwinger (RS) operator as follows. We start with an eigenspinor of the Dirac operator on EAdS$_4$,
\begin{equation}
\label{eq:eigenspinors}
	\slashed{\nabla}\Omega^s_{\ell m} = \frac{\mathrm{i}s}{L}\,\lambda\,\Omega^s_{\ell m} \, ,
\end{equation}
indexed by a continuous eigenvalue $\lambda$, two mode numbers $\ell\in\mathbb{N}$ and $m=1,\ldots,d_\ell$ and a sign $s=\pm$ related to chirality. Here $d_\ell$ is the dimension of the spin-$\ell$ representation of Spin(3). Now define the basis spinors
\begin{equation}
\label{eq:build-eigen-RS}
	\Psi_\mu^{(1)s} = \gamma_\mu\Omega^s \, , \qquad \Psi_\mu^{(2)s} = \nabla_\mu\Omega^s \, ,
\end{equation} 
where we momentarily suppress the mode indices. Consider the RS operator for a spin-3/2 field of mass $M$ on EAdS$_4$,
\begin{equation}
\label{eq:RS-op}
	(D_\pm)^{\mu\nu} = \gamma^{\mu\nu\rho}\nabla_\rho \pm M\gamma^{\mu\nu} \, ,
\end{equation}
where the sign factor encodes chirality. This operator acts on~\eqref{eq:build-eigen-RS} as
\begin{equation}
	\begin{split}
	(D_\pm)_\mu{}^\nu\Psi_\nu^{(1)s} =&\; \bigg(\pm3M - \frac{2\mathrm{i}s}{L}\,\lambda\bigg)\Psi_\mu^{(1)s} + 2\,\Psi_\mu^{(2)s} \, , \\
	(D_\pm)_\mu{}^\nu\Psi_\nu^{(2)s} =&\; \bigg(\frac{R}{8} \pm \frac{\mathrm{i}s}{L}\,M\,\lambda\bigg)\Psi_\mu^{(1)s} \mp M\,\Psi_\mu^{(2)s} \, ,
	\end{split}
\end{equation}
where we have used the defining relation~\eqref{eq:eigenspinors}. This shows that appropriate linear combinations of the spinors $\Psi^{(1)s}_\mu$ and $\Psi^{(2)s}_\mu$ generate a (non-orthonormal) basis for the eingenfunctions of the RS operator. In particular, consider the following linear combination, 
\begin{equation}
	\Psi^s_{\mu\ell m}(\lambda) = \mathcal{N}^s_\ell\bigg(\nabla_\mu + \frac{s}{2}\,M\gamma_\mu\bigg)\Omega^s_{\ell m} \, .
\end{equation}
Here, the normalization constant $\mathcal{N}^s_\ell$ should be fixed by demanding 
\begin{equation}
	\langle\Psi^s_{\mu\ell m}(\lambda),\Psi^{\mu s'}_{\ell'm'}(\lambda')\rangle = \delta_{ss'}\delta_{\ell\ell'}\delta_{mm'}\delta(\lambda - \lambda') \, ,
\end{equation}
where we define the Euclidean inner product as
\begin{equation}
	\langle f,g\rangle = \int d^4 x \sqrt{g}\,f^\dagger g \, .
\end{equation}
For a massless gravitino field in EAdS$_4$ with $M=1/L$, it is easy to see that $\Psi^s_{\mu\ell m}$ is in fact a zero-mode of the RS operator:
\begin{equation}
(D_s)_\mu{}^\nu \Psi^s_{\nu\ell m}(\lambda) = 0 \, ,
\end{equation}
where the sign factors are now correlated. Thus, it appears that quantum fluctuations of a spin-3/2 field can include zero modes of the RS operator on EAdS$_4$. We should now ask whether such modes are square-integrable with
\begin{equation}
	\langle\Psi^s_{\mu\ell m}(\lambda),\Psi^{\mu s}_{\ell m}(\lambda)\rangle < \infty\, .
\end{equation}
To answer this question, we can use the known expression for the Dirac eigenspinors $\Omega^s_{\ell m}$ in~\cite{Camporesi:1995fb}. We give details in Appendix~\ref{App:RS-non-local}, where we show that the zero modes $\Psi^s_{\mu\ell m}$ either have vanishing norm or are \emph{not} square-integrable and must therefore be discarded.\\

The upshot of this analysis is that, on EAdS$_4$, the non-local contribution \eqref{eq:C-non-local} comes only from 2-forms as
\begin{equation}
\label{eq:Cnon-local-AdS4}
	C_\text{non-local} = \sum_{\text{massless 2-forms}}(-1)^F(-j-1) \, .
\end{equation}
A common feature of all 4d KK supergravity theories we will consider is that the spectrum does not include $p$-forms with $p > 2$. As a result, any 2-form is necessarily bosonic and at ghost level $j=0$ (see Footnote~\ref{foot:ghosts}), in which case the above simplifies to 
\begin{equation}
	C_\text{non-local}(\text{EAdS}_4) = -N_{A_2} \, ,
\end{equation} 
with $N_{A_2}$ the total number of massless 2-form fields. We stress that the non-local contribution~\eqref{eq:Cnon-local-AdS4} to the EAdS$_4$ logarithmic correction is always a pure number, regardless of the field content of the gravitational theory we consider.

\subsubsection{Zero-modes on EAdS$_2 \times \Sigma_\mathfrak{g}$}
\label{sec:sugra:non-local-AdS2xS}

On a product space like EAdS$_2 \times \Sigma_\mathfrak{g}$, the quadratic operator $\mathcal{Q}$ acting on arbitrary field fluctuations can be split into
\begin{equation}
	\mathcal{Q}_{4\text{d}} = \mathcal{Q}_{\text{EAdS}_2} + \mathcal{Q}_{\Sigma_{\mathfrak{g}}} \, .
\end{equation}
We will always consider compact Riemann surfaces, which implies that the operator $\mathcal{Q}_{\Sigma_{\mathfrak{g}}}$ has real non-negative eigenvalues. We will see that, as originally shown in \cite{Camporesi:1994ga,camporesi1994plancherel,Camporesi:1995fb}, normalizable modes on EAdS$_2$ also have non-negative eigenvalues. Therefore, we can study the modes that are in the kernel of both two-dimensional operators separately to obtain the zero-modes on the 4d space.

The EAdS$_2 \times \Sigma_{\mathfrak{g}>1}$ solution to \eqref{eom} with the corresponding graviphoton reads 
\begin{equation}
\label{eq:NHG-Romans-ZM}
\begin{split}
	ds^2 & = \frac{L^2}{4}\bigg[(\rho^2-1)\,d\tau^2 + \frac{d\rho^2}{\rho^2-1} + 2\,\frac{dx^2 + dy^2}{y^2}\bigg] \, ,\\
	dA & = \frac{L}{2y^2}\,dx \wedge dy \, .
\end{split}
\end{equation}
We restrict ourselves to a Riemann surface with genus $\mathfrak{g} > 1$ since this is the situation arising from the near-horizon limit of supersymmetric extremal black holes in AdS$_4$, see Section~\ref{sec:bootstrap:Romans}. On this background, a 4d field can be dimensionally reduced down to the EAdS$_2$ factor with coordinates $(\tau,\rho)$. In this two-dimensional space, the relevant Laplace-type operators for scalar and spinor fields are the usual 2d scalar Laplacian and the square of the Dirac operator. Their eigenvalues are well-known,
\begin{equation}
\begin{split}
	-\square_{\text{EAdS}_2}\phi =&\; \Bigl(\lambda^2 + \frac14\Bigr)\phi \, , \qquad \lambda \in \mathbb{R} \, , \\
	-\slashed{\nabla}^2_{\text{EAdS}_2}\psi =&\; \lambda^2\,\psi \, , \qquad\qquad\;\;\; \lambda > 0 \, ,
\end{split}
\end{equation}
which shows that there are no zero modes for these 2d fields.  In contrast, 2d fields with spin $1\leq s \leq 2$ have normalizable zero modes whose explicit expressions can be found in~\cite{Sen:2012cj}. These zero modes are indexed by a non-zero integer $\ell$ whose range depends on the spin, and they satisfy
\begin{equation}
	\sum_{|\ell|\geq1} |a^{(\ell)}_\alpha|^2 = \frac{2}{\pi L^2} \, , \qquad \sum_{\ell \geq 1} |\xi^{(\ell)}_\alpha|^2 = \frac{4}{\pi L^2} \, , \qquad \sum_{|\ell|\geq2} |w_{\alpha\beta}^{(\ell)}|^2 = \frac{6}{\pi L^2} \, ,\label{sq:sum}
\end{equation}
for a vector, gravitino and metric zero mode, respectively. Here $\alpha$ denotes a space-time EAdS$_2$ index and the norm is taken with appropriate 2d metric contractions. Multiplying \eqref{sq:sum} by the regularized volume $\text{vol}(\text{EAdS}_2) = -\pi L^2/2$, we obtain the number of zero-modes for each two-dimensional field,
\begin{equation}
\label{eq:n0-H2}
	n^{\text{EAdS}_2}_0 = n^{\text{EAdS}_2}_{1/2} = 0 \, , \qquad n^{\text{EAdS}_2}_1 = -1 \, , \qquad n^{\text{EAdS}_2}_{3/2} = -2 \, , \qquad n^{\text{EAdS}_2}_2 = -3 \, ,
\end{equation}
%

On the Riemann surface, the only zero-mode of the scalar Laplacian is the constant function.\footnote{This can be shown, for instance, by studying the Laplacian on the upper half-plane and quotienting by a discrete subgroup of $PSL(2,\mathbb{R})$ to obtain the spectrum on the compact Riemann surface.} For spin-$1/2$, we must take into account the fact that there is a non-trivial gauge field along $\Sigma_\mathfrak{g}$ as in~\eqref{eq:NHG-Romans-ZM}. This implies that the relevant differential operator acting on spinor fluctuations is the Dirac operator on the compact Riemann surface twisted by the line bundle $\mathcal{L}$ defined by the gauge connection. Denoting this operator by $\slashed{\nabla}_A$, we can make use of the result of~\cite{ALMOROX20062069} for $\mathfrak{g} > 1$, 
\begin{equation}
	\text{dim}\,\text{Ker}\slashed{\nabla}_A = \text{deg}(\mathcal{L}) = 2(\mathfrak{g}-1) \, ,
\end{equation}
Thus, the number of zero modes for scalar and spinor fields on the Riemann surface are
\begin{equation}
\label{eq:n0-S}
	n^{\Sigma}_0 = 1 \, , \qquad n^{\Sigma}_{1/2} = 2(\mathfrak{g} - 1) \, .
\end{equation}
We will not need the number of zero modes for fields of spin $1 \leq s \leq 2$ since they always come tensored with scalar or spinor zero modes on the EAdS$_2$ factor, which as we saw above do not exist.

We can now decompose all 4d fields according to their 2d spins and use the above results to compute the number of 4d zero modes. The 4d scalar and spinor fields decompose into combinations of scalars and fermions on EAdS$_2$ and the Riemann surface, so they have no zero modes by~\eqref{eq:n0-H2}. A vector field in 4d decomposes into a spin-1 field on EAdS$_2$ and a scalar on $\Sigma_{\mathfrak{g}}$, so we have $-1$ such zero-modes. A 4d gravitino field decomposes into a combination of spin-3/2 and spin-1/2 two-dimensional fields, producing a total of $-4(\mathfrak{g}-1)$ gravitino zero-modes in 4d. For the metric, the absence of globally defined Killing vectors on the Riemann surface of genus $\mathfrak{g}>1$ implies that the 4d mode decomposes into a combination of metric and scalar modes on the 2d spaces. Thus,
\begin{equation}
	n^{4\text{d}}_0 = n^{4\text{d}}_{1/2} = 0 \, , \qquad n^{4\text{d}}_1 = -1 \, , \qquad n^{4\text{d}}_{3/2} = -4(\mathfrak{g} - 1) \, , \qquad n^{4\text{d}}_2 = -3 \, .\label{n:EAdS2}
\end{equation}
Observe that the dependence on the genus comes from the fact that the gravitino is charged under the background gauge field that defines the line bundle $\mathcal{L}$.\\

The non-local contribution to the logarithmic correction from the above zero modes is obtained by substituting \eqref{eq:beta} and \eqref{n:EAdS2} into \eqref{eq:C-non-local}. The result is
\begin{equation}
\label{eq:C:non-local-AdS2xS-gen}
	C_{\text{non-local}} = \Bigl[\sum_{s=1}(-1)^F j\Bigr] + 4(1-\mathfrak{g})\Bigl[\sum_{s=3/2}(-1)^F(2-j)\Bigr] - 3\Bigl[\sum_{s=2}(-1)^F(1-j)\Bigr] \, ,
\end{equation}
where the sums run over the massless fields of spin $s$. The 4d KK supergravity theories we want to consider do not contain ghost fields of spin $s>1$, so that the last two sums are restricted to the $j=0$ sector and have definite fermion numbers. Gauge-fixing in the spin-2 fluctuation sector requires the introduction of a pair of vector ghosts (see also Appendix~\ref{App:coeffs-graviton}). If we assume that this gauge-fixing is the only source of $s=1$ ghosts, the above simplifies to
\begin{equation}
\label{eq:C:non-local-AdS2xS}
	C_\text{non-local}(\text{EAdS}_2\times\Sigma_{\mathfrak{g}>1}) = 8N_{3/2}(\mathfrak{g}-1) - 4N_{2} \, ,
\end{equation}
where $N_{3/2}$ and $N_2$ denote the number of massless gravitino and metric fluctuations in the spectrum. Note that the contribution~\eqref{eq:C:non-local-AdS2xS-gen} to the logarithmic correction is always integer, regardless of the KK theory under consideration. Moreover, it depends on the parameters of the background geometry only through the genus of the Riemann surface, which arises from the non-trivial charge of the gravitino fluctuations under the background gauge field.

\subsubsection{Generic zero-modes}
\label{sec:sugra:non-local-gen}

The results~\eqref{eq:Cnon-local-AdS4} and~\eqref{eq:C:non-local-AdS2xS} illustrate how the spectral problem of finding the zero modes of the $\mathcal{Q}$ operators depends on the supergravity background. In particular, when the latter includes a factor of $d$-dimensional Euclidean AdS, one can leverage known results in the literature to study the zero modes. Unfortunately, the task is much more arduous on generic backgrounds. Nevertheless, our examples make it clear that the non-local contributions to the logarithmic corrections on a general $\mathcal{M}_4$ will always take the form of pure numbers related to the counting of zero modes and possibly a simple dependence on discrete parameters of topological origin like the rank of some fiber bundles. In particular, we do not expect that $C_{\text{non-local}}$ can depend on continuous parameters specifying the 4d supergravity background.

Before closing this section, we would like to point out that the discussion above was focused on the calculation of the Euclidean gravitational path integral in the so-called grand canonical ensemble of fixed temperature and chemical potentials. If one is interested in a different thermodynamic ensemble, for example in order to calculate the entropy of a black hole, an appropriate Laplace transform according to the standard rules of thermodynamics should be performed. This change of ensemble can lead to additional contribution to the logarithmic corrections of the partition function, see \cite{Sen:2012dw}. Notably, these additional terms in the coefficient of the logarithmic correction are always pure numbers independent of continuous parameters in the supergravity background of interest.

\section{Bootstrapping the local contributions}
\label{sec:bootstrap}

Following the procedure described in Section \ref{sec:sugra}, one can in principle compute the logarithmic coefficient $C$ in the 4d Euclidean path integral \eqref{eq:sugra-saddle} around any asymptotically EAdS$_4$ Einstein-Maxwell background in a given 4d KK supergravity theory. As we have pointed out, there are a few technical hurdles:
\begin{enumerate}
	\item Computing the local contribution $C_\text{local}$ for generic Einstein-Maxwell backgrounds with a non-vanishing graviphoton remains out of reach, since we have restricted our results in Table~\ref{tab:coeffs} to pure Einstein backgrounds, and we therefore do not have access to the $b_2$ coefficient in~\eqref{eq:a4-bulk} for every KK supergravity field.

	\item While imposing conformal boundary conditions on the metric fluctuations ensures that the differential operator $\mathcal{Q}_h$ is elliptic, this choice falls outside of the class for which explicit surface contributions to the SdW coefficient are known.
	
	\item Computing the non-local contribution $C_\text{non-local}$ involves a spectral problem that we cannot solve in general for arbitrary Einstein-Maxwell backgrounds.
\end{enumerate}
In view of these difficulties, it would seem that a first-principle computation of $C$ for various 4d supergravity backgrounds still eludes us. However, we will now leverage holography and explain how the various logarithmic coefficients $\mathcal{C}$ on the field theory side (see Table~\ref{tab:CFT-log}) can be used to ``bootstrap'' the heat kernel coefficients $(a_E,c,b_1,b_2)$ in the bulk. This will allow us to elegantly circumvent the first two issues in the above list. The main idea is to study the logarithmic corrections as a function of various backgrounds supporting arbitrary quadratic field fluctuations, rather than studying individual fields in the KK spectrum evolving on a fixed background. Before we illustrate how this procedure works in detail on several examples, we note that from the expressions for the SdW coefficient in \eqref{eq:a4} and \eqref{eq:a4-bulk}, it is clear that the contribution to $C_\text{local}$ from the $a_E$ coefficient comes accompanied by the integrated Euler density $\int E_4$. This integral yields a topological invariant of the 4d Euclidean manifold $\mathcal{M}$ used as a supergravity background, i.e. its Euler characteristic~$\chi(\mathcal{M})$, and thus the contribution of $a_E$ to $C_\text{local}$ is always a pure number independent of any continuous parameters that may be present in the supergravity background. As we show below this general expectation is realized in all examples we study.

\subsection{EAdS$_4$}
\label{sec:bootstrap:EAdS4}

We start with the pure EAdS$_4$ solution to \eqref{eom} given by
\begin{equation}
\label{sol:EAdS4}
	ds^2 = \frac{L^2}{L^2 + r^2}\,dr^2 + r^2\,d\Omega_3^2 \, , \qquad A = 0 \, ,
\end{equation}
where $d\Omega_3^2$ is the metric on the round $S^3$ with unit radius. This space is conformally flat and so has vanishing Weyl tensor. Introducing appropriate counter-terms, the regularized Euler characteristic and $R^2$ invariant are given by
\begin{equation}
\label{eq:chi-AdS}
	\chi = \frac{1}{32\pi^2}\int d^4 x \sqrt{g}\,E_4 = 1 \, , \qquad \frac{1}{32\pi^2} \int d^4 x \sqrt{g}\,R^2 = 6 \, .
\end{equation}
Thus, we find that an arbitrary fluctuation of a field $\phi$ (of general mass and spin) with corresponding kinetic operator $\mathcal{Q}_\phi$ propagating in this background produces a bulk contribution to the integrated SdW coefficient~\eqref{eq:a4-bulk} given by
\begin{equation}
\label{eq:a4-bulk-round}
	\int d^4x \sqrt{g}\,a_4^{\text{bulk}}(\mathcal{Q}_\phi) = 2\bigl[6\,b_1(\phi) - a_E(\phi)\bigr] \, .
\end{equation}
Observe that the coefficient of $a_E$ equals $-2\,\chi$ with $\chi$ given in~\eqref{eq:chi-AdS}. In addition, the total derivative contribution~\eqref{eq:a4-tot-der} vanishes trivially since the bulk Ricci scalar is constant and $A=0$. It remains to analyze the boundary contributions \eqref{eq:a4-details:bdry}. The extrinsic curvature of EAdS$_4$ in the coordinate system \eqref{sol:EAdS4} reads 
\begin{equation}
	K_{cd} = \frac{\sqrt{L^2 + r_b^2}}{L\,r_b}\,\delta_{cd}  = \frac{1}{L}\,\delta_{cd} +\mathcal{O}(r_b^{-2}) \, ,
\end{equation}
where $r_b$ is the location of the boundary in the radial direction. This in turn determines the projectors $\Pi_\pm$ and $S$ specifying the boundary conditions. Note that following the discussion in Section~\ref{sec:sugra:local:bdry}, the latter can be parameterized as $S = S' K$ where the $S'$ field space matrix is constant. Working in a radial expansion with a large cutoff $r_b$, we find the following surface contribution to the integrated SdW coefficient:
\begin{equation}
\label{eq:a4-bdry-AdS4}
	\begin{split}
		\int_{\partial\mathcal M}&d^3y\,\sqrt{\gamma}\,a_4^{\text{bdry}}(\mQ_\phi) \\
		=&\; \frac{1}{360}\Tr\Bigl[45L^2 E(1 - 6 S') - 2(\Pi_- - 29\Pi_+) + 1620S'^2(1 - S') - 93\Bigr]\,\frac{r_b^3}{L^3} \\[1mm]
		&\;+ \frac{1}{240}\Tr\Bigl[15L^2 E(1 - 6 S') - 2(\Pi_- - 29\Pi_+) - 180S'(2 - 9S' + 9S'^2) - 31\Bigr]\,\frac{r_b}{L} \\[1mm]
		&\;+ \mathcal{O}(r_b^{-1}) \, ,
	\end{split}
\end{equation}
where $\Pi_\pm$, $S'$ and $E$ are all independent of the radial cutoff. This boundary contribution can be holographically renormalized upon introducing the counter-term
\begin{equation}
\label{eq:CT}
	\int_{\partial \mathcal{M}} d^3y \,\sqrt{\gamma}\,\bigl(\mathfrak{c}_1 + \mathfrak{c}_2 \mathcal{R}\bigr) \, ,
\end{equation}
where $\mathcal{R}$ denotes the Ricci scalar of the boundary. The constant $\mathfrak{c}_1$ can be chosen so as to cancel the cubic divergence in~\eqref{eq:a4-bdry-AdS4} while $\mathfrak{c}_2$ can be tuned to cancel the linear term. Importantly, this renormalization scheme does not introduce any constant term on EAdS$_4$. In this way, the boundary contribution to the SdW coefficient vanishes by holographic renormalization. With this prescription, the total local contribution \eqref{eq:C-local} is
\begin{equation}
	C_{\text{local}} = 2\sum_\phi (-1)^F\,\bigl[6\,b_1(\phi) - a_E(\phi)\bigr] \, .\label{eq:C-local:EAdS4}
\end{equation}
Here we have taken the sum of the contributions of all fields propagating in the background weighted by their fermion number $F$.\\

Let us make the following comment on our result. Using the heat kernel coefficients $(a_E,b_1)$ given in Table~\ref{tab:coeffs} and expressing the masses of the bulk fields in terms of conformal dimensions of the dual operators using Table~\ref{tab:conf-m}, we find that the summand in~\eqref{eq:C-local:EAdS4} can be written as
\begin{equation}
6\,b_1(\phi) - a_E(\phi) = \frac{2s+1}{48}\,\Bigl[\Bigl(\Delta - \frac32\Bigr)^4 - \Bigl(s + \frac12\Bigr)^2\Bigl(2\Bigl(\Delta - \frac32\Bigr)^2 + \frac16\Bigr) - \frac7{240}\Bigr] \, ,
\end{equation}
for any bosonic field of spin $s \in \{0,1,2\}$, and as
\begin{equation}
6\,b_1(\phi) - a_E(\phi) = \frac{2s+1}{48}\,\Bigl[\Bigl(\Delta - \frac32\Bigr)^4 - \Bigl(s + \frac12\Bigr)^2\Bigl(2\Bigl(\Delta - \frac32\Bigr)^2 - \frac13\Bigr) + \frac1{30}\Bigr] \, ,
\end{equation}
for any fermionic field of spin $s \in \{1/2,3/2\}$. In other words, we have
\begin{equation}
\label{eq:Cloc-S3-G}
	C_{\text{local}} = \sum_\phi\,G_s\Bigl(\Delta - \frac32\Bigr) \, , 
\end{equation}
where the function $G_s$ is defined by
\begin{equation}
	G_s(x) = \begin{cases} \frac{2s+1}{24}\,\bigl(x^4 - \bigl(s + \frac12\bigr)^2\bigl(2x^2 + \frac16\bigr) - \frac7{240}\bigr) \quad &s \in \{0,1,2\} \\[2mm] -\frac{2s+1}{24}\,\bigl(x^4 - \bigl(s + \frac12)^2\bigl(2x^2 - \frac13\bigr) + \frac1{30}\bigr) \quad &s \in \{1/2,3/2\} \end{cases} \, .
\end{equation}
The appearance of this function is not an accident. Rather, as shown in~\cite{Camporesi:1993mz,Liu:2016dau,Binder:2021euo}, the function $G_s$ is precisely the one that controls the logarithmic divergence in the free energy $F_{\Delta,s}$ of a free field in EAdS$_4$ dual to an operator of spin $s$ and conformal dimension $\Delta$,
\begin{equation}
\label{eq:3-sphere-log}
F_{\Delta,s} = G_s\Bigl(\Delta - \frac32\Bigr)\log(L/\epsilon) + \text{finite} \, ,
\end{equation}
where $\epsilon$ is a UV cutoff. The result~\eqref{eq:3-sphere-log} has been obtained by a direct spectral analysis on pure Euclidean AdS$_4$, and what we have shown is that our heat kernel coefficients precisely reproduce the expected logarithmic behavior. Note that to achieve this agreement, it was crucial to use the precise values of the $a_E$ and $b_1$ coefficients given in Table~\ref{tab:coeffs}, which in particular take into account the contribution from St\"uckelberg fields when $1 \leq s \leq 2$. Therefore, we can view~\eqref{eq:Cloc-S3-G} as a non-trivial check of our computations in Appendix~\ref{App:coeffs}. Moreover, we can now argue on the basis of~\eqref{eq:3-sphere-log} that there cannot be local contributions to the EAdS$_4$ logarithmic correction due to the oblique nature of boundary conditions on the metric fluctuations, in agreement with our assumption at the end of Section~\ref{sec:sugra:local:bdry}.\footnote{Recall from~\eqref{eq:Cnon-local-AdS4} that the non-local contribution vanishes in the absence of bulk massless 2-forms in the spectrum, which is assumed in deriving~\eqref{eq:3-sphere-log}.} In the simple EAdS$_4$ background, we can therefore address all three points raised at the beginning of the present section and conclude that the logarithmic correction to the path integral is entirely controlled by $C=C_{\text{local}}$ given in~\eqref{eq:C-local:EAdS4}. To compare with the CFT coefficient $\mathcal{C}$, we still need to explicitly sum over the field content of the relevant KK supergravity theory and use a suitable regularization. We will discuss this in Section~\ref{sec:KK}, but for the time being we move on to other minimal supergravity backgrounds.

\subsection{Euclidean Romans}
\label{sec:bootstrap:Romans}

Next we consider the Euclidean Romans solution to~\eqref{eom} given by \cite{Romans:1991nq,Bobev:2020pjk}
\begin{equation}
	\label{eq:romans}
	\begin{split}
		ds^2 =&\; U(r)\,d\tau^2 + \frac{dr^2}{U(r)} + r^2 ds^2_{\Sigma} \, , \qquad U(r) = \bigg(\frac{r}{L} + \frac{\kappa L}{2r}\bigg)^2 - \frac{Q^2}{4r^2} \, , \\
		dA =&\;  \frac{Q}{2r^2}\,d\tau\wedge dr + \frac{\kappa L}{2}\,V_{\Sigma} \, ,
	\end{split}
\end{equation}
where $\kappa\in\{0,\pm1\}$ is the normalized curvature of the Riemann surface $\Sigma$ and $Q$ is a free, continuous parameter. Evaluating the relevant curvature-squared terms with appropriate counter-terms, we find that the bulk contribution \eqref{eq:a4-bulk} for a field $\phi$ of general spin and mass is given by
\begin{equation}
\label{eq:a4-bulk:Romans}
	\begin{split}
	\int d^4x \sqrt{g}\,a_4^{\text{bulk}}(\mathcal{Q}_\phi) =&\; -4(1 - \mathfrak{g})a_E(\phi) + \frac{3(|Q| + \kappa L)^2}{10\pi L|Q|}\text{vol}(\Sigma)\,c(\phi) \\
	&\;- \frac{3(|Q| - \kappa L)^2}{2\pi L |Q|}\text{vol}(\Sigma)\,b_1(\phi) - \frac{3(Q^2 + \kappa^2 L^2)}{4\pi L |Q|}\text{vol}(\Sigma)\,b_2(\phi) \, .
	\end{split}
\end{equation}
The coefficient of $a_E$ equals $-2\,\chi$ where the regularized Euler characteristic is $\chi = 2(1 - \mathfrak{g})$. We stress that these results are valid for quadratic fluctuations of an arbitrary field $\phi$ around the Romans background. Using the same holographic renormalization prescription as in the case of pure EAdS$_4$, the total derivative and boundary contributions to $C_{\text{local}}$ can be shown to vanish. This implies that the only contribution to \eqref{eq:C-local} reads
\begin{equation}
\label{eq:C-local:Romans}
	C_\text{local} = \sum_\phi (-1)^F\int d^4x\,\sqrt{g}\,a_4^\text{bulk}(\mQ_\phi) \, ,
\end{equation}
with the integrated $a_4^\text{bulk}(\mQ_\phi)$ given in \eqref{eq:a4-bulk:Romans} and the sum is taken over all fields propagating in the background weighted by their fermion number $F$.

The 4d Euclidean path integral around the Romans background is holographically dual to the TTI of the boundary SCFT on $S^1 \times \Sigma_\mathfrak{g}$. The logarithmic coefficient $\mathcal{C}$ extracted from the large $N$ limit of the TTI is given in Table~\ref{tab:CFT-log} for various SCFTs, and one readily checks that this coefficient is always a pure number. On the bulk side, we have argued in Section~\ref{sec:sugra:non-local-gen} that the non-local contribution to $C$ is a pure number for arbitrary backgrounds. Therefore, holography dictates that $C_{\text{local}}$ given in~\eqref{eq:C-local:Romans} should also be a pure number, and in particular should be independent of the continuous parameter $Q$ of the background. This requirement then translates into a linear constraint on the heat kernel coefficients summed over the spectrum,
\begin{equation}
\label{constraints:1}
\begin{split}
	\sum_{\phi}(-1)^F\bigl[2\,c(\phi) - 10\,b_1(\phi) - 5\,b_2(\phi)\bigr] = 0 \, .
\end{split}
\end{equation}
We stress that this should be true in any four-dimensional KK supergravity theory whose SCFT dual falls in one of the classes discussed in Table~\ref{tab:CFT-log}. Indeed, this constraint must be statisfied in order to fit the expectations of the AdS/CFT correspondence at order $\mathcal{O}(\log N)$ in the large $N$ expansion.

The background~\eqref{eq:romans} admits a supersymmetric limit as $Q \rightarrow 0$. For hyperbolic Riemann surfaces, i.e. for $\mathfrak{g}>1$, the resulting geometry has a Lorentzian interpretation as a magnetically charged Reissner-Nordström (RN) black hole in AdS$_4$. In Section~\ref{sec:log-conj} we will comment on the implications of our results for the thermodynamics of AdS black holes.

\subsection{U(1)$\times$U(1) squashing}
\label{sec:bootstrap:U(1)-squash}

Another solution to \eqref{eom} we can consider is one where the conformal boundary is a squashed 3-sphere. A particular squashing that preserves supersymmetry and a U(1)$\times$U(1) subgroup of the ${\rm SO}(4)$ isometry group can be obtained in the bulk by turning on an anti-self-dual graviphoton compared to the pure EAdS$_4$ case. In a convenient coordinate system, this squashed background takes the form~\cite{Martelli:2011fu}
\begin{align}
	ds^2 =&\; f_1(x,y)^2\,dx^2 + f_2(x,y)^2\,dy^2 + L^2\,\frac{(x^2 - 1)(y^2 - 1)}{\mathrm{s}^2 - 1}\,d\phi_1^2 + L^2\,\frac{(\mathrm{s}^2 - x^2)(y^2 - \mathrm{s}^2)}{\mathrm{s}^2(\mathrm{s}^2-1)}\,d\phi_2^2 \, , \nonumber \\
	A =&\; \frac{L}{2(x+y)}\bigg(\frac{\mathrm{s}^2 + xy}{\mathrm{s}}\,d\phi_2 - (1 + xy)\,d\phi_1\bigg) \, ,
\end{align}
where the squashing parameter is $\mathrm{s}$ (not to be confused with the spin $s$ of a field), and 
\begin{equation}
	f_1(x,y)^2 = L^2\frac{y^2 - x^2}{(x^2 - 1)(\mathrm{s}^2 - x^2)} \, , \qquad f_2(x,y)^2 = L^2\frac{y^2 - x^2}{(y^2 - 1)(y^2 - \mathrm{s}^2)} \, .
\end{equation}
Computing the relevant curvature invariants, we obtain the bulk contribution
\begin{equation}
\label{eq:a4-bulk-U1-squash}
	\int d^4x \sqrt{g}\,a_4^\text{bulk}(\mQ_\phi) = 2\bigl[6\,b_1(\phi) - a_E(\phi)\bigr] - \frac{3(\mathrm{s}-1)^2}{2\,\mathrm{s}}\,b_2(\phi) \, ,
\end{equation}
for any field $\phi$ of general mass and spin propagating on this background. Again, the coefficient of $a_E$ in this expression is minus twice the Euler characteristic of the squashed background, which is topologically indistinguishable from the pure EAdS$_4$ geometry.

When $\mathrm{s}=1$, we recover the result for pure EAdS$_4$ with a round 3-sphere boundary. In addition, our renormalization prescription ensures that there are no other contributions to $C_{\text{local}}$ besides~\eqref{eq:a4-bulk-U1-squash}. Following the logic described in the previous subsection, AdS/CFT requires that the above integrated SdW coefficient be independent of the squashing parameter after summing over all fields in the spectrum.\footnote{The universal CFT coefficient $\mathcal{C}$ in Table~\ref{tab:CFT-log} has been obtained analytically in the large $N$ limit of the squashed 3-sphere partition function for $\mathrm{s}=1$ \cite{Fuji:2011km,Marino:2011eh} and $\mathrm{s}=3$ \cite{Hatsuda:2016uqa}, and conjectured for a generic squashing parameter in \cite{Bobev:2022eus,Bobev:2023lkx}. One-loop calculations in the dual 11d backgrounds, however, make the universality of the logarithmic coefficient clear for any value of $\mathrm{s}$, since in this approach the logarithmic coefficient is entirely controlled by the number of 11d zero-modes \cite{Bhattacharyya:2012ye} that should not depend on any continuous parameter.} This immediately imposes
\begin{equation}
	\sum_{\phi}(-1)^F b_2(\phi) = 0 \, ,
\end{equation}
which shows that while we do not have access to the $b_2$ coefficient for a given field in general, universality of the logarithmic correction and holography imply a strong constraint on the total contribution after summing over the spectrum. This addresses the first point raised at the beginning of this section, at least partially. We can combine the above constraint with~\eqref{constraints:1} to obtain a more stringent relation among the remaining heat kernel coefficients,
\begin{equation}
\label{constraints:2}
	\sum_{\phi}(-1)^F\bigl[c(\phi) - 5\,b_1(\phi)\bigr] = 0 \, .
\end{equation}

\subsection{SU(2)$\times$U(1) squashing }
\label{sec:bootstrap:SU(2)-squash}

We now consider another supersymmetric bulk solution whose conformal boundary is a squashed 3-sphere, with the squashing preserving an SU(2)$\times$U(1) isometry. The minimal supergravity background is given by \cite{Martelli:2011fw}
\begin{equation}
	\begin{split}
		ds^2 =&\; \frac{r^2 - \mathrm{s}^2}{\Omega(r)}\,dr^2 + (r^2 - \mathrm{s}^2)(\sigma_1^2 + \sigma_2^2) + \frac{4\mathrm{s}^2\Omega(r)}{r^2 - \mathrm{s}^2}\,\sigma_3^2 \, ,\\
		A =&\; \mathrm{s}\,\sigma_3\,\frac{r-\mathrm{s}}{r+\mathrm{s}}\sqrt{\frac{4\mathrm{s}^2}{L^2} - 1} \, ,
	\end{split}
\end{equation}
where
\begin{equation}
	\Omega(r) = (r - \mathrm{s})^2\bigg(1 + \frac{(r-\mathrm{s})(r+3\mathrm{s})}{L^2}\bigg) \, ,
\end{equation}
$\sigma_i$ are the SU(2) left-invariant one-forms, and $\mathrm{s}$ is the squashing parameter. We can proceed as before and compute the renormalized bulk contribution \eqref{eq:a4-bulk} which now takes the form 
\begin{equation}
\begin{split}
	\int d^4x \sqrt{g}\,a_4^\text{bulk}(\mQ_\phi) =&\; -2\,a_E(\phi) + \frac{2(L^2 - 4\mathrm{s}^2)^2}{L^4}\,c(\phi) \\
&\;+ \frac{96\mathrm{s}^2(L^2 - 2\mathrm{s}^2)}{L^4}\,b_1(\phi) + \frac{24\mathrm{s}^2(L^2 - 4\mathrm{s}^2)}{L^4}\,b_2(\phi) \, ,
\end{split}
\end{equation}
and the surface terms again vanish by holographic renormalization. Since the logarithmic term in the dual squashed sphere partition function is independent of the squashing parameter, we can apply the same logic as before and deduce that AdS/CFT imposes a new constraint on heat kernel coefficients summed over the spectrum. Together with the previous constraints \eqref{constraints:1} and \eqref{constraints:2}, we arrive at the final conclusion that
\begin{equation}
	\label{eq:no-Cloc}
	\sum_{\phi}(-1)^Fc(\phi) = \sum_{\phi}(-1)^Fb_1(\phi) = \sum_{\phi}(-1)^Fb_2(\phi) = 0 \, .
\end{equation}
In other words, the universal character of the logarithmic corrections to the large $N$ limit of SCFT partition functions imposes, via holography, that the sum of the heat kernel coefficients $(c,b_1,b_2)$ over the spectrum of the theory should vanish. This must be true \emph{regardless of the particular field content of the 4d KK supergravity theory under consideration}. As we will see below, these constraints have important consequences. Recall also that the~$a_E$ coefficient is not subject to any constraint since the Euler characteristic $\chi(\mathcal{M})$ cannot depend on continuous parameters.

\subsection{AdS-Taub-NUT}
\label{sec:AdSTNsol}

The bootstrap result (\ref{eq:no-Cloc}) is obtained by studying the Euclidean path integrals around three distinct supersymmetric backgrounds holographically dual to 3d $\mN=2$ SCFT partition functions whose logarithmic coefficients have been argued to be pure numbers by supersymmetric localization and the index theorem. As we discuss in Section~\ref{sec:Kundera}, one can derive similar constraints on heat kernel coefficients without the need to assume $\mN=2$ supersymmetry. To illustrate this point we now consider two different non-BPS supergravity backgrounds.

As a first example, we study the following Euclidean AdS-Taub-NUT background~\cite{Emparan:1999pm}
\begin{equation}
ds^2 = 4n^2 V(r)\,\sigma_3^2 + V(r)^{-1}dr^2 + (r^2 - n^2)(\sigma_1^2 + \sigma_2^2) \, ,
\end{equation}
where
\begin{equation}
V(r) = \frac{r^2 + n^2 - 2mr + L^{-2}(r^4 - 6n^2r^2 - 3n^4)}{r^2 - n^2} \, .
\end{equation}
The above metric solves the Einstein equations in \eqref{eom} with a vanishing Maxwell field, and the solution fully breaks supersymmetry. The mass parameter $m$ is related to the NUT charge $n$ as
\begin{equation}
m = n - \frac{4n^3}{L^2} \, ,
\end{equation}
and the boundary at $r\rightarrow \infty$ is a squashed 3-sphere with metric
\begin{equation}
ds^2_\infty = r^2\Bigl(\sigma_1^2 + \sigma_2^2 + \frac{1}{1+\mathrm{s}}\sigma_3^2\Bigr) \, .
\end{equation}
The squashing parameter is related to the NUT charge as
\begin{equation}
n = \frac{L}{2\sqrt{1+\mathrm{s}}} \, .
\end{equation}
With this at hand, we compute the bulk part of the integrated fourth SdW coefficient:
\begin{align}
\int d^4 x \sqrt{g}\,a_4^\text{bulk}(\mathcal{Q}_\phi) =&\; \frac{8}{\sqrt{1+\mathrm{s}}}\bigl[6\,b_1(\phi) - a_E(\phi)\bigr]\,\frac{r_b^3}{L^3} - \frac{6}{(1+\mathrm{s})^{3/2}}\bigl[6\,b_1(\phi) - a_E(\phi)\bigr]\,\frac{r_b}{L} \nonumber \\
&\;+ \frac{2}{(1+\mathrm{s})^2}\Bigl[6\,b_1(\phi) - (1 + \mathrm{s}^2)\,a_E(\phi) + \mathrm{s}^2\,c(\phi)\Bigr] \\[1mm]
&\;+ \mathcal{O}(r_b^{-1}) \, , \nonumber
\end{align}
where we have introduced a boundary cut-off $r_b$. Using the counter-term~\eqref{eq:CT}, the divergent terms above and the surface part of the integrated SdW coefficient can both be holographically renormalized away, and we find that the local contribution to the logarithmic correction fo the AdS-Taub-NUT free energy $F(\mathrm{s})$ is given by
\begin{equation}
\label{eq:Cloc-TN}
	C_{\text{local}}(\mathrm{s}) = -2\sum_{\phi}(-1)^F\Bigl[a_E(\phi) - \frac{1}{(1+\mathrm{s})^2}\Bigl(\mathrm{s}^2 c(\phi) + 6(1 + 2\mathrm{s})\,b_1(\phi)\Bigr)\Bigr] \, ,
\end{equation}
where as usual we need to add up the heat kernel coefficients of all bulk fields in the gravitational effective theory. In addition, we do not expect the non-local contribution to depend on the squashing parameter since they arise from zero modes. The correction~\eqref{eq:Cloc-TN} translates into a logarithmic correction to the stress tensor two-point function coefficient $C_T$ in the boundary CFT via the relation \cite{Bobev:2017asb}
\begin{equation}
\label{eq:CT-def}
	C_T = -\frac{48}{\pi^2}\,\partial_{\mathrm{s}}^2 F(\mathrm{s})\big\vert_{\mathrm{s} = 0} \, .
\end{equation}
Using the above, we thus find that $C_T$ generically contains a term
\begin{equation}
\label{eq:CT-gen}
	C_T \ni \frac{192}{\pi^2}\sum_{\phi}(-1)^F\bigl[6\,b_1(\phi) - c(\phi)\bigr] \log N \, .
\end{equation}
If we now assume that the stress tensor two-point function coefficient in any holographic CFT does not contain a logarithmic term in its large $N$ expansion, we arrive at the bootstrap constraint 
\begin{equation}
	\sum_{\phi}(-1)^F\bigl[6\,b_1(\phi) - c(\phi)\bigr] = 0 \, .
\end{equation}
This is compatible with our previous results using supersymmetric backgrounds and SCFT observables. We will discuss our assumption about the absence of log terms in $C_T$ and further implications of this result in Section~\ref{sec:Kundera}.

\subsection{Kerr-Newman}
\label{sec:KNsol}

Lastly, we come to the Euclidean incarnation of the 4d electric Kerr-Newman (KN) black hole solution in AdS. The metric and Maxwell field take the form~\cite{Bobev:2019zmz,Caldarelli:1998hg,Cassani:2019mms}
\begin{align}
	\label{eq:KN-AdS}
            ds^2 =&\; \frac{\Delta_r}{V}\left( d\tau + \frac{\alpha}{\Xi}\sin^2\theta d\phi \right)^2 + V \left( \frac{dr^2}{\Delta_r} + \frac{d\theta^2}{\Delta_\theta}\right) + \frac{\Delta_\theta\sin^2\theta}{V}\left( \alpha d\tau - \frac{\tilde{r}^2 - \alpha^2}{\Xi} d\phi \right)^2\, ,\nonumber\\
            A =&\; \mathrm{i}\,m \sinh(2\delta) \frac{\tilde{r}}{V} \left( d\tau + \frac{\alpha}{\Xi} \sin^2\theta d\phi \right)\, , 
\end{align}
where
\begin{equation}
\label{eq:KN-param}
\begin{split}
            \tilde{r} =&\; r + 2m \sinh^2 \delta \, ,\quad \Xi = 1 + \frac{\alpha^2}{L^2}\, ,\quad V = \tilde{r}^2 - \alpha^2 \cos^2\theta\, ,\\
            \Delta_r =&\; r^2 - \alpha^2 - 2mr + \frac{\tilde{r}^2}{L^2}(\tilde{r}^2 - \alpha^2)\, ,\quad \Delta_\theta = 1 + \frac{\alpha^2}{L^2}\cos^2\theta\, .
\end{split}
\end{equation}
The parameters $(m,\alpha,\delta)$ are related to the energy, angular momentum and electric charge of the solution. After a lengthy computation, we find that the bulk contribution to the integrated SdW coefficient can be holographically renormalized to give
\begin{equation}
\begin{split}
\int d^4 x \sqrt{g}\,a_4^{\text{bulk}}(\mathcal{Q}_\phi) =&\; - 4\,a_E(\phi) + f_1(m,\alpha,\delta)\,c(\phi) \\
&\; + f_2 (m,\alpha, \delta)\,b_1(\phi) + f_3 (m,\alpha, \delta)\,b_2(\phi)\, ,
\end{split}
\end{equation}
where the various functions on the right-hand side are given by
\begin{align}
\label{eq:fi-KN-AdS}
	f_1 =&\; \frac{\beta  m^2}{2\pi \alpha ^5 \Xi}\Biggl\{\frac{3 m^2 c^4 s^4}{\tilde{r}_{+}^4}(\alpha ^4-\tilde{r}_{+}^4) \log\Bigl(\frac{\tilde{r}_{+}+\alpha}{\tilde{r}_{+}-\alpha }\Bigr) -\frac{16 \alpha^5 m c^2 s^2 (c^2+s^2) \tilde{r}_{+}^3 (3 \tilde{r}_{+}^2+\alpha ^2)}{\tilde{r}_{+}^3(\tilde{r}_{+}^2-\alpha ^2)^3} \nonumber \\
&\;\; +\frac{2 \alpha(m^2 c^4 s^4(3 \tilde{r}_{+}^8-8 \alpha ^6 \tilde{r}_{+}^2+42 \alpha ^4 \tilde{r}_{+}^4-8 \alpha ^2 \tilde{r}_{+}^6+3 \alpha ^8)+4 \alpha ^4 (c^2+s^2)^2 \tilde{r}_{+}^4 (\tilde{r}_{+}^2+\alpha ^2))}{\tilde{r}_{+}^3(\tilde{r}_{+}^2-\alpha ^2)^3}\Biggr\}\, , \nonumber\\[1mm]
	f_2 =&\; \frac{12 \beta}{\pi L^4 \Xi}  \left(\alpha ^2 \tilde{r}_{+}-\tilde{r}_{+}^3+mL^2(c^2 + s^2)\right) \, , \nonumber \\[3mm]
	f_3 =&\; \frac{24 \beta m^2  c^2 s^2 \tilde{r}_+}{\pi \Xi (\tilde{r}_{+}^2-\alpha^2)} \, . 
\end{align}
In these expressions we have used the shorthand $s=\sinh\delta$ and $c = \cosh\delta$. The quantity $\tilde{r}_+ = r_+ + 2m\sinh^2\delta$ is given in~\eqref{eq:KN-param} with $r_+$ being the largest real root of $\Delta_{r}(r_+) = 0$, and $\beta$ is the periodicity of the Euclidean time circle. Regularity of the solution at $r=r_+$ fixes the value of $\beta$ in terms of the solution parameters as
\begin{equation}
\beta = 4\pi(\tilde{r}^2 - \alpha^2)\Bigl[\frac{d\Delta_r}{dr}\Bigr]^{-1}\Big\vert_{r=r_+} \, .
\end{equation}
The mass parameter $m$ can also be related to the value of the radial coordinate $r_+$ at which the $\tau$ circle shrinks to zero size, see e.g.~\cite{Cassani:2019mms}.

Some interesting special cases can be studied from the above general expressions. First, if we set
\begin{equation}
\label{eq:KN-BPS}
\alpha = \mathrm{i}L\bigl(\coth(2\delta) - 1\bigr) \, ,
\end{equation}
the solution~\eqref{eq:KN-AdS} preserves two real supercharges. In this supersymmetric limit, it is not hard to see from~\eqref{eq:fi-KN-AdS} that the bulk contribution to the integrated SdW coefficient will in general depend on the value of the continuous parameter $\delta$. The holographically dual description of this gravitational solution is in terms of the superconformal index of the dual SCFT, i.e. the supersymmetric partition function on $S^1\times_\omega S^2$ with the angular fugacity $\omega$ related to $\delta$. As summarized in Table~\ref{tab:CFT-log}, the logarithmic contribution to the superconformal index is independent of $\omega$ and therefore $C_{\text{local}}$ should not depend on $\delta$. This in turn leads to the following constraints for the total values of the heat kernel coefficients for fields propagating on the AdS-KN background
\begin{equation}
	\label{eq:no-ClocKN}
\sum_{\phi}(-1)^Fc(\phi) = \sum_{\phi}(-1)^Fb_1(\phi) = \sum_{\phi}(-1)^Fb_2(\phi) = 0 \, .
\end{equation}
This confirms the result we have obtained previously in~\eqref{eq:no-Cloc}. We can also consider the non-BPS limit of vanishing electric charge $\delta \rightarrow 0$ in which case the above solution reduces to the Euclidean version of the AdS-Kerr black hole. In this limit,
\begin{equation}
\label{eq:fi-K-AdS}
\begin{split}
	f_1(m,\alpha,0) =&\; \frac{\beta (r_{+}^2+L^2)^2 (r_{+}^2+\alpha ^2)}{\pi  L^4 \Xi \,r_{+} (r_{+}^2-\alpha^2)}\, , \\[1mm]
	f_2(m,\alpha,0) =&\; -\frac{6 \beta (r_{+}^2-L^2) (r_{+}^2-\alpha^2)}{\pi  L^4 \Xi  \,r_{+}} \, , \\[1mm]
	f_3(m,\alpha,0) =&\; 0 \, .
\end{split}
\end{equation}
The AdS-Kerr solution is holographically dual to the grand canonical thermal partition function of a CFT placed on $S_\beta^1 \times S^2$, where the circle has size $\beta$ which sets the (inverse) temperature. This QFT observable can be studied in a saddle-point approximation which defines the so-called thermal effective action of the theory, see~\cite{Benjamin:2023qsc}. The thermal effective action of a given holographic CFT may contain a logarithmic term in the large $N$ limit. If we assume that the coefficient of this $\log N$ term does not depend on the temperature or the spin fugacity, then by our general reasoning the dependence of $C_{\text{local}}$ on the dual continuous parameters arising from~\eqref{eq:fi-K-AdS} can only be suppressed if we impose 
\begin{equation}
\sum_{\phi}(-1)^Fc(\phi) = \sum_{\phi}(-1)^Fb_1(\phi) = 0 \, . 
\end{equation}
This example illustrates how the strong constraints on the spectrum of the gravitational theory we have obtained are not necessarily a consequence of supersymmetry. Rather, they can be implied more generally by the topological nature of logarithmic terms in the large $N$ limit of holographic CFT observables. We are not aware of a general proof that the $\log N$ term in the thermal effective action of any holographic CFT cannot depend on chemical potentials, so our statements in the non-supersymmetric setting are weaker than the ones we made using supersymmetric observables. It would be very interesting to sharpen these statements using QFT arguments.

\section{Explicit KK supergravity examples}
\label{sec:KK}

We have just learned that AdS/CFT dictates that the sum over the spectrum of individual heat kernel coefficients $(c,b_1,b_2)$ must vanish as in~\eqref{eq:no-Cloc}. In this section, we will explicitly check this property by studying the logarithmic coefficient $C$ in the semi-classical expansion of the 4d Euclidean path integral \eqref{eq:sugra-saddle} in certain supergravity theories for which the KK spectrum is known. We will focus in the present section on supergravity theories with at least $\mathcal{N}=2$ supersymmetry. Generically, supersymmetry requires the presence of non-minimal couplings that enter the quadratic expansion of the action around an arbitrary background. Thus, it would seem that we cannot make use of the results we have obtained in Table~\ref{tab:coeffs} for minimally coupled fields. However, we can adapt a useful trick put forward in~\cite{Sen:2012kpz} in the context of asymptotically flat space-times to the AdS setting. Rather than repeating the trace computations in Appendix~\ref{App:coeffs} with non-minimal couplings, we will instead supersymmetrize the $C_{\text{local}}$ term~\eqref{eq:C-local} directly. To do so, we use the results of~\cite{Bobev:2021oku} to relate the four-derivative invariants controlling the fourth SdW coefficient. On any background of $\mathcal{N}=2$ minimal supergravity, this relation reads
\begin{equation}
\label{eq:tshirt}
W^2\vert_{\mathcal{N}=2} = E_4 + \frac{4}{L^2}\,\Bigl[R + \frac{6}{L^2} - F_{\mu\nu}F^{\mu\nu}\Bigr] \, ,
\end{equation}
modulo terms that vanish using the equations of motion~\eqref{eom}. Here the subscript on the left-hand side denotes the supersymmetrization of the Weyl-squared density~\eqref{eq:W2}, see~\cite{Bobev:2021oku} for the explicit expression. The relation~\eqref{eq:tshirt} implies that the bulk part of the fourth SdW coefficient~\eqref{eq:a4-bulk} in all theories that can be expressed in an $\mathcal{N}=2$ language is straightforwardly supersymmetrized:
\begin{equation}
(4\pi)^2a_4^{\text{bulk}}\vert_{\mathcal{N}\geq 2} = (c - a_E)\,E_4 + \Bigl(b_1 - \frac{c}{6}\Bigr)\,R^2 + \Bigl(b_2 + \frac{c}{3}\Bigr)\,RF_{\mu\nu}F^{\mu\nu} \, .
\end{equation}
Since all non-minimal couplings required by supersymmetry have been taken into account in this formula, we can use the minimally coupled results of Table~\ref{tab:coeffs} to infer the value of the heat kernel coefficients $(a_E,c,b_1)\vert_{\mathcal{N}\geq2}$ in supersymmetric theories,
\begin{equation}
\label{eq:hk-non-min}
a_E\vert_{\mathcal{N}\geq2} = a_E - c \, , \qquad c\vert_{\mathcal{N}\geq2} = 0 \, , \qquad b_1\vert_{\mathcal{N}\geq2} = b_1 - \frac{c}{6} \, .
\end{equation}
In the rest of this paper, we will systematically compute heat kernel coefficients as if all fields were minimally coupled according to Table~\ref{tab:coeffs}, and only restore the effect of non-minimal couplings using~\eqref{eq:hk-non-min} when necessary. We also note that if a particular spectrum produces a vanishing total $\sum_\phi (-1)^F c(\phi) = 0$, the relation~\eqref{eq:hk-non-min} immediately shows that $\mathcal{N}\geq2$ non-minimal couplings have no bearing on the other heat kernel coefficients. We will make extensive use of this fact  below when studying specific KK spectra.

\subsection{KK supergravity on $S^7$}
\label{sec:KK:S7}

As a first example, we consider the simple case of eleven-dimensional supergravity compactified on the round $S^7$. Since the length scales of the internal and external spaces are of the same order, this compactification gives rise to maximal gauged supergravity in 4d along with an infinite KK tower of massive fields. All fields can be arranged into 4d $\mathcal{N}=8$ supermultiplets, and the complete spectrum was obtained long ago in~\cite{Freedman:1983na,Sezgin:1983ik,Biran:1983iy,Casher:1984ym,Duff:1986hr}. In what follows, we will denote the SO(8) $R$-symmetry representations by their Dynkin labels $(\alpha_1,\alpha_2,\alpha_3,\alpha_4)$ and follow the convention of Appendix A in~\cite{Bobev:2021wir} for the SO(8) triality frame. The massless and massive field content of the KK supergravity theory is indexed by a single positive integer $k$ that labels the KK level and also controls the mass of the field. The spectrum is summarized in Table~\ref{tab:S7} and Table~\ref{tab:S7-m}. \\
\begin{table}[t]
	\centering
	\renewcommand*{\arraystretch}{1.4}
	\begin{tabular}{|c|c|c|c|}
		\hline
		spin & SO(8) irrep & Dynkin label & $\Delta$ \\
		\hline
		2 & $\mathbf{1}$ & $(0,0,0,0)$ & 3 \\
		\hline
		3/2 & $\mathbf{8}_s$ & $(0,0,0,1)$ & $\frac52$ \\
		\hline
		1 & $\mathbf{28}$ & $(0,1,0,0)$ & 2 \\
		\hline
		1/2 & $\mathbf{56}_s$ & $(1,0,1,0)$ & $\frac32$ \\
		\hline
		$0_+$ & $\mathbf{35}_v$ & $(2,0,0,0)$ & 1 \\
		\hline
		$0_-$ & $\mathbf{35}_c$ & $(0,0,2,0)$ & 2 \\
		\hline
	\end{tabular}
	\caption{The massless $\mathcal{N}=8$ supermultiplet.\label{tab:S7}}
\end{table}
\begin{table}[t]
	\centering
	\renewcommand*{\arraystretch}{1.4}
	\begin{tabular}{|c|c|c|}
		\hline
		spin & Dynkin label & $\Delta$ \\
		\hline
		2 & $(k,0,0,0)_{k\geq0}$ & $3 + \frac{k}{2}$ \\
		\hline
		\multirow{2}{*}{3/2} & $(k,0,0,1)_{k\geq0}$ & $\frac52 + \frac{k}{2}$ \\[-1mm] 
		& $(k-1,0,1,0)_{k\geq1}$ & $\frac72 + \frac{k}{2}$ \\
		\hline
		\multirow{3}{*}{1} & $(k,1,0,0)_{k\geq0}$ & $2 + \frac{k}{2}$ \\[-1mm] 
		& $(k-1,0,1,1)_{k\geq1}$ & $3 + \frac{k}{2}$ \\[-1mm]
		& $(k-2,1,0,0)_{k\geq2}$ & $4 + \frac{k}{2}$ \\
		\hline
		\multirow{4}{*}{1/2} & $(k+1,0,1,0)_{k\geq0}$ & $\frac32 + \frac{k}{2}$ \\[-1mm] 
		& $(k-1,1,1,0)_{k\geq1}$ & $\frac52 + \frac{k}{2}$ \\[-1mm]
		& $(k-2,1,0,1)_{k\geq2}$ & $\frac72 + \frac{k}{2}$ \\[-1mm]
		& $(k-2,0,0,1)_{k\geq2}$ & $\frac92 + \frac{k}{2}$ \\
		\hline
		\multirow{3}{*}{$0_+$} & $(k+2,0,0,0)_{k\geq0}$ & $1 + \frac{k}{2}$ \\[-1mm] 
		& $(k-2,2,0,0)_{k\geq2}$ & $3 + \frac{k}{2}$ \\[-1mm]
		& $(k-2,0,0,0)_{k\geq2}$ & $5 + \frac{k}{2}$ \\
		\hline
		\multirow{2}{*}{$0_-$} & $(k,0,2,0)_{k\geq0}$ & $2 + \frac{k}{2}$ \\[-1mm] 
		& $(k-2,0,0,2)_{k\geq2}$ & $4 + \frac{k}{2}$ \\
		\hline
	\end{tabular}
	\caption{Massive $\mathcal{N}=8$ supermultiplets at KK level $k$. For $k=0$ we recover Table~\ref{tab:S7}.\label{tab:S7-m}}
\end{table}

To compute the logarithmic correction to the path integral around a given 4d background $\mathcal{M}$ in this KK theory, we use the general formulae of Section~\ref{sec:sugra}. The local contribution \eqref{eq:C-local} is controlled by the SdW coefficient, which receives a bulk contribution of the form~\eqref{eq:a4-bulk}. In Appendix~\ref{App:KK:S7}, we explain how to compute the heat kernel coefficients $(a_E,c,b_1)$ by combining the results summarized in Table~\ref{tab:coeffs} together with the KK spectrum given in Tables~\ref{tab:S7} and~\ref{tab:S7-m}. For the $c$ and $b_1$ coefficients, we find a non-trivial cancellation ``level-by-level'',\footnote{This result is reminiscent of the level-by-level cancellation in the one-loop beta function of $\mN=8$ gauged supergravity observed in~\cite{Gibbons:1984dg}.}
\begin{equation}
	\label{eq:c-b1-S7}
	c(k) = 0 \, , \qquad b_1(k) = 0 \, , \qquad \forall \; k \geq 0 \, .
\end{equation}
Since the sum over the spectrum reduces in this simple case to a sum over the KK level $k$, we obtain
\begin{equation}
	c^{\text{tot}} \equiv \sum_{k\geq 0} (-1)^F c(k) = 0 \, , \quad \text{and} \quad b_1^{\text{tot}} \equiv \sum_{k\geq 0} (-1)^F b_1(k) = 0 \, .
\end{equation}
Since $c^{\text{tot}}$ vanishes,~\eqref{eq:hk-non-min} implies that the vanishing of $b_1^{\text{tot}}$ also holds for non-minimally coupled fields around any background $\mathcal{M}$. Likewise, the vanishing of $\sum_k (-1)^F c(k)\vert_{\mathcal{N}=8}$ automatically follows from~\eqref{eq:hk-non-min}. These results are in perfect agreement with the bootstrap constraints in~\eqref{eq:no-Cloc} applied to KK supergravity on the 7-sphere. Having knowledge of the full spectrum also allows us to compute the $a_E$ coefficient, for which no such level-by-level cancellation occurs. Instead we find that its contribution to $C_{\text{local}}$ is given by
\begin{equation}
\label{eq:aE-S7}
	-\frac{1}{72}\,\chi(\mathcal{M})\sum_{k\geq 0}(k+1)(k+2)(k+3)^2(k+4)(k+5) \, ,
\end{equation}
where $\chi(\mathcal{M})$ is the regularized Euler characteristic of the 4d background. For a generic background $\mathcal{M}$, another contribution to $C_{\text{local}}$ is of the form
\begin{equation} 
\label{eq:b2-S7}
	\frac{1}{16\pi^2}\Bigl(\int_\mathcal{M} d^4x \sqrt{g}\,R\,F_{\mu\nu}F^{\mu\nu}\Bigr)\,b_2^{\text{tot}} \, ,
\end{equation}
where $b_2^{\text{tot}} = \sum_k (-1)^F b_2(k)$, but this term vanishes according to~\eqref{eq:no-Cloc}. Lastly, the local part of the logarithmic correction also receives contributions from surface terms in the expression of the SdW coefficient~\eqref{eq:a4}. Since these cannot be written compactly for generic backgrounds and boundary conditions, we generically denote them by $S(\mathcal{M})$ after summing over the spectrum. With this notation, the full local contribution is
\begin{equation}
\label{eq:Cloc-S7-M}
	C_{\text{local}}(\mathcal{M}) = S(\mathcal{M}) - \frac{1}{72}\,\chi(\mathcal{M})\sum_{k\geq 0}(k+1)(k+2)(k+3)^2(k+4)(k+5) \, .
\end{equation}
Clearly this expression contains divergent sums that need to be appropriately regularized. Before doing so, we recall that yet another contribution to the logarithmic term is given by $C_{\text{non-local}}$ in~\eqref{eq:C-non-local}. As discussed in Section~\ref{sec:sugra:non-local}, the corresponding spectral problem for generic backgrounds $\mathcal{M}$ remains out of reach at present. So from now on, we will focus on specific backgrounds of the $S^7$ KK supergravity theory where we can make progress.\\

An interesting (and tractable) case is to come back to the situation where $\mathcal{M}$ is pure EAdS$_4$. There, we argued in Section~\ref{sec:bootstrap} that the surface terms could be holographically renormalized away prior to summing over the spectrum of the KK theory. As a result, we have $S(\text{EAdS}_4) = 0$. In addition, the KK spectrum does not include any 2-form fields, which by~\eqref{eq:Cnon-local-AdS4} implies that the non-local contribution vanishes. In total, we arrive at the logarithmic correction to the path integral
\begin{equation}
\label{eq:C-tot-AdS4}
C(\text{EAdS}_4) = -\frac{1}{72}\,\sum_{k\geq 0}(k+1)(k+2)(k+3)^2(k+4)(k+5) \, ,
\end{equation}
where we have used that $\chi(\text{EAdS}_4) = 1$. Note that we could have arrived at this result starting from~\eqref{eq:Cloc-S3-G} and explicitly implementing the sum over the spectrum by keeping track of the dimensions of the SO(8) representations and the conformal dimensions of the dual operators at a given KK level. Indeed,~\eqref{eq:C-tot-AdS4} is compatible with~\cite{Camporesi:1993mz,Liu:2016dau,Binder:2021euo}. 

What remains is to discuss a possible regularization of the infinite series. For instance, the method used in~\cite{Liu:2016dau} consists of introducing a $z^k$ regulator to the summand with $|z| < 1$ and defining the regularized series as the finite term in the expansion around $z=1$. This attaches the value
\begin{equation}
\label{eq:C-tot-AdS4-reg-Liu}
	C(\text{EAdS}_4) \stackrel{?}{=} 0 \, ,
\end{equation} 
to the logarithmic coefficient in the one-loop EAdS$_4$ free energy. In Appendix~\ref{app:reg}, we present two other methods based on zeta-function regularization that yield the same result. However,~\eqref{eq:C-tot-AdS4-reg-Liu} should give us pause. Indeed, if correct, we are then forced to conclude that the bulk computation of the logarithmic coefficient cannot possibly agree with the holographically dual ABJM result
\begin{equation}
\label{eq:C-M2-round}
\mathcal{C}(S^3_{b=1}) = -\frac14 \, ,
\end{equation}
given in Table~\ref{tab:CFT-log}. At this stage, it seems that there are a couple of possible resolutions for this discrepancy. Recall that in 11d supergravity on EAdS$_4 \times S^7$~\cite{Bhattacharyya:2012ye}, the logarithmic term comes entirely from the zero mode of the 2-form ghost required to quantize the 3-form potential of the theory. After dimensional reduction, we have insisted that the KK supergravity fields arrange themselves into $\mathcal{N}=8$ multiplets that do not accommodate 2-form fields. As a result, the non-local contribution in the 4d theory vanishes on EAdS$_4$ according to~\eqref{eq:Cnon-local-AdS4}. To remedy this, we could opt to dualize some (or all) of the massless scalars in the KK theory to 2-forms and hope to obtain a non-zero contribution to the $C_{\text{non-local}}$ part of the logarithmic correction. While this change of duality frame explicitly breaks the $\mathcal{N}=8$ multiplet structure, it may be necessary to obtain a match between bulk and boundary contributions at the one-loop level. However, a short computation shows that this cannot resolve the discrepancy. The heat kernel coefficient $a_E$ for fluctuations of a massless 2-form can be obtained by the same trace computations that led to the results in Table~\ref{tab:coeffs}. It was shown in~\cite{Larsen:2015aia} that there is a simple relation between the $a_E$ coefficient for a massless scalar field and a massless 2-form,
\begin{equation}
	a_E(\text{2-form}) = a_E(\text{scalar}) - \frac12 \, .
\end{equation}
This shows that dualizing $q$ scalar fields into 2-forms shifts the local contribution to the logarithmic coefficient as
\begin{equation}
	C_{\text{local}} \longrightarrow C_{\text{local}} + q \, .
\end{equation}
On the other hand, the non-local contribution also acquire a shift after dualizing,
\begin{equation}
C_{\text{non-local}} \longrightarrow C_{\text{non-local}} - q \, ,
\end{equation}
where we have used~\eqref{eq:Cnon-local-AdS4} with $j=0$. In this case, the two shifts precisely cancel and we see that the total logarithmic coefficient $C(\text{EAdS}_4)$ does not depend on the duality frame. This shows that dualizing massless fields cannot help in resolving the discrepancy between bulk and boundary computations. A second possible resolution of the tension between~\eqref{eq:C-tot-AdS4-reg-Liu} and~\eqref{eq:C-M2-round} would be to find a regularization scheme that yields a non-zero result for the total $\sum_k (-1)^F a_E(k)$ contribution to $C_{\text{local}}$. Although we have discussed three distinct methods to regularize the series and arrive at a vanishing result, it is interesting to note that the regularization
\begin{equation}
\label{eq:good-reg}
\sum_{k\geq 0}(k+1)(k+2)(k+3)^2(k+4)(k+5) = 24 \, ,
\end{equation}
is compatible with AdS/CFT expectations. Indeed, this would imply 
\begin{equation}
C(\text{EAdS}_4) = -\frac13 \, , 
\end{equation}
which together with the holographic dictionary for class I SCFTs~\eqref{eq:holo-dict-gen}, $L^2/G_N \sim N^{\frac32}$, precisely matches~\eqref{eq:C-M2-round}. Unfortunately, we are not aware of a rigorous way to arrive at the result~\eqref{eq:good-reg}. If it does exist, it will be interesting to understand why holography seems to favor this putative regularization method over the ones regularizing the infinite sum to zero.\\

We can provide arguments in favor of the regularization~\eqref{eq:good-reg} by considering supergravity backgrounds $\mathcal{M}$ that are not pure EAdS$_4$. For the 4d background whose conformal boundary is the U(1)$\times$U(1) squashed 3-sphere, we have also explained previously how the surface terms can be renormalized away. The Euler characteristic of this background is the same as pure EAdS$_4$ since $\chi$ cannot depend on the squashing parameter, and therefore~\eqref{eq:good-reg} implies that the Euclidean path integral around the squashed background contains a logarithmic term $-\frac14\log N$, again in agreement with the ABJM results in Table~\ref{tab:CFT-log}.

Let us also consider the EAdS$_2 \times \Sigma_{\mathfrak{g}}$ near-horizon region of the Romans solution with genus $\mathfrak{g} > 1$. As we have explained in Section~\ref{sec:bootstrap:Romans} the surface term can be holographically renormalized to zero. In addition, we have $\chi(\text{EAdS}_2 \times \Sigma_{\mathfrak{g}}) = 2(1 - \mathfrak{g})$ and putting this together with the non-local contribution in~\eqref{eq:C:non-local-AdS2xS}, we obtain
\begin{equation}
C(\text{EAdS}_2 \times \Sigma_{\mathfrak{g}}) = (\mathfrak{g}-1)\Bigl[64 + \frac{1}{36}\sum_{k\geq 0}(k+1)(k+2)(k+3)^2(k+4)(k+5)\Bigr] - 4 \, .
\end{equation}
Regardless of the regularization chosen for the infinite series, this cannot match the logarithmic coefficient in the TTI of the boundary theory given in Table~\ref{tab:CFT-log},
\begin{equation}
\label{eq:C-M2-TTI}
\mathcal{C}(S^1\times\Sigma_{\mathfrak{g}}) = \frac12(\mathfrak{g} - 1) \, .
\end{equation}
However, the comparison in this case is more subtle since there could be degrees of freedom localized outside the near-horizon region of the Romans solution that contribute to the logarithmic term seen from the asymptotic boundary. In fact, it was argued in~\cite{Liu:2017vbl} that taking into account the full Romans geometry modifies the zero-mode counting in a drastic way compared to Section~\ref{sec:sugra:non-local-AdS2xS}. In particular, they argued that on the full geometry,
\begin{equation}
\label{eq:Cnon-local-Romans}
	C_{\text{non-local}}(\text{Romans}) = 2(1 - \mathfrak{g})\sum_{\text{massless 2-form}}(-1)^F(-j-1) \, .
\end{equation}
In the absence of two-forms in the spectrum, we therefore find
\begin{equation}
C(\text{Romans}) = \frac{2}{3}(\mathfrak{g} - 1) \, ,
\end{equation}
after using~\eqref{eq:good-reg}. With the holographic dictionary relevant for M2-branes, this again perfectly matches~\eqref{eq:C-M2-TTI} found in the ABJM TTI.

One last example we can consider is the supersymmetric AdS-KN background holographically dual to the SCI of the boundary theory. Because we can again set $S(\text{KN}) = 0$ using holographic renormalization, combining $\chi(\text{KN}) = 2$ together with the regularization~\eqref{eq:good-reg} yields
\begin{equation}
C(\text{KN}) = -\frac23 + C_{\text{non-local}}(\text{KN}) \, .
\end{equation}
For this to match the ABJM result
\begin{equation}
\mathcal{C}(S^1 \times_\omega S^2) = -\frac12 \, ,
\end{equation} 
upon using $L^2/G_N \sim N^{\frac32}$, the zero-modes of various Laplacian operators on the AdS-KN background must be such that the non-local contribution~\eqref{eq:C-non-local} vanishes. This is a highly non-trivial prediction for the spectral problem of Laplace-type differential operators on this complicated background.

\subsection{A conjecture for $C$ and black hole entropy}
\label{sec:log-conj}

At this stage, we assume that the regularization~\eqref{eq:good-reg} can be made rigorous and we will use it to predict the logarithmic correction of general 4d backgrounds. For this, we observe that in all bulk geometries holographically dual to the squashed 3-sphere, TTI and SCI observables of the ABJM theory, the non-local contribution vanishes and the logarithmic term comes entirely from the heat kernel coefficient $a_E^{\text{tot}}$ after summing over the spectrum. The situation should be contrasted with the 11d computations in~\cite{Bhattacharyya:2012ye, Liu:2017vbl} where the coefficient of the log correction can only come from zero modes since the heat kernel vanishes in odd dimensions. We now would like to conjecture that, in all minimal gauged supergravity background that are relevant for AdS$_4$/CFT$_3$, a non-local contribution to $C$ can only  arise from the presence of massless 2-forms in the spectrum of the theory. In KK supergravity on $S^7$, maximal supersymmetry in 4d does not accommodate such 2-forms (in a duality frame where we keep all 70 massless scalars), which then becomes the reason behind the fact that $C_{\text{non-local}}$ vanishes. This conjecture is certainly compatible with~\eqref{eq:Cnon-local-AdS4} and~\eqref{eq:Cnon-local-Romans}, and it would be most interesting to derive similar formulae for other asymptotically locally AdS$_4$ backgrounds. 

Assuming that the general picture holds, we can go ahead and compute the logarithmic correction in the Euclidean path integral around any $S^7$ KK supergravity background $\mathcal{M}$. The result for $C$ takes a very simple form:
\begin{equation}
\label{eq:C-S7-gen}
	C(\mathcal{M}) = -\frac13\,\chi(\mathcal{M}) \, ,
\end{equation}
where we have also assumed that the boundary contribution $S(\mathcal{M})$ can be set to zero using holographic renormalization on $\mathcal{M}$ with the counter-term~\eqref{eq:CT}. Recall that this was checked explicitly for all backgrounds of Section~\ref{sec:bootstrap}. We note that~\eqref{eq:C-S7-gen} takes the same functional form as the one derived in~\cite{Hristov:2021zai} based on 4d $\mathcal{N}=2$ gauged supergravity localization. In that work, the specific field content of the theory was left arbitrary and the dependence of the logarithmic correction on the Euler characteristic of $\mathcal{M}$ was derived using an index theorem. In this framework, each multiplet of the 4d $\mathcal{N}=2$ theory contributes a fixed amount to $C$, and that amount was computed for vector and hypermultiplets in~\cite{Hristov:2021zai}. However, the contribution from the graviton and KK multiplets was left unspecified. Our heat kernel result~\eqref{eq:C-S7-gen} effectively resums these contributions in the case of KK supergravity on $S^7$. Moreover, it predicts a $\log N$ term in the large $N$ limit of the dual ABJM free energy on $\partial\mathcal{M}$ with a coefficient
\begin{equation}
\label{eq:C-M2-gen}
	\mathcal{C}(\partial\mathcal{M}) = -\frac14\,\chi(\mathcal{M}) \, .
\end{equation}
This holographic formula agrees with the SCFT results in Table~\ref{tab:CFT-log} but is also valid regardless of the amount of supersymmetry, so that we can use it to make predictions for non-BPS backgrounds with interesting CFT duals. One such example is the AdS-Taub-NUT solution discussed in Section~\ref{sec:AdSTNsol}. We have already shown in~\eqref{eq:c-b1-S7} that $c^{\text{tot}} = b_1^{\text{tot}} = 0$ in KK supergravity on $S^7$. According to~\eqref{eq:CT-gen}, this implies that there is no logarithmic term in the large $N$ expansion of the stress tensor two-point function coefficient $C_T$ in ABJM theory. This is in agreement with our general formula~\eqref{eq:C-M2-gen}, since the Euler characteristic of AdS-Taub-NUT is a constant equal to one, and taking derivatives with respect to the squashing parameter or the NUT charge as in~\eqref{eq:CT-def} trivially gives zero.

Another non-BPS background we can consider is the so-called AdS soliton~\cite{Horowitz:1998ha}. The Euclidean metric reads
\begin{equation}
\label{eq:AdS-soliton}
ds^2 = \frac{r^2}{L^2}\Bigl[\Bigl(1 - \frac{r_0^3}{r^3}\Bigr)\,d\tau^2 + dx^2 + dy^2\Bigr] + \frac{L^2}{r^2}\Bigl(1 - \frac{r_0^3}{r^3}\Bigr)^{-1}dr^2 \, ,
\end{equation}
and the asymptotic boundary at $r\rightarrow \infty$ is the three-manifold $S^1_\beta \times \mathbb{R}^2$. Here the inverse temperature $\beta$ is related to the parameter $r_0$ by demanding absence of conical singularity in the bulk, which leads to $r_0 = 4\pi L^2/3\beta$. When the ABJM theory is put on this thermal background, the free energy computes the one-point function of the stress tensor which is non-vanishing since conformal invariance is broken by the finite temperature.\footnote{For a review of thermal observables in CFT see~\cite{Iliesiu:2018fao}, and for a discussion of higher-derivative effects in the stress tensor one-point function for general $\mathcal{N}=2$ holographic CFTs see~\cite{Bobev:2023ggk}.} Using~\eqref{eq:AdS-soliton}, it is straightforward to show that the Euler characteristic of the AdS soliton vanishes. Thus, by~\eqref{eq:C-M2-gen}, we predict that the large $N$ expansion of the stress tensor one-point function coefficient does not contain a $\log N$ term,
\begin{equation}
\mathcal{C}(S^1_\beta \times \mathbb{R}^2) = 0 \, .
\end{equation}
To derive this result from a QFT analysis seems arduous, because it pertains to strongly-coupled theories at finite temperature. Our holographic formula~\eqref{eq:C-M2-gen} elegantly sidesteps this difficulty and allows us to make a strong prediction.\\

In light of~\eqref{eq:C-S7-gen}, it is also instructive to consider non-BPS asymptotically AdS black hole backgrounds embedded in 11d supergravity on $S^7$. The simplest such example is the Euclidean AdS-Schwarzschild solution
\begin{equation}
ds^2 = \Bigl(\frac{r^2}{L^2} + 1 - \frac{m}{r}\Bigr)\,d\tau^2 + \Bigl(\frac{r^2}{L^2} + 1 - \frac{m}{r}\Bigr)^{-1}dr^2 + r^2d\Omega_2^2 \, .
\end{equation}
The location of the outer horizon $r_+$ is related to the mass parameter as $m = r_+(r_+^2/L^2 + 1)$. The regularized Euler characteristic is given by $\chi(\text{AdS-Sch}) = 2$ which, by~\eqref{eq:C-S7-gen}, implies that the logarithm of the Euclidean path integral in the semi-classical expansion contains a $\log L$ term:
\begin{equation}
\label{eq:IonshelSch}
I = I_0 + \frac13\log(L^2/G_N) + \mathcal{O}(1) \, .
\end{equation}
Here $I_0$ is the classical on-shell action of the solution. We can now make use of the quantum statistical relation~\cite{Gibbons:1976ue,Gibbons:2004ai}
\begin{equation}
I = \beta M - S \, ,
\end{equation}
to find that the AdS-Schwarzschild black hole entropy receives a logarithmic correction at one-loop
\begin{equation}
S_{\text{AdS-Sch}} = \frac{A_{\rm H}}{4G_N} - \frac13 \log(A_{\rm H}/G_N) + \mathcal{O}(1) \, ,
\end{equation}
with $A_{\rm H} = 4\pi r_+^2$ the horizon area. To obtain this result, we have assumed that we have a large black hole in AdS, i.e. the length scale $r_+$ that sets the size of the black hole horizon is comparable to the AdS scale $L$. In the logarithmic term, this means we can trade $L^2 \sim r_+^2$ up to $\mathcal{O}(1)$ terms.

Using the same approach we can also calculate the correction to the entropy and on-shell action of the general Kerr-Newmann black hole in AdS$_4$ presented in Section~\ref{sec:KNsol}. The quantum statistical relation in this case reads
\begin{equation}
I = \beta M - S - \beta\Phi Q-\beta\omega J \, ,
\end{equation}
where $Q$ and $J$ are the electric charge and angular momentum and $\Phi$ and $\omega$ are the corresponding electric chemical potential and angular velocity. The logarithmic correction to the on-shell action takes the same form as in \eqref{eq:IonshelSch} and since the black hole background is not changed when computing this correction, we conclude that the Bekenstein-Hawking entropy of the AdS-KN black hole is corrected as follows
\begin{equation}\label{eq:SKNlogcorr}
S_{\text{AdS-KN}} = \frac{A_{\rm H}}{4G_N} - \frac13 \log(A_{\rm H}/G_N) + \mathcal{O}(1) \, .
\end{equation}
The explicit expressions for the regularized on-shell action $I_0$ and the horizon area $A_{\rm H}$, as well as those for the black hole charges and fugacities, are quite lengthy and can be found in \cite{Cassani:2019mms,Bobev:2021oku}. Importantly this result for the logarithmic correction to the AdS-KN entropy is valid in a thermodynamic ensemble of fixed $(T,\Phi,\omega)$ since this is the natural ensemble used in the evaluation of the on-shell action. One is of course free to change to a different thermodynamic ensemble by employing a suitable Legendre transformation. However, as emphasized by Sen \cite{Sen:2012dw}, upon such a change in ensemble one should carefully track how the black hole charges and fugacities scale with the large parameter in the theory that controls the logarithmic corrections, i.e. the rank $N$ of the gauge group in our M-theory examples, since this could lead to additional (constant) contributions to the logarithmic coefficient.  

Finally, we note that the AdS-KN black hole admits a supersymmetric limit as discussed in Section~\ref{sec:KNsol}. By taking the $Q\rightarrow 0$ limit of the Romans solution presented in Section~\ref{sec:bootstrap:Romans} with $\mathfrak{g}>1$, we can also study the supersymmetric AdS-Reissner-Nordström (AdS-RN) black hole. In both of these examples we find that the supersymmetric black hole entropy receives the same type of logarithmic correction as in \eqref{eq:SKNlogcorr}, i.e. $- \frac{\chi}{6} \log(A_{\rm H}/G_N)$. For theories arising from M2-branes this can be compared to the logarithmic corrections in the large $N$ limit of the superconformal index (for AdS-KN) or the topologically twisted index (for AdS-RN). Comparing the supergravity logarithmic corrections to the corresponding $\log N$ entries for the $S^1 \times_{\omega} S^2$ and $S^1\times\Sigma_{\mathfrak{g}}$ path integrals in Table~\ref{tab:CFT-log}, we indeed find perfect agreement. This constitutes an important precision test of the microscopic counting of supersymmetric black hole entropy in AdS$_4$ using our 4d supergravity approach.

\subsection{Other KK supergravity examples}
\label{sec:KK:other}

We now proceed with examples of KK compactifications of 11d supergravity to four dimensions with less supersymmetry. We focus on three examples for which the KK spectrum is known in detail and comment on a fourth example that illustrates some challenges and subtleties.

\subsubsection{KK supergravity on $S^7/\mathbb{Z}_k$}
\label{sec:KK-S7-k}

The KK supergravity spectrum of the AdS$_4\times S^7/\mathbb{Z}_k$ 11d supergravity solution\footnote{We caution the reader that here $k$ refers to the order of the orbifold and should not be confused with the label of the KK level in the previous section.} can be obtained by branching the $\mN=8$ KK spectrum labeled by $\mathfrak{so}(8)$ representations into the $\mN=6$ KK spectrum labeled by the $\mathfrak{so}(6)\oplus\mathfrak{u}(1)$ representations and keeping the supermultiplets with vanishing U(1) charges modulo $k$ only \cite{Liu:2016dau}. In Appendix~\ref{App:KK:S7modk} we present the resulting $\mN=6$ KK spectrum following this procedure, and then evaluate the heat kernel coefficients. The KK level is now labelled by two integers $h$ and $r$ that satisfy $h - |r| \geq 0$ and $k\,|\,2r$. We again find a non-trivial cancellation level-by-level for the $c$ and $b_1$ coefficients, which immediately leads to 
\begin{equation}
	c^\text{tot}=b_1^\text{tot}=0 \, .
\end{equation}
This provides another non-trivial test of our bootstrap results in Section~\ref{sec:bootstrap}. The total $a_E$ coefficient is obtained as the following sum over the spectrum,
\begin{equation}
		a_E^{\text{tot}}=\sum_{h=1}^\infty\fft{1+2h}{24}\sum_{\ell=-\left\lfloor h/k'\right\rfloor}^{\left\lfloor h/k'\right\rfloor}\Bigl[h(1+h)(-4+5h+5h^2)+(7-10h-10h^2)(\ell k')^2+5(\ell k')^4\Bigr] \, ,
\end{equation}
where we have defined $k'$ as
\begin{equation}
	k'=\begin{cases}
		k/2 &\text{for} \;\; k\in2\mathbb Z \\
		k &\text{for} \;\; k\in2\mathbb Z+1
	\end{cases} \, ,
\end{equation}
and imposed the constraints $h - |r| \geq 0$ and $k\,|\,2r$ to translate the sum over $r$ to a finite sum over an auxiliary label $\ell$. Computing this finite sum explicitly, we obtain:
\begin{align}
\label{N=6:aEtot:k-finite}
		a_E^{\text{tot}} =&\; \sum_{h=1}^\infty\fft{(1+2h)(1+2\left\lfloor h/k'\right\rfloor)}{72}\bigg[3h(1+h)(-4+5h+5h^2) \\
		&\quad+\left\lfloor h/k'\right\rfloor\left(1+\left\lfloor h/k'\right\rfloor\right)k^2(7-10h+10h^2-k^2)+3\left\lfloor h/k'\right\rfloor^2\left(1+\left\lfloor h/k'\right\rfloor\right)^2k^4\bigg]\,. \nonumber
\end{align}
In addition,~\eqref{eq:no-Cloc} imposes that $b_2^{\text{tot}} = 0$ and we arrive at the conclusion that
\begin{equation}
\label{eq:Cloc-S7-modk}
	C_{\text{local}}(\mathcal{M}) = -2\,\chi(\mathcal{M})\,a_E^{\text{tot}} \, ,
\end{equation}
modulo surface terms $S(\mathcal{M})$ that can be renormalized to zero on all 4d backgrounds of interest. According to our zero mode conjecture in Section~\ref{sec:log-conj},~\eqref{eq:Cloc-S7-modk} is in fact the complete contribution to the logarithmic coefficient in the Euclidean path integral. AdS/CFT then demands that the regularized value of the infinite series~\eqref{N=6:aEtot:k-finite} must be independent of $k$ and equal to $1/6$. Once again, we are not aware of a regularization method that would yield such a simple result. In fact, the situation in KK supergravity on $S^7/\mathbb{Z}_k$ is richer than for $k=1$, since we can now consider the limit where $k\rightarrow \infty$ and study logarithmic corrections to supersymmetric observables in the IIA limit of the ABJM theory. In this limit, we have
\begin{equation}
\label{N=6:aEtot:k-infinite}
	\lim_{k\to\infty}a_E^\text{tot}=\sum_{h=1}^\infty\fft{h(1+h)(1+2h)(-4+5h+5h^2)}{24} \, ,
\end{equation}
and the corresponding supergravity logarithmic coefficient is
\begin{equation}
C(\mathcal{M}) = -2\,\chi(\mathcal{M})\sum_{h=1}^\infty\fft{h(1+h)(1+2h)(-4+5h+5h^2)}{24} \, .
\end{equation}
For this to be compatible with the dual $S^3$ free energy, we must require that the series be regularized to $1/12$ so that $\mathcal{C}(S^3) = -1/6$, after using the dictionary~\eqref{eq:holo-dict-gen} for SCFTs whose dual uplifts to massless Type IIA. However, this regularization then fails to give the correct log coefficient for the IIA limit of the ABJM TTI given in Table~\ref{tab:CFT-log}. We will not dwell on this further, but simply view it as a clear indication that a holographic match heavily depends on the choice of regularization when summing over the infinitely many fields of the dual KK spectra. Moreover, one should keep in mind that in the Type IIA limit there could also be $\log\lambda$ corrections to the free energy, where $\lambda = N/k$ is the 't Hooft coupling. Such terms could compete with the $\log N$ term we study above and disentangling these two contributions could lead to additional subtleties in the bulk supergravity calculation and the holographic comparison.

\subsubsection{mABJM}
\label{subsubsec:mABJM}

We now consider the gravity dual of a 3d $\mathcal{N}=2$ SCFT sometimes referred to as the mABJM theory. The theory can be obtained by focusing on the $k=1,2$ ABJM model and adding a particular superpotential mass term that involves also the light monopole operators responsible for the enhancement to  $\mathcal{N}=8$ supersymmetry in the ABJM theory. The RG flow that ensues ends at an IR fixed point which has been studied with various QFT methods, see \cite{Benna:2008zy,Jafferis:2011zi,Bobev:2018uxk} for more details. Notably the known $\log N$ corrections to various partition functions in the mABJM theory coincide with the ones in the parent ABJM model. 

The holographic studies of the mABJM theory are facilitated by the presence of an explicitly known AdS$_4$ dual background. This AdS solution was first found as a supersymmetric critical point of the potential in the 4d $\mathcal{N}=8$ ${\rm SO}(8)$ gauged supergravity in \cite{Warner:1983vz}. The 4d solution was then uplifted to a background of 11d supergravity in \cite{Corrado:2001nv} which we will refer to as the CPW solution. Importantly the full KK spectrum of the CPW solution was explicitly presented in \cite{Klebanov:2008vq,Malek:2020yue}. This allows us to use the results for the heat kernel coefficients in Table~\ref{tab:coeffs} and calculate the total contributions from all KK modes. The calculation proceeds in an analogous way to the one for the AdS$_4\times S^7$ solution presented in Section~\ref{sec:KK:S7}. There are a number of technical differences due to the lower amount of supersymmetry and the treatment of long multiplets. We discuss all this in Appendix~\ref{App:KK:mABJM} where we present the details of this analysis. While the calculation is somewhat arduous, the final result is easy to state and simple. The $c$ and $b_1$ coefficients once again vanish level-by-level,  and therefore we have
\begin{equation}
c^{\text{tot}} = b_1^{\text{tot}} =0 \, ,
\end{equation}
in agreement with~\eqref{eq:no-Cloc}. The $a_E$ coefficient is non-vanishing and one finds that the sum over KK levels is given by
\begin{equation}
\label{eq:aE-S7}
a_E^{\text{tot}} = \frac{1}{144}\sum_{n\geq 0}(n+1)(n+2)(n+3)^2(n+4)(n+5) \, .
\end{equation}
Notably, this is the same sum we encountered in \eqref{eq:C-tot-AdS4} for the $a^{\text{tot}}_E$ contribution from all KK modes on the round $S^7$. We thus arrive at an interesting observation, namely that the total $(a_E^{\text{tot}},c^{\text{tot}},b_1^{\text{tot}})$ coefficients for the KK spectrum on AdS$_4\times S^7$ are the same as those for the CPW solution. While we do not have a complete explanation for this result, it is reasonable to suspect that this is due to the fact that the two AdS$_4$ solutions are connected by a smooth gravitational domain wall, i.e. a holographic RG flow. Along this flow the spin of all KK modes does not change. The mass of these modes does change, but since they are organized into multiplets of the $\mathcal{N}=2$ supersymmetry preserved by the flow, the results in Table~\ref{HeatKernelCoefficientsMultiplets} dictate that the heat kernel coefficients for these multiplets are independent of the mass. Nevertheless, this line of reasoning does not immediately lead to the equivalence of the two sets of coefficients since the $(a_E,c,b_1)$ coefficients of the KK multiplets of the CPW background depend on the R-charges. Perhaps a more conceptual explanation of the equivalence between the UV and IR heat kernel coefficients may be given by some type of anomaly matching à la 't Hooft, but we were unable to make this precise at this stage. It will certainly be most interesting to understand this phenomenon better and to uncover whether it is a general feature of AdS vacua connected by smooth domain walls.

Under the same assumptions as in the $S^7$ case, we then arrive at the same logarithmic correction in the KK supergravity theory dual to mABJM, namely
\begin{equation}
	C(\mathcal{M}) = -\frac13\chi(\mathcal{M}) \, .
\end{equation}
This once again matches the results obtained in the dual CFT. Indeed, the logarithmic term in the large $N$ limit of the mABJM free energy on $S^3_b$, $S^1 \times \Sigma_\mathfrak{g}$ and $S^1 \times_\omega S^2$ can be obtained using supersymmetric localization by starting from the same UV quiver as that of the ABJM theory. Supersymmetry ensures that the log term in these observables is protected along the flow to the IR, and therefore the boundary logarithmic coefficient $\mathcal{C}(\partial\mathcal{M})$ is expected to be the same as that of the ABJM theory given in Table~\ref{tab:CFT-log}. We refer to~\cite{Bobev:2018wbt,Bobev:2023lkx} for additional details on mABJM supersymmetric observables.

\subsubsection{$N^{0,1,0}$}
\label{subsubsec:N010}

Another example we can study is the Freund-Rubin AdS$_4 \times N^{0,1,0}$ solution of 11d supergravity, see \cite{Castellani:1983yg}. The internal 7d space $N^{0,1,0}$ is Einstein and Tri-Sasakian and thus the dual 3d SCFT preserves $\mathcal{N}=3$ supersymmetry. The KK spectrum for this background has been computed in \cite{Fre:1999gok,Termonia:1999cs} where it was also organized in multiplets of the 3d $\mathcal{N}=3$ superconformal algebra and representations of the additional ${\rm SU}(3)$ isometry of $N^{0,1,0}$, which plays the role of a flavor symmetry in the dual SCFT. To perform this arduous calculation the authors of \cite{Fre:1999gok,Termonia:1999cs} used the fact that $N^{0,1,0}$ is a coset space and therefore the spectrum of various operators on it can be calculated using group theory techniques. The 3d $\mathcal{N}=3$ SCFT dual to the AdS$_4 \times N^{0,1,0}$ solution has been studied in \cite{Gaiotto:2009tk,Hohenegger:2009as,Hikida:2009tp,Cheon:2011th}.

The details of this explicit example of an AdS/CFT dual pair of theories will not be of great importance for our discussion. What is crucial is that on the field theory side one can employ supersymmetric localization to compute the partition function of the SCFT on various compact manifolds, including the omnipresent $\log N$ term of interest to us. We refer to Table~\ref{tab:CFT-log} for more details and a list of relevant references. To access the $\log N$ terms we can follow the logic outlined in Section~\ref{sec:KK:S7} for the AdS$_4\times S^7$ solution of 11d supergravity. To this end we use the explicit results on the KK spectrum from \cite{Fre:1999gok,Termonia:1999cs} to find the contributions of each KK mode to the heat kernel coefficients. The details of this calculation are important but not very illuminating and we present them in Appendix~\ref{App:KK:N010}. Here we will only summarize the results. 

Using the organization of the KK spectrum into multiplets of the $\mathcal{N}=3$ superconformal algebra, we find that each multiplet has vanishing $b_1$ coefficient. Thus, the total contribution from all KK modes is 
\begin{equation}
\label{eq:b1-N3}
b_1^{\rm tot} = 0 \, . 
\end{equation}
The situation is more interesting for the total $c^{\rm tot}$ coefficient. The long superconformal multiplets have vanishing $c$ coefficients, while the various short multiplets have a non-zero value for $c$ that depends on their R-charge. Since we have an infinite number of short multiplets in the KK spectrum, we arrive at a divergent sum indexed by a single integer parametrizing the R-charge. Interestingly, in Appendix~\ref{App:KK:N010}, we find a regularization method based on~\cite{friedman2012special} that yields a vanishing $c^{\text{tot}}$. Combining this with~\eqref{eq:b1-N3} and the discussion around~\eqref{eq:hk-non-min}, we come to the conclusion that $c^{\text{tot}}\vert_{\mathcal{N}=3} = b_1^{\text{tot}}\vert_{\mathcal{N}=3} = 0$ for the $N^{0,1,0}$ spectrum, in accordance with the bootstrap constraints~\eqref{eq:no-Cloc}.\footnote{In contrast, we have checked that the regularization method of~\cite{Liu:2016dau} for $c^{\text{tot}}$ yields results for $b_1^{\text{tot}}\vert_{\mathcal{N}=3}$ that are not compatible with~\eqref{eq:no-Cloc}.}  To arrive at this result, we use a ``spin-by-spin'' prescription where the contributions from short graviton, short gravitini and short vector multiplets are regularized individually using the method in Appendix~\ref{app:reg} before assembling them together. We hasten to stress that the method we use to regularize the infinite sums in each spin sector fails to produce the result~\eqref{eq:good-reg} for $a_E^{\text{tot}}$ in KK supergravity on $S^7$, as we have mentionned in Section~\ref{sec:KK:S7}. It is therefore clear that we currently lack a good understanding of the prescriptions needed to obtain results that are compatible with the AdS/CFT correspondence in all situations described so far. Nevertheless, it is encouraging to find some prescription ensuring that the bootstrap constraints~\eqref{eq:no-Cloc} are compatible with a somewhat standard way of regularizing divergent series.

The total $a_E^{\rm tot}$ coefficient is even more involved, since we need to take into account the contribution of all long and short superconformal multiplets in the KK spectrum. Once the dust settles, we find various contributions that are spelled out in detail in Appendix~\ref{App:KK:N010}. In particular, there are double series indexed by two integers coming from the long sector, and we do not know how to systematically regularize such expressions. What we can do is once again combine the various series contributing to $a_E^{\text{tot}}$ with our zero mode conjecture to eventually attach a finite value to all the divergent sums. That value must be compatible with the dual~$N^{0,1,0}$ CFT results presented in Table~\ref{tab:CFT-log}. However, this exercise is not very illuminating and we will not go through it explicitly here.

\subsubsection{$Q^{1,1,1}$}
\label{subsubsec:Q111}

The last example we will study is the AdS$_4 \times Q^{1,1,1}$ solution of 11d supergravity. The dual 3d SCFT preserves $\mathcal{N}=2$ supersymmetry and has SU(2)$^3$ flavor group. The spectrum of the KK theory has been organized into $\mathcal{N}=2$ multiplets in~\cite{Merlatti:2000ed} by leveraging the coset structure of the $Q^{1,1,1}$ internal Sasaki-Einstein space. In general there are short and long multiplets, and they are indexed by the three quantum numbers of SU(2)$^3$ and by the R-charge. We denote the tuple of quantum numbers by $(j_1,j_2,j_3,k)$. In the short sector, the sum over the spectrum reduces to a sum over $k \in \mathbb{N}$ and using the tables provided in~\cite{Merlatti:2000ed} we find
\begin{equation}
\label{eq:HK-Q111-short}
\begin{split}
a_E^\text{short} =&\; \frac18\sum_{k\geq0}(k+1)(13k^2+26k-31) \, , \\
c^\text{short} =&\;  -\frac14\sum_{k\geq0}(k+1)(17k^2 + 58k +109) \, , \\ 
b_1^\text{short} =&\; \frac{1}{24}\sum_{k\geq0}(k+1)(3k^4 + 16k^3 + 84k^2 + 204k + 109) \, .
\end{split}
\end{equation}
We assume a regularization scheme that assigns a finite value to these infinite sums. The long sector is more involved and depends on the SU(2)$^3$ quantum numbers. However, using the spectrum in~\cite{Merlatti:2000ed}, we find that the contributions to the $(a_E, c, b_1)$ coefficients from the long multiplets take a simple form:
\begin{equation}
\label{eq:Q111-long}
(a_E^\text{long},c^\text{long},b_1^\text{long}) = \Bigl(\frac94,-\frac14,-\frac{1}{24}\Bigr) \sum_{k,j_1,j_2,j_3} \text{dim}(j_1,j_2,j_3) \, ,
\end{equation}
where $\text{dim}(j_1,j_2,_3)$ is the dimension of the SU(2)$^3$ representation. We do not know of a rigorous way to regularize the infinite sums on the right-hand side. However, we can turn the logic around and see what can be learned from the bootstrap constraints~\eqref{eq:no-Cloc}. In particular, we see from~\eqref{eq:hk-non-min} and~\eqref{eq:Q111-long} that $b_1^{\text{long}}\vert_{\mathcal{N}=2} = b_1^{\text{long}} - c^{\text{long}}/6 = 0$. Thus, the bootstrap constraints imply that the spectrum in the short sector must be such that
\begin{equation}
\label{eq:HK-Q111-long}
c^\text{short} = 6\,b_1^\text{short}
\end{equation}
to achieve $b_1^\text{short}\vert_{\mathcal{N}=2}=0$. The regularization scheme used for~\eqref{eq:HK-Q111-short} should satisfy the above constraint. It is straightforward to check that using the method in Appendix~\ref{app:reg} to regularize $c^{\text{short}}$ and $b_1^{\text{short}}$, we do not find finite values that satisfy~\eqref{eq:HK-Q111-long}. Since our computation is highly sensitive to the full spectrum of the KK theory, this contradiction could point to issues in the analysis of~\cite{Merlatti:2000ed}. We note that this possibility has been raised independently in~\cite{Cheon:2011th} and further emphasized in \cite{Eager:2013mua}. Of course, we cannot exclude that there exists some renormalization scheme for the sums in the short sector that is compatible with the bootstrap constraints~\eqref{eq:HK-Q111-long}. At present, we are unable to make a definite statement, but we feel that this example highlights how sensitive our log computations are to the intricate details of the KK spectrum in a given supergravity compactification.

\section{The unbearable lightness of the KK scale}
\label{sec:Kundera}

So far we have discussed how to compute logarithmic corrections to the gravitational path integral on asymptotically AdS$_4$ backgrounds. Our focus was on 4d KK supergravity theories that preserve some amount of supersymmetry and arise from consistent truncations of 11d supergravity on a 7d manifold. In these explicit top-down constructions the size of the internal manifold is comparable to the size of the AdS$_4$ vacuum. This in turn implies that the four-dimensional gravitational theory does not really fall into the framework of standard EFT with finitely many fields, i.e. there is an infinite tower of KK modes that have masses of the same order that is parametrically smaller than the Planck scale. We now turn our attention to a more agnostic, or bottom-up, approach to EFTs coupled to gravity in AdS and their holographic physics in the context of the logarithmic corrections of interest in this work.

In order to be as general as possible we will work with a 4d gravitational theory (with at least one AdS$_4$ vacuum) coupled to a finite number of massive matter fields with spin up to 2 that obey some form of positive energy condition. We will assume that this theory has a consistent UV completion into a quantum gravitational theory, or to use modern parlance, that it is not in the Swampland. We will also assume that the 4d gravitational EFT is valid up to some energy scale which for concreteness we take to be the 4d Planck scale, although our arguments are general and can be adapted to other UV cutoff scales of interest. Lastly, we assume that the UV completion of the theory is such that it admits an equivalent holographic description in terms of a consistent unitary 3d CFT. As usual in AdS/CFT, the semi-classical gravity approximation is controlled by a large parameter which we call $N$. We emphasize that, at this somewhat abstract level of discourse, $N$ is just a label for a quantity much bigger than~1 that measures the number of degrees of freedom in the CFT, or rather the sequence of CFTs. For concreteness we can think of $N$ as being defined by the free energy of the CFT on $S^3$. Of course, in concrete and familiar AdS/CFT examples, $N$ can often be thought of as the rank of a gauge group. By evaluating the regularized on-shell action of the Euclidean AdS$_4$ vacuum of the gravitational theory we can find a map between the 4d Newton constant $G_N$, the scale $L$ of AdS$_4$, and the large parameter $N$. By dimensional analysis, this map will take the form
\begin{equation}
\frac{L^2}{G_N} = \beta N^{\gamma}\,,
\end{equation}
where the equality is to be understood to leading order in the large $N$ limit, the parameters $\beta$ and $\gamma$ are positive real numbers, and in all known AdS/CFT examples $\gamma$ is rational.

The path integral of the CFT on a compact Euclidean manifold should admit a large $N$ expansion of the form we already discussed around \eqref{logZ}, i.e.
\begin{equation}
	\log Z_{\text{CFT}} = F_0 + \mathcal{C} \log N + \mathcal{O}(N^0) \,.
\end{equation}
Here $F_0$ contains all positive powers of $N$ while the terms in $\mathcal{O}(N^0)$ are constant, have negative powers of $N$, or are exponentially suppressed at large $N$. These terms may have a complicated dependence on various continuous parameters of the theory including geometric deformations of the Euclidean manifold and dependence on relevant or exactly marginal couplings. Importantly, we will assume that the quantity $\mathcal{C}$ does not depend on any continuous parameters, i.e. it is a fixed real number for any given large $N$ CFT. Since this is a crucial assumption in our reasoning it is important to scrutinize it. We are not aware of any example of a sequence of CFTs (holographic or not) parametrized by $N$ for which, in the large $N$ limit, one finds that $\mathcal{C}$ depends on continuous parameters. One way to understand this may be to look at the 3d CFT placed on a squashed $S^3$ background and study the implications of any potential dependence in $\mathcal{C}$ on the continuous squashing parameter(s). If there is such a dependence one can expand the path integral in the limit of small squashing and deduce that there is a $\log N$ term in the large $N$ expansion of integrated correlators of the CFT on the round $S^3$. In particular, there will be a $\log N$ term in the $C_T$ coefficient of the two-point function of the energy momentum tensor.\footnote{See \eqref{eq:CT-def} for an explicit example of such a relation between $C_T$ and a squashed sphere partition function.} We are not aware of an example of a sequence of 3d CFTs where such a term appears, and it does not seem compatible with the usual large $N$ diagrammatics of gauge theories à la 't Hooft. The absence of any dependence of $\mathcal{C}$ on continuous parameters is also manifest in various explicit examples of partition functions of 3d $\mathcal{N}=2$ SCFTs (holographic or not) that can be studied by supersymmetric localization. It will be very interesting to construct a more robust argument or a proof that $\mathcal{C}$ does not depend on continuous parameters. One possible avenue to do this is to follow the approach in \cite{Cassani:2021fyv} and study the supersymmetric effective action of a 3d $\mathcal{N}=2$ SCFT on $S^1\times_{\omega} S^2$ with supersymmetric boundary conditions on $S^1$ and with angular momentum fugacity $\omega$, in the limit where the radius of the $S^1$ is small. It was shown in \cite{Cassani:2021fyv} that for 4d SCFTs the logarithmic term in the analogous effective action does not depend on the continuous parameter $\omega$ and has a topological nature. Furthermore, one can try to break supersymmetry and study the question more generally in the context of the thermal path integral of the 3d CFT on $S^1\times S^2$ which is controlled by the thermal effective action, see \cite{Benjamin:2023qsc} for a recent discussion. It may be possible to use this formalism to prove that no logarithmic terms can appear in $n$-point functions. In the absence of a solid proof, from now on we will assume that $\mathcal{C}$ is independent of continuous parameters and will proceed to study the implications of this assumption for the holographically dual 4d gravitational theory. Before doing so, we would like to emphasize that the essence of the discussion above is not specific to the $\log N$ terms in CFT path integrals and local correlation functions. If the CFT at hand has any dimensionless parameter, like an exactly marginal coupling $\lambda$, we also do not expect the coefficient of a $\log \lambda $ term in the path integral to depend on continuous parameters like squashing deformations or chemical potentials. 

We can now proceed in a similar manner as in Section~\ref{sec:bootstrap}. Studying the logarithmic corrections to the gravitational path integral on various explicit gravitational backgrounds,\footnote{We expect that the path integral of a standard 4d gravitational EFT allows for a standard diagrammatic expansion with respect to the large parameter $L^2/G_N$. The coefficient of the logarithmic correction $\log\relax(L^2/G_N)$ that is dual to $\mathcal{C}$ in the field theory side is then anticipated to be insensitive to the details of the UV cutoff and thereby determined unambiguously in the diagrammatic expansion. In particular, this requires the absence of exotic contributions like $[\relax(L^2/G_N)]^p[\log\relax(L^2/G_N)]^q$ with $p\geq 0$, $q>1$ in the expansion. In the discussion below we assume that such contributions indeed do not arise.} we find that the coefficient of $\log N$ (or $\log \lambda$) will depend on continuous parameters unless the total contribution to the $(c,b_1,b_2)$ heat kernel coefficients from the finite number of matter fields in the effective gravitational theory vanishes, i.e. we find the constraint
\begin{equation}
c^\text{tot}=b_1^\text{tot}=b_2^\text{tot}=0 \, . \label{HKE:Einstein-Maxwell}
\end{equation}
We emphasize that this conclusion is independent of the supersymmetry preserved by the gravitational theory or the background. This result imposes very strong constraint on any candidate consistent effective theory of gravity in AdS with finitely many fields. We now illustrate this with several examples.

Consider the 4d $\mathcal{N}=8$ ${\rm SO}(8)$ gauged supergravity theory constructed in \cite{deWit:1982bul}, i.e. the nonlinear theory for the fields in the massless $\mathcal{N}=8$ gravity multiplet presented in Table~\ref{tab:S7}. Using \eqref{eq:hk-non-min} and the results summarized in Section~\ref{sec:KK:S7} we find that the heat kernel coefficients for the maximally supersymmetric AdS$_4$ vacuum in this theory read\footnote{We note that if we choose to break $\mathcal{N}=8$ supersymmetry we can dualize some of the 35 pseudoscalars in the $\mathcal{N}=8$ gravity multiplet into 2-forms which will in turn yield different values for the $a_E$ coefficient. However, the $c$ and $b_1$ coefficients remain invariant under dualization~\cite{Larsen:2015aia}.}
\begin{equation}\label{eq:N=8sugraabc}
a_E^{\mathcal{N}=8} = \frac{5}{2}\,, \qquad c^{\mathcal{N}=8}=b_1^{\mathcal{N}=8}=0\,.
\end{equation}
Following the arguments above we therefore cannot conclude that this gauged supergravity is not a consistent holographic theory of quantum gravity, see also \cite{Alday:2022ldo,Montero:2022ghl} for recent discussion on the viability of the 4d $\mathcal{N}=8$ ${\rm SO}(8)$ gauged supergravity as a consistent quantum gravity theory. Arguments, similar in spirit to ours, have been used to exclude 4d $\mathcal{N}=8$ ungauged supergravity as a consistent quantum gravitational theory \cite{Green:2007zzb}.

We can proceed in a similar manner to study 4d $\mathcal{N}=4$ and $\mathcal{N}=2$ minimal gauged supergravity, i.e. the gauged supergravity theory of the gravity multiplet with the respective amount of supersymmetry. We find the following results:
\begin{equation}
\begin{alignedat}{3}
a_E^{\mathcal{N}=4} &= \frac12\,,& \qquad c^{\mathcal{N}=4} &= 0\,,& \qquad b_1^{\mathcal{N}=4}&=-\frac1{12}\,,\\[1mm]
a_E^{\mathcal{N}=2} &= -\frac{11}{24}\,,& \qquad c^{\mathcal{N}=2} &= 0 \,,& \qquad b_1^{\mathcal{N}=2}&= - \frac{13}{36}\,.
\end{alignedat}
\end{equation}
We therefore conclude that our logarithmic constraint excludes these two models as consistent quantum gravitational holographic theories. At this point one may be emboldened by the relative simplicity of our arguments and proceed to analyze all matter coupled (super)gravity theories and check whether they stand a chance of acting as consistent effective theories of gravity. This analysis is however complicated by the fact that we are interested in gravitational theories, like gauged supergravity, in which the scalar fields generically have highly non-trivial potentials. In such models it is a very involved task to study the space of AdS$_4$ vacua and the mass spectrum of linear fluctuations around them. As an illustration, consider the 4d $\mathcal{N}=8$ gauged supergravity with an $\rm{ISO} (7)$ gauging. This theory has the same matter content as the maximal ${\rm SO}(8)$ gauged supergravity, i.e. the fields in Table~\ref{tab:S7}, but a very different scalar potential. In particular, none of the 219 known AdS$_4$ vacua in this theory, see \cite{Bobev:2020qev}, is located at the origin of the scalar manifold where the full gauge group and maximal supersymmetry is preserved. This means that the heat kernel coefficients for any of the AdS vacua of this theory are not simply given by \eqref{eq:N=8sugraabc}. To find the correct logarithmic corrections to the gravitational path integral, one first needs to choose one of the vacua of the gauged supergravity theory, compute the mass spectrum of excitations around it (as was done in \cite{Bobev:2020qev}) and only then use the results in Table~\ref{tab:coeffs} above to calculate the total $(a_E,c,b_1)$ coefficients. Given these considerations we are left with the conclusion that the validity of a given effective theory of gravity in AdS with finitely many fields, supersymmetric or not, has to be studied on a cases-by-case basis. This can be done systematically for any given theory with fields with spin up to 2 and a known mass spectrum by using the results in Table~\ref{tab:coeffs} and checking whether the total heat kernel coefficients obey the constraints in \eqref{HKE:Einstein-Maxwell}.\\

In several corners of string theory the notion of a scale-separated AdS$_4$ vacuum appears, see for instance \cite{Kachru:2003aw,Balasubramanian:2005zx,DeWolfe:2005uu,Polchinski:2009ch} for several well-known constructions of this kind and \cite{Coudarchet:2023mfs} for a recent review. These are proposed consistent backgrounds of string or M-theory of the schematic form AdS$_4\times Y$ where there may be a warp factor in front of the AdS$_4$ metric and the internal space $Y$ is either completely geometric, i.e. a smooth manifold with properly quantized R-R and NS-NS fluxes, or has some mild singularities allowed by string theory such as orbifolds and orientifolds. The defining feature of such backgrounds is that the length scale $\ell_Y$ associated to $Y$ is parametrically smaller than the scale $L$ of AdS$_4$,
\begin{equation}
\ell_Y\ll L\,.
\end{equation}
We hasten to add that there could be some ambiguity in the definition of the scale $\ell_Y$. It can be associated to the volume or the diameter of $Y$, or be defined in terms of the eigenvalues of some appropriate differential operator on $Y$ used in the determination of the KK spectrum. What is important for our discussion is that below the scale set by $\ell_Y$ one should have a consistent effective gravitational theory in AdS$_4$ with finitely many matter fields of spin up to 2. In this setup, we can employ our results for the logarithmic corrections to the gravitational path integral to shed new light on the consistency of such scale-separated AdS$_4$ vacua. More specifically, we conclude that the spectrum of low-lying excitations around a given scale-separated AdS$_4$ vacuum, i.e. all fields with masses up to the scale set by $\ell_Y$, should be such that their total contribution to the $(c,b_1,b_2)$ coefficients vanishes as in \eqref{HKE:Einstein-Maxwell}. We note that these strong constraints should be viewed as an addition to the constraints and obstructions for the existence of scale-separated vacua discussed previously in the literature, see \cite{Gautason:2015tig,Alday:2019qrf,Lust:2019zwm,DeLuca:2021mcj,DeLuca:2021ojx,Collins:2022nux} and references thereof.\\

Since many AdS$_4$ vacua discussed in the literature, including scale-separated constructions, preserve $\mathcal{N}=1$ supersymmetry, we finish this section with a short discussion on the calculation of the heat kernel coefficients for 4d $\mathcal{N}=1$ matter multiplets. This should facilitate the exploration of the consistency condition in \eqref{HKE:Einstein-Maxwell} for any given AdS$_4$ vacuum with a known spectrum of light modes. To organize our results we first present the possible 4d $\mathcal{N}=1$ multiplets. We use notation similar to the one used for 4d $\mathcal{N}=2$ multiplets in Appendix~\ref{App:KK:mABJM} and find the following list of multiplets
   \begin{itemize}
        \item LGRAV = $L_1\left[E_0, \frac{3}{2}\right],\quad $ SGRAV = $A_1\left[ \frac{5}{2}, \frac{3}{2}\right],\quad $ 
        
        \item LGINO = $L_1\left[E_0, 1 \right],\quad $ SGINO = $A_1\left[ 2, 1 \right],\quad $

        \item LVEC = $L_1\left[E_0, \frac{1}{2}\right],\quad $ SVEC = $A_1\left[ \frac{3}{2}, \frac{1}{2}\right],\quad $

        \item LSCA = $L_2\left[E_0, 0 \right],\quad $ SSCA = $A_2\left[ \frac{1}{2}, 0 \right].\quad $
    \end{itemize}
On the right-hand side we have indicated how each supergravity multiplet can be related to the 3d $\mathcal{N}=1$ SCFT multiplets tabulated in Section 2.1.1 of \cite{Cordova:2016emh}. This facilitates the comparison with the holographically dual CFT and also allows one to consult the explicit field content of each multiplet using the detailed presentation in \cite{Cordova:2016emh}. We note that in the list above we have not included the identity multiplet denoted by $B_1[0,0]$ in that reference. The list above contains long multiplets (LGRAV, LGINO, LVEC, LSCA), for which all fields with $s\ge 1$ are massive, as well as short multiplets (SGRAV, SGINO, SVEC, SSCA), for which the fields with $s\ge 1$ are massless. Finally, $E_0$ is equal to the dimension $\Delta$ of the dual CFT primary operator and the mass of the field in AdS$_4$ can be found by using the relations between $\Delta$ and $m$ in Table~\ref{tab:conf-m}. The heat kernel coefficients $(a_E, c, b_1)$ of each 4d $\cN = 1$ multiplets are presented in Table~\ref{HeatKernelCoefficientsN=1Multiplets}.
%
        \begin{table}[h]
    \renewcommand{\arraystretch}{1.5}
        \centering
        \begin{tabular}{|c|c|c|c|}\hline
             & $a_E$ & $c$ & $b_1$  \\\hline 
            LGRAV & $\frac{23}{12}$ & $\frac{7}{3}$ & $\frac{5}{16}-\frac{1}{12}(E_0 - 1)^2$\\\hline
            SGRAV & $\frac{113}{48} $ & $\frac{77}{24}$ & $ 0 $ \\\hline
            LGINO & $-\frac{7}{16}$ & $-\frac{7}{8}$ & $-\frac{1}{8}+\frac{1}{16}(E_0 - 1)^2$\\\hline
            SGINO & $- \frac{31}{48}$ & $ - \frac{25}{24}$ & $ 0 $\\\hline
            LVEC & $\frac{5}{24}$ & $\frac{1}{6}$ & $\frac{1}{32}-\frac{1}{24}(E_0 - 1)^2$\\\hline
            SVEC & $\frac{3}{16}$ & $\frac{1}{8}$ & $0 $\\\hline
            LSCA  & $\frac{1}{48}$ & $\frac{1}{24}$ & $ \frac{1}{48}(E_0 - 1)^2 $\\\hline
             SSCA & $\frac{13}{720}$ & $\frac{1}{30}$ & $\frac{23}{4608}$\\\hline
         \end{tabular}
        \caption{\rm The heat kernel coefficients for 4d $\cN = 1$ multiplets. }
        \label{HeatKernelCoefficientsN=1Multiplets}
    \end{table}

We stress that for all entries in Table~\ref{HeatKernelCoefficientsN=1Multiplets} we have already incorporated the $(-1)^F$ that accounts for the fermion number of a given field in sums like the one in \eqref{eq:C-local:EAdS4}. Therefore, for a given AdS$_4$ vacuum with a known mass spectrum organized into 4d $\cN = 1$ multiplets, one simply needs to add the contributions from the table above to check whether the constraint in \eqref{HKE:Einstein-Maxwell} is obeyed. Since most multiplets in Table~\ref{HeatKernelCoefficientsN=1Multiplets} have positive $c$ coefficients, it is clear that imposing \eqref{HKE:Einstein-Maxwell} will result in non-trivial constraints. It will be very interesting to explore these constraints in detail for many of the well-known AdS$_4$ vacua in string and M-theory.

\section{Outlook}
\label{sec:ccl}

In this work we studied several aspects of logarithmic corrections to gravitational path integrals in AdS$_4$ and their relation to holography. In addition to a compendium of new technical results on this subject, we can extract at least two general lessons from our analysis. First, for a given explicit AdS$_4$ vacuum of string or M-theory, it is crucial to study the logarithmic corrections in the proper setup, i.e. in the higher-dimensional supergravity theory or in the 4d consistent truncation accompanied by a proper regularization scheme for the infinite tower of KK modes that contribute to the heat kernel. Second, we find that the form of the logarithmic corrections to the path integral on asymptotically locally AdS$_4$ backgrounds impose non-trivial constraints on the possible consistent 4d effective gravitational theories. In particular, assuming the validity of holography and the 4d gravitational EFT up to some cutoff scale, we conclude that either the total $(c,b_1,b_2)$ coefficients of the matter fields in the theory need to vanish, or that the dual 3d CFTs necessarily contain certain exotic logarithmic terms in their local correlation functions and thermal observables.

Our analysis points to several interesting open questions and directions for future work which we now briefly discuss.

\begin{itemize}

\item As emphasized in Section~\ref{sec:KK:S7}, the 4d KK supergravity calculation of the $\log N$ correction to the free energy on AdS$_4 \times S^7$ does not agree with the field theory result or the 11d supergravity analysis in \cite{Bhattacharyya:2012ye} if one chooses a regularization scheme in which $a_E=0$. It may be possible to resolve this discrepancy by using that the non-trivial contribution to $\log N$ in the 11d calculation comes from a zero mode of a 2-form ghost field needed for the proper quantization of the 3-form in 11d supergravity. To this end, one may contemplate a KK reduction of all ghost fields used in the quantization of the 11d graviton, gravitino, and 3-form around AdS$_4 \times S^7$ which can perhaps be organized into 4d $\mathcal{N}=8$ ``ghost KK multiplets''. With this at hand one should repeat our 4d KK supergravity analysis to account for the contribution of these ``KK ghosts'' to the coefficient of the $\log N$ term. Even if this calculation ends up successfully accounting for the correct $\frac14\log N$ contribution to the free energy, it is still unclear to us why such an analysis is justified. It will be interesting to understand this point better and draw lessons for the analysis of similar logarithmic corrections in more general gravitational backgrounds.

\item As discussed in Section~\ref{sec:log-conj}, if we adopt a particular regularization scheme for the infinite tower of KK modes on $S^7$ we can find an expression for the total $a_E$ heat kernel coefficient that agrees with the $\frac14\log N$ corrections to the AdS$_4\times S^7$ path integral. If in addition we assume the absence of zero modes other than for 2-forms in asymptotically locally AdS$_4$ spaces, we find results for the logarithmic corrections to the gravitational path integral that are consistent with the supersymmetric localization results in the ABJM theory summarized in Table~\ref{tab:CFT-log}. Moreover, this allows us to calculate the logarithmic corrections to the Bekenstein-Hawking entropy of non-supersymmetric black holes in AdS$_4$, generalizing the work of Sen \cite{Sen:2012dw} for asymptotically flat black holes. It is clearly important to put these calculations and assumptions on a more solid footing and find efficient methods to treat the contributions of zero modes to the logarithmic corrections in asymptotically AdS$_4$ backgrounds. We should also emphasize that we carried the calculation of the $b_2$ heat kernel coefficient in Appendix~\ref{App:coeffs} for minimally coupled charged scalars and spin-$1/2$ fermions and relied on the bootstrap argument in Section~\ref{sec:bootstrap} in most of our analysis. It will be nice to fill this gap and compute the $b_2$ coefficient for matter fields of arbitrary spin, mass, and charge.

\item The methods we employed here can be generalized to study logarithmic corrections to path integrals in de Sitter space, see \cite{Anninos:2020hfj,Bobev:2022lcc} for some recent explorations in this direction. It is important to develop this topic further and calculate both logarithmic corrections to the entropy of empty de Sitter space as well as to the entropy of asymptotically de Sitter black holes.

\item Supersymmetric localization has proven to be an efficient method for the calculation of path integrals in supersymmetric theories. It may be possible to use localization calculations in gauged supergravity as an efficient method to compute the Euclidean gravitational path integral and access the logarithmic corrections discussed above. This has been explored in \cite{Dabholkar:2014wpa,Hristov:2018lod,Hristov:2019xku,Hristov:2021zai} and it is important to address the subtleties and open questions that accompany this method. We hope that the results presented in this work will prove helpful in this regard. To this end, it is encouraging to note the fact that the 4d KK supergravity regularization discussed in Section~\ref{sec:log-conj} yields the same $\chi \log (L^2/G_N)$ structure of the logarithmic correction as the one found in \cite{Hristov:2021zai} using supergravity localization and index theorems.

\item An interesting generalization of the topologically twisted index of 3d $\mathcal{N}=2$ SCFTs on $S^1 \times \Sigma_{\mathfrak{g}}$ is the supersymmetric partition function on the Seifert manifold $\mathcal{M}_{\mathfrak{g},p}$ given by a degree $p$ fibration of the $S^1$ over $\Sigma_{\mathfrak{g}}$. This path integral was discussed in detail in~\cite{Closset:2017zgf} for general SCFTs where it was shown how it can be computed using supersymmetric localization. The holographic description of this path integral is given by the supersymmetric Euclidean AdS-Taub-Bolt solution presented in \cite{Alonso-Alberca:2000zeh,Toldo:2017qsh} and was studied for theories arising from M2- and M5-branes in \cite{Toldo:2017qsh} and \cite{Gang:2019uay,Benini:2019dyp,Bobev:2020zov}, respectively. It is interesting to apply the conjecture in Section~\ref{sec:log-conj} to the AdS-Taub-Bolt background. As shown in \cite{Bobev:2020zov}, the Euler number of this 4d supergravity background is independent of $p$ and is given by $\chi = 2(1-\mathfrak{g})$. This implies that the logarithmic correction to the SCFT free energy on $\mathcal{M}_{\mathfrak{g},p}$ does not depend on $p$. Indeed, as shown in \cite{Gang:2019uay,Bobev:2020zov}, this is true for class $\mathcal{R}$ theories arising from M5-branes. This is yet another piece of evidence that supports the conjectural results in Section~\ref{sec:log-conj}. Using this we can then conclude that the free energy on $\mathcal{M}_{\mathfrak{g},p}$ for the ABJM theory, discussed to leading $N^{3/2}$ and subleading $N^{1/2}$ order in \cite{Toldo:2017qsh} and \cite{Bobev:2020egg,Bobev:2021oku}, respectively, should receive a $\frac{1}{2}(1 - \mathfrak{g})\log N$ correction. Confirming this prediction by an explicit matrix model calculation based on the results of \cite{Toldo:2017qsh} will amount to a very non-trivial test of our results. We hope to report on this in the near future.

\item A central assumption underlying the discussion in Section~\ref{sec:Kundera} is the independence of the coefficients of the $\log N$ and $\log\lambda$ terms in CFT partition functions on continuous parameters like squashing deformations or relevant operator couplings. This can also be phrased as the absence of $\log N$ and $\log\lambda$ terms in correlation functions of local operators. Given the strong constraint this assumption imposes on gravitational theories in AdS, it is important to scrutinize it and understand whether it can be put on a more rigorous footing.

\item In many of the discussions above we focused on the $\log N$ terms in the path integrals of 3d $\mathcal{N}=2$ holographic SCFTs in M-theory. There can of course be $\log \lambda$ corrections to holographic SCFTs path integrals if a suitable notion of a 't Hooft coupling exists in the theory. These corrections should then correspond to $\log L/\ell_s$ corrections to the AdS$_4$ path integral where $\ell_s$ is the string length. Studying these corrections in 10d supergravity is in principle possible with the heat kernel methods discussed here and will be very interesting to pursue. However, a technical complication arises since one will first need to calculate the 10d heat kernel coefficients of the various supergravity fields in order to compute the local contribution to the coefficient of $\log L/\ell_s$. It will be very interesting to pursue this analysis, contrast the result with a 4d KK supergravity approach along the lines presented in this work, and ultimately compare the outcome of these calculations with the available 3d SCFT results for the relevant path integrals. More generally, even in the absence of supersymmetry, when there are dimensionful parameters like temperature, electric charges, and angular momenta present in the gravitational path integral, one can combine them with the AdS scale to form dimensionless parameters that can be scaled in various ways. It will be interesting to explore these scaling limits systematically and understand whether one can compare the coefficient of the logarithmic terms with dual 3d CFT results. Moreover, it will be nice to understand whether this analysis can lead to any interesting constraints on effective gravitational theories in AdS$_4$ along the lines of Section~\ref{sec:Kundera}.

\item The methods we used in this work can be applied to logarithmic corrections of AdS path integrals in other space-time dimensions. It will be particularly interesting to perform such an analysis for even-dimensional asymptotically AdS backgrounds since then there are non-trivial local contributions to the logarithmic corrections and it should be possible to deduce constraints on gravitational theories in AdS as we did in Section~\ref{sec:Kundera}.

\item In Section~\ref{sec:Kundera} we presented general constraints on the spectrum of light excitations around a given AdS$_4$ vacuum arising from a UV-complete quantum gravitational theory. It is important to understand whether these constraints are obeyed by the many known AdS$_4$ vacua in string and M-theory. This is of particular interest in the context of scale-separated AdS vacua where these new constraints can either rule out some a priori admissible backgrounds, or point to exotic features in the holographically dual 3d CFTs. It will also be interesting to explore the interplay between the constraints on the matter fields presented in Section~\ref{sec:Kundera} and the species bound discussed in the work of Dvali \cite{Dvali:2007hz}. Finally, we note that in at least one example, discussed in Section~\ref{sec:KK}, we found that the total heat kernel coefficients of two AdS$_4$ solutions connected by a gravitational domain wall are the same. It is important to understand whether this is a general feature of AdS vacua connected by domain walls. If such a property is generally true it can be used in conjunction with the constraints from Section~\ref{sec:Kundera} to derive even stronger consistency conditions on gravitational theories in AdS.

\end{itemize}

\bigskip

\noindent\textbf{Acknowledgments}

\medskip

\noindent We are grateful to Arash Ardehali, Chris Beem, Alejandro Cabo Bizet, Guillaume Bossard, Alejandra Castro, Richard Eager, Alfredo Gonzalez, Rajesh Kumar Gupta, Kiril Hristov, Zohar Komargodski, Mark Mezei, Eric Perlmutter, Silviu Pufu, Thomas Van Riet, Yifan Wang, and in particular to Ashoke Sen for useful discussions. This research is supported by the FWO projects G003523N, G0E2723N, and G094523N. NB, MD, JH and XZ are also supported in part by the KU Leuven C1 grant ZKD1118 C16/16/005 and by Odysseus grant G0F9516N from the FWO. MD and JH are Postdoctoral Fellows of the Research Foundation - Flanders. VR is supported by a public grant as part of the Investissement d'avenir project, reference ANR-11-LABX-0056-LMH, LabEx LMH. NB and VR are grateful to the ENS Paris for warm hospitality during part of this project. VR is partly supported by a Visibilit\'e Scientifique Junior Fellowship from LabEx LMH and is grateful to the CCPP at New York University for hospitality during the final stages of this project.

\appendix
\addtocontents{toc}{\protect\setcounter{tocdepth}{1}}

\section{Euclidean spinors}
\label{App:Euclid-spinors}

In this appendix we briefly summarize the 4d Euclidean convention for spinors and gamma matrices, see \cite{VanProeyen:1999ni,deWit:2017cle} for details. We use hermitian gamma matrices satisfying
\begin{equation}
	(\gamma_a)^\dagger=\gamma_a\,,\qquad\{\gamma_a,\gamma_b\}=2\delta_{ab}\,,
\end{equation}
and define the 5th gamma matrix $\gamma^5$ as
\begin{equation}
	\gamma^5=\gamma_1\gamma_2\gamma_3\gamma_4\,.
\end{equation}
The charge conjugation matrix $\Omega$ satisfies
\begin{equation}
	\Omega = -\Omega^\mathrm{T}\,,\qquad \Omega\gamma_a\Omega^{-1} = -(\gamma_a)^\mathrm{T}\,,\label{eq:calC}
\end{equation}
and also the unitarity relation $\Omega^\dagger = \Omega^{-1}$.

In 4d Euclidean signature, Majorana spinors are not irreducible and instead we must use symplectic-Majorana spinors. A symplectic-Majorana spinor $\lambda^i$ satisfies the constraint
\begin{equation}
	\Omega^{-1}(\bar{\lambda}_{i})^{\mathrm{T}} = \varepsilon_{ij}\,\lambda^j\,, \label{eq:sympMaj}
\end{equation}
written in terms of the charge conjugation matrix $\Omega$ and the antisymmetric tensor $\varepsilon_{ij}$ with $\varepsilon_{12} = 1$, 
where the bar is defined as $\bar{\lambda}_i := (\lambda^i)^\dagger$ and SU(2) indices $i,j\in\{1,2\}$ are raised and lowered by complex conjugation. The symplectic-Majorana condition \eqref{eq:sympMaj} yields the following useful identities on bilinears of \emph{anti-commuting} symplectic-Majorana spinors:
\begin{equation}
\label{eq:sM-bil-aC}
	\bigl(\bar{\lambda}_j\,\Gamma^\dagger\,\varphi^i\bigr)^\dagger = \bar{\varphi}_i\,\Gamma\,\lambda^j = -\varepsilon_{ik}\,\varepsilon^{jl}\,\bar{\lambda}_l\,\Omega^{-1}\Gamma^{\mathrm{T}}\Omega\,\varphi^k \, ,
\end{equation}
where $\Gamma$ is any combination of gamma matrices. For \emph{commuting} symplectic-Majorana spinors, we have instead
\begin{equation}
\label{eq:sM-bil-C}
	\bigl(\bar{\lambda}_j\,\Gamma^\dagger\,\varphi^i\bigr)^\dagger = \bar{\varphi}_i\,\Gamma\,\lambda^j = \varepsilon_{ik}\,\varepsilon^{jl}\,\bar{\lambda}_l\,\Omega^{-1}\Gamma^{\mathrm{T}}\Omega\,\varphi^k \, ,
\end{equation}
where the additional minus sign comes from $(AB)^\mathrm{T} = -B^\mathrm{T}A^\mathrm{T}$ for two Grassmann-odd matrices $A$ and $B$.

\section{Trace computations}
\label{App:coeffs}

In this appendix, we study the quadratic operator $\mathcal{Q}$ and the associated bulk contribution to the fourth SdW coefficient in~\eqref{eq:a4-bulk}. We do so for all massless and massive quantum fluctuations of fields with spin $0\leq s\leq 2$. For $s=0$ and $s=1/2$ we study the fluctuations around a generic Einstein-Maxwell background satisfying the equations of motion \eqref{eom}, while for spin $1\leq s\leq 2$ we turn off the background graviphoton and focus on Einstein backgrounds for simplicity.

\subsection{Scalar fluctuations}
\label{App:coeffs-scal}

The Euclidean action for scalar fluctuations to quadratic order reads
\begin{equation}
	S = \int d^4x \sqrt{g}\,\phi\bigl[-\mathcal{D}^\mu\mathcal{D}_\mu + m^2\bigr]\phi \, ,\label{S:0}
\end{equation}
where the covariant derivative is
\begin{equation}
\mathcal{D}_\mu\phi = \nabla_\mu\phi - \mathrm{i}\,q\,A_\mu\,\phi \, .
\end{equation}
Here, $q$ is the charge of the scalar field with respect to the background graviphoton. The quadratic operator can be read off from the action \eqref{S:0} as
\begin{equation}
	\mathcal{Q}_{\text{scal}} = \mD^\mu\mD_\mu - m^2 \, . 
\end{equation}
The associated quantities for the trace computations are defined in Section~\ref{sec:sugra:local} and are given by
\begin{equation}
	E = -m^2 \, , \qquad \Omega_{\mu\nu} = -\mathrm{i}\,q\,F_{\mu\nu} \, .\label{Q-component:0}
\end{equation}
Substituting \eqref{Q-component:0} into the trace formula \eqref{eq:a4-details:bulk} and then rewriting the result in terms of the 4-derivative quantities as in \eqref{eq:a4-bulk}, we obtain the heat kernel coefficients
\begin{equation}
	a_E = \frac{1}{360} \, , \quad c = \frac{1}{120} \, , \quad b_1 = \frac{1}{288}\bigl((mL)^2 + 2\bigr)^2 \, , \quad b_2 = \frac{1}{144}\,(q L)^2 \, ,\label{coeffi:0}
\end{equation}
where we have also used the background equations of motion \eqref{eom}. The heat kernel coefficients for the massless case can be obtained simply by setting $m^2=-2/L^2$ in \eqref{coeffi:0}.

\subsection{Spinor fluctuations}
\label{App:coeffs-ferm}

The Euclidean action for massive symplectic-Majorana spinors is given by \cite{deWit:2017cle}\footnote{The Euclidean Dirac action in a similar form can be derived by generalizing the Wick rotation to spinors~\cite{vanNieuwenhuizen:1996tv}. A proper derivation of the Euclidean spinor action from conformal supergravity was implemented in~\cite{deWit:2017cle}.}
\begin{equation}
	S = \int d^4x \sqrt{g}\,\bar{\psi}_i\gamma^5\bigr[\delta^i{}_j\,\slashed{\mathcal{D}} - m\,\sigma_3{}^i{}_j\bigr]\psi^j \equiv \int d^4x \sqrt{g}\,\bar{\psi}\,\mathbb{D}\,\psi \, ,
\end{equation}
where the covariant derivative is
\begin{equation}
	\mathcal{D}_\mu\Omega^i = \nabla_\mu\Omega^i - \mathrm{i}\,q\,\sigma_3{}^i{}_j\,A_\mu\Omega^j \, ,
\end{equation}
and $\sigma_3$ is the third Pauli matrix. To obtain a Laplace-type differential operator, we square the massive Dirac operator to $\mathcal{Q}_{\text{ferm}} = -\mathbb{D}^2$, see e.g. \cite{Vassilevich:2003xt}, and use
\begin{equation}
	\log\det\mathbb{D} = \frac12\log\det\mathcal{Q}_\text{ferm} \, .
\end{equation}
The resulting second-order operator reads
\begin{equation}
\label{eq:Laplace-ferm}
	\mathcal{Q}_{\text{ferm}} = \delta^i{}_j\bigg(\mD^\mu\mD_\mu - \frac14\,R - m^2\bigg) \, ,
\end{equation}
where the Ricci scalar term arises due to the Lichnerowicz formula based on the identity $\gamma^{ab}\gamma^{cd}R_{abcd}=-2R$. This operator is diagonal in the SU(2) indices. Thus, to obtain the contribution to the heat kernel coefficients from a single symplectic-Majorana spinor, we can work with the above $\mathcal{Q}_{\text{ferm}}$ and divide the final result by four: one factor of two for the SU(2) indices and another factor of two for the symplectic-Majorana condition. From~\eqref{eq:Laplace-ferm} we read off the matrices
\begin{equation}
\begin{split}
	E =&\; - \frac14\bigl(R + 4m^2\bigr)\delta^i_j - \frac12\mathrm{i}\,q\,F_{ab}\gamma^{ab}\,\sigma_3{}^i{}_j \, , \\
	\Omega_{\mu\nu} =&\; \frac14\,R^{ab}{}_{\mu\nu}\gamma_{ab}\,\delta^i_j - \mathrm{i}\,q\,F_{\mu\nu}\,\sigma_3{}^i{}_j \, ,
\end{split}\label{Q-component:1/2}
\end{equation}
defined in Section~\ref{sec:sugra:local}. Substituting \eqref{Q-component:1/2} into the trace formula \eqref{eq:a4-details:bulk} and then rewriting the result in terms of the 4-derivative quantities in \eqref{eq:a4-bulk}, we obtain the heat kernel coefficients
\begin{equation}
	a_E = -\frac{11}{720} \, , \quad c = -\frac{1}{40} \, ,\quad b_1 = \frac{1}{144}(mL)^2\bigl((mL)^2 - 2\bigr) \, , \quad b_2 = -\frac{1}{36}(qL)^2 \, ,\label{coeffi:1/2}
\end{equation}
for a single massive spin-1/2 field. The heat kernel coefficients for the massless case can be obtained simply by setting $m=0$ in \eqref{coeffi:1/2}.

\subsection{Vector fluctuations}
\label{App:coeffs-vec}

For a massless Abelian spin-1 field, the Euclidean action reads
\begin{equation}
	S = \int d^4 x \sqrt{g}\,\bigg[-\frac14\,F^{\mu\nu}F_{\mu\nu} - \frac12\,(\nabla^\mu A_\mu)^2\bigg] \, ,
\end{equation}
where we include a gauge-fixing term imposing the gauge $\nabla^\mu A_\mu = 0$. To consider the massive case, we introduce a Stückelberg scalar $B$ and consider the action \cite{Ruegg:2003ps}
\begin{equation}
\label{eq:spin-1-S}
	S = \int d^4 x \sqrt{g}\,\bigg[-\frac14\,F^{\mu\nu}F_{\mu\nu} - \frac12\,\bigl(m\,A_\mu + \nabla_\mu B\bigr)^2 - \frac12\,\bigr(\nabla^\mu A_\mu + m\,B\bigl)^2\bigg] \, .
\end{equation} 
Note that the first two terms are invariant under
\begin{equation}
	\delta A_\mu = \nabla_\mu \lambda \, , \qquad \delta B = - m\,\lambda \, ,
\end{equation}
and that the last term is the new gauge-fixing term for this symmetry. To enforce the gauge at the quantum level, we introduce anticommuting scalar ghost and anti-ghost fields with action
\begin{equation}
	S_{\text{gh}} = \int d^4x \sqrt{g}\,\bar{b} \bigl[-\nabla^\mu\nabla_\mu + m^2\bigr] c \, .
\end{equation}
The gauge-fixing term allows us to write the action \eqref{eq:spin-1-S} in terms of a Laplace-type quadratic operator. This operator is diagonal in field space and reads
\begin{equation}
\label{eq:Q-vec}
	\mathcal{Q}_{\text{vec}} = \begin{pmatrix} g^{\mu\nu}\bigl(\nabla^2 - m^2\bigr) - R^{\mu\nu} \;& 0 \\ 0 \;& \nabla^2 - m^2 \end{pmatrix} \, .
\end{equation}
From this we can read off the following matrices defined in Section~\ref{sec:sugra:local},
\begin{equation}
	E = \begin{pmatrix} -m^2 g^{\mu\nu} - R^{\mu\nu} \;& 0 \\ 0 \;& -m^2 \end{pmatrix} \, , \quad \Omega_{\rho\sigma} = \begin{pmatrix} R^{\mu\nu}{}_{\rho\sigma} \;& 0 \\ 0 \;& 0 \end{pmatrix} \, .
\end{equation}
This leads to the following heat kernel coefficients for the physical fields,
\begin{equation}
	a_E^{\text{ph}} = \frac{13}{72} \, , \qquad c^{\text{ph}} = \frac{1}{8} \, ,\qquad b_1^{\text{ph}} = \frac{1}{288}\bigl(5(mL)^4 - 4(mL)^2 + 12\bigr) \, , \label{coeffi:1-ph}
\end{equation}
based on the trace formula \eqref{eq:a4-details:bulk} and the equivalent expression \eqref{eq:a4-bulk} under the background equations of motion \eqref{eom} with a vanishing background graviphoton. The ghost sector contributes minus twice the coefficients for a real scalar field \eqref{coeffi:0}, namely 
\begin{equation}
	a_E^{\text{gh}} = -\fft{1}{180} \, , \qquad c^{\text{gh}} = -\fft{1}{60} \, ,\qquad b_1^{\text{gh}} = -\fft{1}{144}((mL)^2+2)^2 \, , \label{coeffi:1-gh}
\end{equation}
where the factor of two accounts for the pair of ghost and anti-ghost fields and the minus sign arises due to their anticommuting nature. 

Putting the contribution from the physical sector \eqref{coeffi:1-ph} and ghost sector \eqref{coeffi:1-gh} together, we arrive at the following heat kernel coefficients for a massive Abelian vector field
\begin{equation}
	a_E = \frac{7}{40} \, , \qquad c = \frac{13}{120} \, ,\qquad b_1 = \frac{1}{288}\bigl(3(mL)^4 - 12(mL)^2 + 4\bigr) \, .\label{coeffi:1}
\end{equation}
Note that in these trace computations, the Stückelberg field $B$ effectively adds a simple scalar degree of freedom to the coefficients obtained from the vector and ghost fields. This is a manifestation of the fact that the quadratic operator~\eqref{eq:Q-vec} is diagonal in field space. One can check that the heat kernel coefficients for a massless Abelian vector field presented in \cite{David:2021eoq}
\begin{equation}
	a_E = \frac{31}{180} \, , \qquad c = \frac{1}{10} \, ,\qquad b_1 = 0 \, ,\label{coeffi:1-ml}
\end{equation}
are indeed equivalent to the result obtained by subtracting \eqref{coeffi:0} from \eqref{coeffi:1} with $m=0$.

\subsection{Gravitino fluctuations}
\label{App:coeffs-gravitino}

The Euclidean action for a massive gravitino field reads
\begin{equation}
\label{eq:RS}
	S = \int d^4x \sqrt{g}\,\bigg(\bar{\psi}_{\mu i}\,\gamma^5\bigl[\delta^i{}_j\gamma^{\mu\nu\rho}\mathcal{D}_\nu - m\,\sigma_3{}^i{}_j\,\gamma^{\mu\rho}\bigr]\psi_{\rho}{\!}^j + \frac12\mathrm{i}\,F^{\rho\sigma}\,\bar{\psi}_{\mu i}\gamma^5\gamma_\rho\gamma^{\mu\nu}\gamma_\sigma\psi_\nu{\!}^i\bigg) \, ,
\end{equation}
where $\mathcal{D}_\mu$ denotes the covariant derivative. In the following we will turn off the background gauge field and focus on the gravitino fluctuation around a general Einstein background. The Rarita-Schwinger action above needs to be modified to include a symplectic-Majorana St\"uckelberg field $\chi^i$ as\footnote{See \cite{Porrati:2008ha} and references therein for how the Stückelberg formalism works for a Rarita-Schwinger field in the Lorentzian signature. Here we generalized it to the Euclidean signature based on \cite{deWit:2017cle}. Note also that we use $\nabla_{\mu}$ and not the more general covariant derivative $\mathcal{D}_\mu$ since we are considering backgrounds with vanishing graviphoton.} 
\begin{equation}
\label{eq:RS-S}
	S_\chi = \int d^4x \sqrt{g}\,\Bigl(\bar{\psi}_{\mu i} + \frac{1}{m}\,\nabla_\mu\bar{\chi}_i\,\gamma^5\Bigr)\,\gamma^5\bigl[\delta^i{}_j\,\gamma^{\mu\nu\rho}\nabla_\nu - m\,\sigma_3{}^i{}_j\,\gamma^{\mu\rho}\bigr]\Bigl(\psi_\rho{\!}^j + \frac{1}{m}\,\gamma^5\nabla_\rho\chi^j\Bigr) \, .
\end{equation}
Implementing the field redefinition
\begin{equation}
\label{eq:grav-shift}
	\psi_\mu{\!}^i \to \psi_\mu{\!}^i + \frac12\,\sigma_3{}^i{}_j\,\gamma_\mu\gamma^5\chi^j \, ,
\end{equation}
the modified action can be written in the form
\begin{equation}
\label{eq:start-Stuck}
\begin{split}
	S_\chi = S - \int d^4x \sqrt{g}\,&\bigg(\frac32\,\bar{\chi}_i\gamma^5\slashed{\nabla}\chi^i - 3\,m\,\bigl[\sigma_3{}^i{}_j\,\bar{\chi}_i\gamma^5\chi^j + \bar{\psi}_{\mu i}\gamma^\mu\chi^i\bigr] \\
	&- \frac{1}{4m}\Bigl[R\,\sigma_3{}^i{}_j\,\bar{\chi}_i\gamma^5\chi^j - 4\,G^{\mu\nu}\bar{\psi}_{\mu i}\gamma_\nu\chi^i + \frac{2}{m}\,G^{\mu\nu}\bar{\chi}_i\gamma^5\gamma_\mu\nabla_\nu\chi^i\Bigr]\bigg) \, ,
\end{split}
\end{equation}
where $G_{\mu\nu}$ is the Einstein tensor. After the shift~\eqref{eq:grav-shift}, the action $S$ in~\eqref{eq:RS-S} is invariant under the BRST transformations
\begin{equation}
\begin{alignedat}{2}
	\delta_B \psi_\mu{\!}^i &= \nabla_\mu c^i - \frac{m}{2}\,\sigma_3{}^i{}_j\,\gamma_\mu c{\,}^j \, ,&\qquad \delta_B \chi^i &= -m\gamma^5 c^i \, , \\
	\delta_B b^i &= \mathrm{i}\,\gamma^5\slashed{\nabla} B^i\,,&\qquad \delta_B c^i &= 0 \, , \qquad\qquad \delta_B B^i = 0 \, ,
\end{alignedat}
\end{equation}
where $b^i$ and $c^i$ are commuting symplectic-Majorana ghosts and $B^i$ is an anti-commuting symplectic-Majorana Nakanishi-Lautrup field. To consistently fix the gauge, we introduce the BRST-invariant action
\begin{equation}
\label{eq:3/2-gf-action}
	S_{\text{gf}} = \int d^4x \sqrt{g}\,\Bigl(\mathrm{i}\,\delta_B\bigl[\,\bar{b}_i\,\mathcal{G}^i\,\bigr] + \xi\,\bar{B}_i\gamma^5\slashed{\nabla} B^i\Bigr) \, ,
\end{equation}
and choose the gauge-fixing function
\begin{equation}
\label{eq:harmonic}
	\mathcal{G}^i = \gamma^\mu\psi_\mu{\!}^i \, . 
\end{equation}
Lastly, we introduce the massless commuting Kallosh-Nielsen ghost $d^{\,i}$ with action
\begin{equation}
	S_{\text{KN}} = \frac12\,\mathrm{i}\,\int d^4x \sqrt{g}\,\bar{d}_i \slashed{\nabla} d^{\,i} \, .
\end{equation}
Upon choosing $\xi = -\frac12$ in~\eqref{eq:3/2-gf-action}, the complete action $S_{\text{tot}} = S_\chi + S_{\text{gf}} + S_{\text{KN}}$ can be written in Dirac form
\begin{equation}
	S_{\text{tot}} = \Bigl( \bar{\psi}_{\mu i} \;\; \bar{\chi}_i \;\; \bar{b}_i \;\; \bar{d}_i \Bigr)\,\mathbb{D} \begin{pmatrix} \psi_\nu{}^j \\ \chi^j \\ c^{\,j} \\ d^{\,j} \end{pmatrix} \, ,
\end{equation}
where
\begin{equation}
\label{eq:D}
	\mathbb{D} = \begin{pmatrix} \gamma^5[\delta^i_j\,g^{\mu\nu}\slashed{\nabla} - m\sigma_3{}^i{}_j\gamma^{\mu\nu}] & -M\,\delta^i_j\gamma^\mu & 0 & 0 \\[2mm] -M\,\delta^i_j\gamma^\nu & -\gamma^5[\delta^i_j\slashed{\nabla} - 2m\sigma_3{}^i{}_j] & 0 & 0 \\[1.5mm] 0 & 0 & 2\mathrm{i}[\delta^i_j\slashed{\nabla} - 2m\sigma_3{}^i{}_j] & 0 \\[1.5mm] 0 & 0 & 0 & \mathrm{i}\,\delta^i_j\slashed{\nabla} \end{pmatrix} \, .
\end{equation}
To write the expressions above, we introduced 
\begin{equation}
	M = \sqrt{\frac{\Lambda}{2} + \frac32\,m^2} \, , 
\end{equation} 
and used that $G_{\mu\nu} = -\Lambda\,g_{\mu\nu}$ for an Einstein background. We further rescaled the St\"uckelberg fermion $\chi^i$ and all the ghosts to normalize all kinetic terms. \\

Since the operator~\eqref{eq:D} is block diagonal, we first focus on the physical sector in the $2\times 2$ upper left corner. To extract the heat kernel coefficients, we square this operator to bring it to the Laplace form \eqref{eq:Laplace-op} and define $\mathcal{Q}_{\text{gravitino}} = -\mathbb{D}^2$, as in the spin-1/2 case. Contrary to the spin-1 case, this operator is not diagonal in field space due to the cross terms between the gravitino and the St\"uckelberg fermion:
\begin{equation}
	\mathcal{Q}_{\text{gravitino}} = \begin{pmatrix} \mathcal{Q}_{\psi\psi} & \mathcal{Q}_{\psi\chi} \\[1mm] -\mathcal{Q}_{\psi\chi} & \mathcal{Q}_{\chi\chi} \end{pmatrix} \, .
\end{equation}
Explicitly, we have the following matrix elements:
\begin{equation}
\begin{split}
	\mathcal{Q}_{\psi\psi} =&\; \delta^i_j\,g^{\mu\nu}\slashed{\nabla}^2 - 4m\,\sigma_3{}^i{}_j\,g^{\rho[\mu}\gamma^{\nu]}\nabla_\rho -\delta^i_j(M^2 + 3m^2)g^{\mu\nu} - \delta^i_j(M^2 + 2m^2)\gamma^{\mu\nu} \, , \\[1mm]
	\mathcal{Q}_{\chi\chi} =&\; \delta^i_j\slashed{\nabla}^2 - 4\,\delta^i_j(M^2 + m^2) \, , \\[1mm]
	\mathcal{Q}_{\psi\chi} =&\; 2M\,\delta^i_j\,\gamma^5\nabla^\mu - 5mM\,\sigma_3{}^i{}_j\,\gamma^5\gamma^\mu \, ,
\end{split}
\end{equation}
which leads to the following field space matrix for $E$,
\begin{equation}
\begin{split}
	E_{\psi\psi} =&\; \frac12\,\delta^i_j\,\bigg(R^{\mu\nu}{}_{ab}\gamma^{ab} - \frac12\,R\,g^{\mu\nu} + 6m^2\,g^{\mu\nu} - 2M^2\,\gamma^{\mu\nu}\bigg) \, , \\[1mm]
	E_{\chi\chi} =&\; -\frac14\,\delta^i_j\,\bigl(R + 16m^2\bigr) \, , \\[1mm]
	E_{\psi\chi} =&\; -E_{\chi\psi} = -2mM\,\sigma_3{}^i{}_j\,\gamma^5\gamma^\mu \, ,
\end{split}
\end{equation}
and for $\Omega_{\rho\sigma}$,
\begin{align}
	(\Omega_{\rho\sigma})_{\psi\psi} =&\; \delta^i_j\,\bigg(R^{\mu\nu}{}_{\rho\sigma} + \frac14\,g^{\mu\nu}R_{\rho\sigma ab}\gamma^{ab} + 2m^2\Bigl[\gamma^\mu\gamma_{[\rho}g_{\sigma]}{}^\nu + g^\mu{}_{[\rho}\gamma_{\sigma]}\gamma^\nu - \Bigl(4 + \frac{M^2}{m^2}\Bigr)g^{\mu}{}_{[\rho}g_{\sigma]}{}^\nu\Bigr]\bigg) \, , \nonumber \\[1mm]
	(\Omega_{\rho\sigma})_{\chi\chi} =&\; \frac14\,\delta^i_j\,R_{\rho\sigma a b}\gamma^{ab} \, , \\[1mm]
	(\Omega_{\rho\sigma})_{\psi\chi} =&\; -(\Omega_{\rho\sigma})_{\chi\psi} = 2mM\,\sigma_3{}^i{}_j\,\gamma^5 g^\mu{}_{[\rho}\gamma_{\sigma]} \, . \nonumber
\end{align}
With this at hand, we obtain the heat kernel coefficients for the physical sector
\begin{equation}
	a_E^{\text{ph}} = \frac{109}{36} \, , \qquad c^{\text{ph}} = \frac{25}{6} \, ,\qquad b_1^{\text{ph}} = \frac{1}{18}\bigl(17(mL)^4 - 16(mL)^2 + 11\bigr) \, ,\label{coeffi:3/2:ph}
\end{equation}
based on the trace formula~\eqref{eq:a4-details:bulk} and the equivalent expression~\eqref{eq:a4-bulk} obtained using the background equations of motion \eqref{eom}. The ghost sector does not present additional complications and the corresponding heat kernel coefficients can be evaluated using the $2\times 2$ lower right corner of the operator \eqref{eq:D}. The result is given by
\begin{equation}
	a_E^{\text{gh}} = \frac{11}{60} \, , \qquad c^{\text{gh}} = \frac{3}{10} \, ,\qquad b_1^{\text{gh}} = -\fft49(mL)^2(2(mL)^2-1) \, ,\label{coeffi:3/2:gh}
\end{equation}
where we have taken into account the overall minus sign due to the opposite spin statistics of the ghosts fields (commuting) with respect to the physical fields (anti-commuting). 

Summing the contribution from the physical sector \eqref{coeffi:1-ph} and the ghost sector \eqref{coeffi:1-gh} and dividing the final result by four as in the spin-1/2 case, we obtain the heat kernel coefficients
\begin{equation}
	a_E = \frac{289}{360} \, , \qquad c = \frac{67}{60} \, ,\qquad b_1 = \frac{1}{72}\bigl((mL)^4 - 8(mL)^2 + 11\bigr) \, ,\label{coeffi:3/2}
\end{equation}
for a single massive spin-3/2 field. The heat kernel coefficients for a massless spin-3/2 fluctuation can be computed in the same way but without the contribution from the spin-1/2 St\"uckelberg field. The final result is given by
\begin{equation}
	a_E = \frac{589}{720} \, , \qquad c = \frac{137}{120} \, ,\qquad b_1 = 0 \, .\label{coeffi:3/2-ml}
\end{equation}

\subsection{Graviton fluctuations}
\label{App:coeffs-graviton}

The Euclidean action for a massive spin-2 fluctuation $h_{\mu\nu}$ is given by adding the Pauli-Fierz mass term to the massless action and reads
\begin{equation}
\begin{split}
	S = \int d^4x \sqrt{g}\,\Bigl[&h^{\mu\nu}\nabla^\rho \nabla_\rho h_{\mu\nu}-h\nabla^\mu \nabla_\mu h+2h \nabla_\mu \nabla_\nu h^{\mu\nu}+2\nabla_\nu h^{\mu\nu}\nabla_\rho h_\mu{}^\rho\\
	&+2h^{\mu\nu}h^{\rho\sigma}R_{\mu\rho\nu\sigma}+2h_{\mu\nu}h^{\mu\rho}R^\nu{}_\rho-2h^{\mu\nu}hR_{\mu\nu}-h^{\mu\nu}h_{\mu\nu}R+\frac12h^2R\\
	&+\Lambda(2h_{\mu\nu}h^{\mu\nu}-h^2)-m^2(h_{\mu\nu}h^{\mu\nu}-h^2)\Bigr] \, ,
\end{split}
\end{equation}
where all curvature tensors are those of the background metric $g_{\mu\nu}$ and $h = h^\mu{}_\mu$. As in the previous cases for spin-1 and spin-3/2 fields, the Pauli-Fierz mass term breaks the local gauge invariance. We can restore the symmetry by introducing two Stückelberg fields, namely a scalar $\phi$ and a vector $B_\mu$, and replace the metric fluctuation by
\begin{equation}
	h_{\mu\nu} \to h_{\mu\nu}-\frac12\,g_{\mu\nu}\phi+\frac1m \nabla_\mu\Bigl(B_\nu-\frac{1}{2m}\nabla_\nu\phi\Bigr)+\frac1m \nabla_\nu\Bigl(B_\mu-\frac{1}{2m}\nabla_\mu\phi\Bigr) \, ,
\end{equation}
see for example \cite{Porrati:2008ha} and references therein. The resulting action now enjoys the gauge symmetry
\begin{equation}
	\delta h_{\mu\nu} = \nabla_\mu\xi_\nu + \nabla_\nu\xi_\mu + \lambda\,m\,g_{\mu\nu} \, , \quad \delta B_\mu = \nabla_\mu\lambda - m\,\xi_\mu \, , \quad \delta \phi = 2m\,\lambda \, ,
\end{equation}
for which we add appropriate gauge-fixing terms, 
\begin{equation}
\begin{split}
	S_{1} =&\; -2\int d^4x\sqrt{g}\,\Bigl(\nabla_\nu h^{\mu\nu} - \frac12\nabla^\mu h + m B^\mu\Bigr)^2 \, , \\
	S_{2} =&\; -2\int d^4x\sqrt{g}\,\Bigl(\nabla_\mu B^\mu + \frac{m}{2}(h - 3\phi)\Bigr)^2 \, .
\end{split}
\end{equation}
Then, tedious but straightforward manipulations show that the resulting gauge-fixed action $S_\text{gf} = S + S_1 + S_2$ can be written as
\begin{equation}
\label{eq:spin-2-m-gf}
\begin{split}
	S_{\text{gf}} = \int d^4x \sqrt{g}\,\Bigl\{&h^{\mu\nu}\Bigl[G_{\mu\nu\rho\sigma}(\nabla^2 - m^2) + 2R_{\mu\rho\nu\sigma} - 2g_{\mu\rho}R_{\nu\sigma} + 2\Lambda G_{\mu\nu\rho\sigma}\Bigr]h^{\rho\sigma}\\
	&+ B^\mu\bigl[g_{\mu\nu}(\nabla^2 - m^2 + \Lambda)\bigr]B^\nu + \phi\bigl[\nabla^2 - m^2 - 3\Lambda M^{-2}\bigr]\phi\\
	&+ 2\Lambda M^{-1} h\,\phi -2\sqrt2\Lambda (mM)^{-1}B^\mu\nabla_\mu\phi\Bigr\} \, ,
\end{split}
\end{equation}
where
\begin{equation}
G_{\mu\nu\rho\sigma} = \frac12\bigl(g_{\mu\rho}g_{\nu\sigma} + g_{\mu\sigma}g_{\nu\rho} - g_{\mu\nu}g_{\rho\sigma}\bigr) \, , \qquad M = \sqrt{\frac32 - \frac{\Lambda}{m^2}} \, ,
\end{equation}
and we have rescaled the fields to ensure canonically normalized kinetic terms. In deriving~\eqref{eq:spin-2-m-gf}, we made extensive use of the background equations of motion \eqref{eom} with a vanishing graviphoton. We can now read off the Laplace-type operator \eqref{eq:Laplace-op} from \eqref{eq:spin-2-m-gf},
\begin{equation}
	\mathcal{Q}_{\text{graviton}} = \begin{pmatrix} \mathcal{Q}_{hh} \,& 0 \,& \mathcal{Q}_{h\phi} \\[1mm] 0 \,& \mathcal{Q}_{BB} \,& \mathcal{Q}_{B\phi} \\[1mm] \mathcal{Q}_{h\phi} \,& -\mathcal{Q}_{B\phi} \,& \mathcal{Q}_{\phi\phi} \end{pmatrix} \, .
\end{equation}
Just as in the spin-3/2 case we note that this is not diagonal in field space, indicating non-trivial interactions between the graviton fluctuations and the Stückelberg fields. Extracting the matrices $E$ and $\Omega_{\rho\sigma}$ defined in Section \ref{sec:sugra:local} from the operator $\mQ_\text{graviton}$ is straighforward, and we obtain the following heat kernel coefficients from the physical sector:
\begin{equation}
	a_E^{\text{ph}} = \frac{89}{24} \, , \qquad c^{\text{ph}} = \frac{113}{24} \, ,\qquad b_1^{\text{ph}} = \frac{1}{288}\Bigl(15(mL)^4 + 48(mL)^2 + 256\Bigr) \, ,\label{coeffi:2:ph}
\end{equation}
based on the trace formula \eqref{eq:a4-details:bulk} and the equivalent expression \eqref{eq:a4-bulk}. The gauge-fixing introduces a pair of vector ghosts and a pair of scalar ghosts whose Euclidean action reads
\begin{equation}
\begin{split}
	S_{\text{gh}} = \int d^4x \sqrt{g}\,\Bigl(b_\mu\bigl[g^{\mu\nu}(\nabla^2 - m^2) + R^{\mu\nu}\bigr]c_\nu + b\bigl[\nabla^2 - m^2\bigr]c\Bigr) \, .
\end{split}
\end{equation}
Their contribution to the heat kernel coefficients can be obtained from the spin-1 and spin-0 cases treated above. The result is given by
\begin{equation}
	a_E^{\text{gh}} = -\frac{13}{36} \, , \qquad c^{\text{gh}} = -\fft14 \, ,\qquad b_1^{\text{gh}} = \frac{1}{144}\Bigl(-5(mL)^4 -44(mL)^2 -108\Bigr) \, ,\label{coeffi:2:gh}
\end{equation}
where we have included an overall minus sign due to the anticommuting nature of the ghosts.

Putting the physical sector \eqref{coeffi:2:ph} and the ghost sector \eqref{coeffi:2:gh} together, we arrive at our final result
\begin{equation}
	a_E = \frac{241}{72} \, , \qquad c = \frac{107}{24} \, ,\qquad b_1 = \frac{5}{288}\bigl((mL)^4 - 8(mL)^2 + 8\bigr) \, ,
\end{equation}
for the massive spin-2 fluctuations around a generic Einstein background. The heat kernel coefficients for a massless fluctuation can be computed in the same way but without the contribution from St\"uckelberg fields and the corresponding scalar ghosts. The final result is given by
\begin{equation}
	a_E = \frac{571}{180} \, , \qquad c = \frac{87}{20} \, ,\qquad b_1 = 0 \, .\label{coeffi:2-ml}
\end{equation}

\section{Rarita-Schwinger zero modes on EAdS$_4$} 
\label{App:RS-non-local}

In Section~\ref{sec:sugra:non-local-AdS4}, we discussed how
\begin{equation}
\Psi^s_{\mu\ell m}(\lambda) = \mathcal{N}^s_\ell\Bigl(\nabla_\mu + \frac{s}{2L}\,\gamma_\mu\Bigr)\Omega^s_{\ell m} \, ,
\end{equation}
is a zero mode of the Rarita-Schwinger operator. Here, $\Omega^s_{\ell m}$ is an eigenspinor of the Dirac operator on EAdS$_4$,
\begin{equation}
\slashed{\nabla}\Omega^s_{\ell m} = \frac{\mathrm{i}s}{L}\,\lambda\,\Omega^s_{\ell m} \, .
\end{equation}
We note that using the above, we have
\begin{equation}
\gamma^\mu\Psi_{\mu\ell m}^s(\lambda) = \frac{s}{L}(2 + \mathrm{i}\lambda)\,\mathcal{N}_\ell^s\,\Omega_{\ell m}^s \, .
\end{equation}
Since the path integral over spin-3/2 fluctuations includes a delta function enforcing the harmonic gauge as in~\eqref{eq:harmonic}, we should consider the discrete value
\begin{equation}
\label{eq:lambda-fix}
\lambda = 2\mathrm{i} \, .
\end{equation}
Indeed, the differential operator relevant for non-zero modes is the Laplace-type operator obtained from~\eqref{eq:RS-op} after introducing the proper gauge-fixing term and the corresponding ghost fields, see Appendix~\ref{App:coeffs-gravitino}. To study the zero modes of that operator, we must impose~\eqref{eq:lambda-fix}.\\

For $\lambda \in \mathrm{i}\mathbb{R}$, the Dirac eigenmodes $\Omega^s_{\ell m}$ are not square-integrable on EAdS$_4$~\cite{Camporesi:1995fb}. Using the defining equation~\eqref{eq:eigenspinors}, we can express the norm of the RS zero-modes in terms of the norm of Dirac eigenmodes as follows:
\begin{equation}
\label{eq:Psi-sq}
\langle\Psi^s_{\mu\ell m},\Psi^{\mu s}_{\ell m}\rangle = |\mathcal{N}^s_\ell|^2\Bigl[-\frac{2}{L^2}\langle\Omega^{s}_{\ell m},\Omega^{s}_{\ell m}\rangle + \int d^4x\sqrt{g}\,\nabla^\mu\bigl[(\Omega^{s}_{\ell m})^\dagger \nabla_\mu \Omega^{s}_{\ell m}\bigr]\Bigr]_{\lambda = 2\mathrm{i}} \, .
\end{equation}
Note the appearance of a total derivative term, which can potentially renormalize the divergences that arise since the first term is non-normalizable. For explicit computations, we use a coordinate system in which the metric on EAdS$_4$ reads
\begin{equation}
\label{eq:metric-eta}
ds^2 = L^2(d\eta^2 + \sinh^2\eta\,d\Omega_3^2) \, ,
\end{equation}
where $d\Omega_3^2$ is the line element on $S^3$. In this coordinate system, the unit normal to the $S^3$ boundary is $n^\mu = (L^{-1},0,0,0)$ and the spin-connection has non-vanishing components only along the 3-sphere directions. This reduces the boundary term to
\begin{equation}
\int d^4x\sqrt{g}\,\nabla^\mu\bigl[(\Omega^{s}_{\ell m})^\dagger \nabla_\mu \Omega^{s}_{\ell m}\bigr]_{\lambda = 2\mathrm{i}} = \sinh^3y \int_{S^3}(\Omega^{s}_{\ell m})^\dagger \partial_\eta \Omega^{s}_{\ell m}\big\vert_{\eta = y, \lambda = 2\mathrm{i}} \, ,
\end{equation}
where we put the boundary at $\eta = y$ with $y$ being large.

The Dirac eigenmodes on EAdS$_4$ are given by~\cite{Camporesi:1995fb}
\begin{equation}
\Omega^s_{\ell m} = C_\ell(\lambda)\begin{pmatrix} \phi_{\lambda \ell}(\eta)\,\chi_{\ell m}(\theta) \\ \mathrm{i}\,s\,\psi_{\lambda \ell}(\eta)\,\chi_{\ell m}(\theta)\end{pmatrix} \, ,
\end{equation}
where $\chi_{\ell m}(\theta)$ are eigenspinors of the Dirac operator on $S^3$ (whose coordinates we generically denote as $\theta$), $\phi$ and $\psi$ are radial functions expressed in terms of the hypergeometric function ${}_2F_1$, and $C_\ell(\lambda)$ is a normalization constant. For the continuous case $\lambda\in\mathbb{R}$ the normalization can be found in~\cite{Camporesi:1995fb}, but since we want to consider the discrete mode with $\lambda$ fixed and imaginary, this normalization cannot be obtained from requiring that the modes $\Omega^s_{\ell m}$ provide an orthonormal basis. To proceed we note that the radial functions satisfy the differential equations
\begin{equation}
\label{eq:diff-eq}
\begin{split}
\partial_\eta\phi_{\lambda\ell} =&\; \frac12\Bigl[\ell \coth\tfrac{\eta}{2} + (\ell+1)\tanh\tfrac{\eta}{2}\Bigr]\phi_{\lambda\ell} - \Bigl[\frac{\lambda^2 + (\ell + 2)^2}{\ell + 2}\Bigr]\phi_{\lambda(\ell+1)} \, , \\[1mm]
\partial_\eta\psi_{\lambda\ell} =&\; \frac12\Bigl[(2\ell + 1)\coth\eta + \sinh^{-1}\eta\Bigr]\psi_{\lambda\ell} - \Bigl[\frac{\lambda^2 + (\ell+2)^2}{\ell + 2}\Bigr]\psi_{\lambda(\ell+1)} \, .
\end{split}
\end{equation}
Using this, both the bulk and boundary terms in~\eqref{eq:Psi-sq} can be computed. In terms of 
\begin{equation}
\lVert\chi_{\ell m}\rVert_{S^3} = \int_{S^3}\chi^\dagger_{\ell m}\chi_{\ell m} \, ,
\end{equation}
and $s_y = \sinh y$, the result is
\begin{equation}
\begin{split}
\frac{2}{L^2}\langle\Omega^{s}_{0 m},\Omega^{s}_{0 m}\rangle_{\lambda = 2\mathrm{i}} =&\; L^2|C_0(2\mathrm{i})|^2\Bigl(\frac{e^{4y}}{32} - \frac{e^{2y}}{8} + \frac{3}{16} + \mathcal{O}(e^{-2y})\Bigr)\,\lVert\chi_{0 m}\rVert_{S^3}  \, , \\[1mm]
s_y^3 \int_{S^3}(\Omega^{s}_{0 m})^\dagger \partial_\eta \Omega^{s}_{0 m}\big\vert_{\eta = y,\lambda = 2\mathrm{i}} =&\; L^2|C_0(2\mathrm{i})|^2\Bigl(\frac{e^{4y}}{32} - \frac{e^{2y}}{8} + \frac{3}{16} + \mathcal{O}(e^{-2y})\Bigr)\,\lVert\chi_{0 m}\rVert_{S^3}  \, ,
\end{split}
\end{equation}
for $\ell = 0$, and
\begin{equation}
\begin{split}
\frac{2}{L^2}\langle\Omega^{s}_{1 m},\Omega^{s}_{1 m}\rangle_{\lambda = 2\mathrm{i}} =&\; L^2|C_1(2\mathrm{i})|^2\Bigl(\frac{e^{4y}}{128} - \frac{25e^{2y}}{288} + \frac{475}{576} + \mathcal{O}(e^{-y})\Bigr)\,\lVert\chi_{1 m}\rVert_{S^3}  \, , \\[1mm]
s_y^3 \int_{S^3}(\Omega^{s}_{1 m})^\dagger \partial_\eta \Omega^{s}_{1 m}\big\vert_{\eta = y,\lambda = 2\mathrm{i}} =&\; L^2|C_1(2\mathrm{i})|^2\Bigl(\frac{e^{4y}}{128} - \frac{e^{2y}}{288} + \frac{41}{192} + \mathcal{O}(e^{-y})\Bigr)\,\lVert\chi_{1 m}\rVert_{S^3}  \, ,
\end{split}
\end{equation}
for $\ell = 1$. Results for higher values of $\ell$ can be obtained straightforwardly and follow the same pattern. Namely, we observe a perfect cancellation between bulk and boundary terms when $\ell=0$, while the modes with higher values of $\ell$ only see their leading divergences cancel. Thus, in the large $y$ limit,
\begin{equation}
\begin{split}
L^{-2}\langle\Psi^s_{\mu 0 m},\Psi^{\mu s}_{0 m}\rangle =&\; 0 \, , \\
L^{-2}\langle\Psi^s_{\mu \ell m},\Psi^{\mu s}_{\ell m}\rangle =&\; |\mathcal{N}_\ell^s C_\ell(2\mathrm{i})|^2\,\Bigl[h_1(\ell)\,e^{2y} + h_2(\ell)\Bigr]\lVert\chi_{\ell m}\rVert_{S^3} \quad \text{for} \quad \ell >0 \, ,
\end{split}
\end{equation}
for some $h_1$ and $h_2$ whose first few values are presented in Table~\ref{tab:RS-ZM}.
\begin{table}[h]
\centering
\renewcommand*{\arraystretch}{1.5}
\begin{tabular}{|c||c|c|c|c|c|c|c|c|c|c|c|}
\hline
$\ell$ & 0 & 1 & 2 & 3 & 4 & 5 & 6 & 7 & 8 & 9 & 10 \\
\hline
$h_1(\ell)$ & 0 & $\frac1{12}$ & $\frac{3}{40}$ & $\frac{3}{50}$ & $\frac{1}{21}$ & $\frac{15}{392}$ & $\frac{1}{32}$ & $\frac{7}{270}$ & $\frac{6}{275}$ & $\frac{9}{484}$ & $\frac{5}{312}$ \\
\hline
$h_2(\ell)$ & 0 & $-\frac{11}{18}$ & $-\frac{23}{20}$ & $-\frac{39}{25}$ & $-\frac{118}{63}$ & $-\frac{415}{196}$ & $-\frac{37}{16}$ & $-\frac{1001}{405}$ & $-\frac{716}{275}$ & $-\frac{657}{242}$ & $-\frac{1315}{468}$ \\
\hline
\end{tabular}
\caption{The leading and subleading divergences in the norm of the Rarita-Schwinger zero mode on EAdS$_4$ as a function of the mode number $\ell \geq 0$.\label{tab:RS-ZM}}
\end{table}
This shows that the RS zero mode has vanishing norm for $\ell = 0$, but is not square-integrable for $\ell > 0$. As such, it cannot contribute to the non-local contribution $C_\text{non-local}$.

\section{Heat kernel coefficients for KK supergravities}
\label{App:KK}

In this appendix we compute the contributions $(a_E,c,b_1)$ to the bulk SdW coefficient in \eqref{eq:a4-bulk} for the KK supergravity compactifications discussed in Section \ref{sec:KK}. As explained at the beginning of this section, we will do all computations for minimally coupled fields and use \eqref{eq:hk-non-min} to reinstate the effect of supersymmetric non-minimal couplings when necessary.

\subsection{KK supergravity on $S^7$}
\label{App:KK:S7}

The Weyl dimension formula for the SO(8) representation with Dynkin label $(\lambda_1,\lambda_2,\lambda_3,\lambda_4)$ reads\footnote{We follow the conventions summarized in Appendix A of \cite{Bobev:2021wir}.}
\begin{equation}
	\begin{split}
		&\dim(\lambda_1,\lambda_2,\lambda_3,\lambda_4)\\
		&=\fft{1}{4320}(1+\lambda_1)(1+\lambda_2)(1+\lambda_3)(1+\lambda_4)(2+\lambda_1+\lambda_2)(2+\lambda_2+\lambda_3)(2+\lambda_2+\lambda_4)\\
		&\quad\times(3+\lambda_1+\lambda_2+\lambda_3)(3+\lambda_1+\lambda_2+\lambda_4)(3+\lambda_2+\lambda_3+\lambda_4)\\
		&\quad\times(4+\lambda_1+\lambda_2+\lambda_3+\lambda_4)(5+\lambda_1+2\lambda_2+\lambda_3+\lambda_4)\qquad\qquad\qquad\qquad (\lambda_i\geq0) \, ,
	\end{split}\label{Weyl:dim:so(8)}
\end{equation}
and $\dim(\lambda_1,\lambda_2,\lambda_3,\lambda_4)\equiv0$ if any of the $\lambda_i$'s are negative. We now combine this formula with the data provided in Tables \ref{tab:coeffs}, \ref{tab:conf-m}, \ref{tab:S7}, and \ref{tab:S7-m} to evaluate the heat kernel coefficients for the KK supergravity theory arising from compactification of 11d supergravity on $S^7$.

The $a_E$ coefficient can be computed by adding the contributions from all KK modes of spin $0 \leq s \leq 2$ in the given SO(8) representation. At fixed KK level $k$, we find
\begin{align}
\label{S7:a:bulk}
		a_E(k) =&\; \fft{1}{360}\Bigl(\text{dim}(k+2,0,0,0)+\text{dim}(k-2,2,0,0)+\text{dim}(k-2,0,0,0) \nonumber \\
		&\qquad\;\; +\text{dim}(k,0,2,0)+\text{dim}(k-2,0,0,2)\Bigr) \nonumber \\
		&+\fft{11}{720}\Bigl(\text{dim}(k+1,0,1,0)+\text{dim}(k-1,1,1,0) +\text{dim}(k-2,1,0,1)+\text{dim}(k-2,0,0,1)\Bigr) \nonumber \\[1mm]
		&+\Bigl[\fft{31}{180}+\fft{1}{360}\Theta(k-1)\Bigr]\Bigl(\text{dim}(k,1,0,0)+\text{dim}(k-1,0,1,1)+\text{dim}(k-2,1,0,0)\Bigr) \nonumber\\[1mm]
		&-\Bigl[\fft{589}{720}-\fft{11}{720}\Theta(k-1)\Bigr]\Bigl(\text{dim}(k,0,0,1)+\text{dim}(k-1,0,1,0)\Bigr) \nonumber \\[1mm]
		&+\Bigl[\fft{571}{180}+\Bigl(\fft{31}{180}+\fft{1}{360}\Bigr)\Theta(k-1)\Bigr]\text{dim}(k,0,0,0) \, ,
\end{align}
where $\Theta(x \geq 0) = 1$ and $\Theta(x < 0) = 0$. This step function takes into account the fact that the contributions from St\"uckelberg fields are present only for massive fields with $k\geq1$. Using~\eqref{Weyl:dim:so(8)}, the total heat kernel coefficient $a_E^{\text{tot}} = \sum_{k\geq0} a_E(k)$ is given by
\begin{equation}
a_E^{\text{tot}} = \fft{1}{144}\sum_{k=0}^\infty(k+1)(k+2)(k+3)^2(k+4)(k+5) \, .
\end{equation}

The heat kernel coefficient $c$ is computed in a similar manner. Here, we find a perfect cancellation at each KK level:
\begin{align}
\label{S7:c:bulk}
		c(k) =&\; \fft{1}{120}\Bigl(\text{dim}(k+2,0,0,0)+\text{dim}(k-2,2,0,0)+\text{dim}(k-2,0,0,0) \nonumber\\
		&\qquad\;\;+\text{dim}(k,0,2,0)+\text{dim}(k-2,0,0,2)\Bigr) \nonumber\\
		&+\fft{1}{40}\Bigl(\text{dim}(k+1,0,1,0)+\text{dim}(k-1,1,1,0)+\text{dim}(k-2,1,0,1)+\text{dim}(k-2,0,0,1)\Bigr) \nonumber\\[1mm]
		&+\Bigl[\fft{1}{10}+\fft{1}{120}\Theta(k-1)\Bigr]\Bigl(\text{dim}(k,1,0,0)+\text{dim}(k-1,0,1,1)+\text{dim}(k-2,1,0,0)\Bigr) \nonumber\\[1mm]
		&-\Bigl[\fft{137}{120}-\fft{1}{40}\Theta(k-1)\Bigr]\Bigl(\text{dim}(k,0,0,1)+\text{dim}(k-1,0,1,0)\Bigr) \nonumber\\[1mm]
		&+\Bigl[\fft{87}{20}+\Bigl(\fft{1}{10}+\fft{1}{120}\Bigr)\Theta(k-1)\Bigr]\text{dim}(k,0,0,0)\\[1mm]
		=&\; 0 \, . \nonumber
\end{align}
We therefore conclude that $c^{\text{tot}} = \sum_k c(k)$ vanishes after summing over the KK tower.

For the $b_1$ coefficient, we note that since the lowest-lying KK modes with $k=0$ are massless, Table~\ref{tab:coeffs} implies that $b_1(0)=0$. For the contribution from higher KK modes with $k\geq 1$, we use the dictionary between mass and conformal dimension in Table~\ref{tab:conf-m} and obtain
\begin{align}
\label{S7:b:bulk}
		b_1(k) =&\; s_0\Big(\sqrt{(\fft{k}{2}+1)(\fft{k}{2}-2)}\Big)\text{dim}(k+2,0,0,0)+s_0\Big(\sqrt{(\fft{k}{2}+2)(\fft{k}{2}-1)}\Big)\text{dim}(k,0,2,0)\nonumber\\
		&+s_0\Big(\sqrt{(\fft{k}{2}+3)\fft{k}{2}}\Big)\text{dim}(k-2,2,0,0)+s_0\Big(\sqrt{(\fft{k}{2}+4)(\fft{k}{2}+1)}\Big)\text{dim}(k-2,0,0,2)\nonumber\\
		&+s_0\Big(\sqrt{(\fft{k}{2}+5)(\fft{k}{2}+2)}\Big)\text{dim}(k-2,0,0,0)\nonumber\\
		&+s_{\fft12}\Big(\fft{k+3}{2}-\fft32\Big)\text{dim}(k+1,0,1,0)+s_{\fft12}\Big(\fft{k+5}{2}-\fft32\Big)\text{dim}(k-1,1,1,0)\nonumber\\
		&+s_{\fft12}\Big(\fft{k+7}{2}-\fft32\Big)\text{dim}(k-2,1,0,1)+s_{\fft12}\Big(\fft{k+9}{2}-\fft32\Big)\text{dim}(k-2,0,0,1)\nonumber\\
		&+s_1\Big(\sqrt{(\fft{k}{2}+1)\fft{k}{2}}\Big)\text{dim}(k,1,0,0)+s_1\Big(\sqrt{(\fft{k}{2}+2)(\fft{k}{2}+1)}\Big)\text{dim}(k-1,0,1,1)\nonumber\\
		&+s_1\Big(\sqrt{(\fft{k}{2}+3)(\fft{k}{2}+2)}\Big)\text{dim}(k-2,1,0,0)\nonumber\\
		&+s_{\fft32}\Big(\fft{k+5}{2}-\fft32\Big)\text{dim}(k,0,0,1)+s_{\fft32}\Big(\fft{k+7}{2}-\fft32\Big)\text{dim}(k-1,0,1,0)\nonumber\\
		&+s_2\Big(\sqrt{(\fft{k}{2}+3)\fft{k}{2}}\Big)\text{dim}(k,0,0,0) \, , 
\end{align}
where we have introduced the functions
\begin{equation}
	\begin{split}
		s_0(x)&=\fft{1}{288}(x^2+2)^2\,,\\
		s_{\fft12}(x)&=-\fft{1}{144}x^2(x^2-2)\,,\\
		s_1(x)&=\fft{1}{288}(4-12x^2+3x^4)\,,\\
		s_{\fft32}(x)&=-\fft{1}{72}(11-8x^2+x^4)\,,\\
		s_2(x)&=\fft{5}{288}(8-8x^2+x^4)\,.
	\end{split}\label{s:spins}
\end{equation}
Remarkably, we again find a perfect cancellation
\begin{equation}
b_1(k) = 0 \, ,
\end{equation}
implying that the total heat kernel coefficient $b_1^{\text{tot}} = \sum_k b_1(k) = 0$.

\subsection{KK supergravity on $S^7/\mathbb{Z}_k$}
\label{App:KK:S7modk}

In the notation of \cite{Cordova:2016emh}, the $\mathcal{N}=8$ supermultiplet whose field content is given in Tables~\ref{tab:S7} and~\ref{tab:S7-m} is denoted by
\begin{equation}
B_1[0]^{(n+2,0,0,0)}_{\frac{n}{2}+1} \, .
\end{equation}
Note that we temporarily change notation and denote the KK level by $n \geq 0$ to reserve the notation $k$ for the order of the orbifold action $\mathbb{Z}_k$, as is common in the literature.

The spectrum of 11d supergravity compactified on $S^7/\mathbb{Z}_k$ can be obtained from the spectrum of the $S^7$ compactification as follows. First, we use the $\mathfrak{so}(8)\to\mathfrak{so}(6)\oplus\mathfrak{u}(1)$ branching rule~\cite{Liu:2016dau}
\begin{equation}
	B_1[0]_{\fft{n}{2}+1}^{(n+2,0,0,0)}=B_1[0]_{\fft{n}{2}+1,-n-2}^{(0,0,n+2)}\oplus B_1[0]_{\fft{n}{2}+1,n+2}^{(0,n+2,0)}\oplus\sum_{i=0}^nB_1[0]_{\fft{n}{2}+1,-n+2i}^{(0,i+1,n-i+1)}\,,\label{N=8to6}
\end{equation}
where the subscripts on the right-hand side denote the scaling dimension of the primary in the given representation and the U(1) charge, respectively, and the superscripts are the SO(6) Dynkin label $(a,b,c)$. We must then select the multiplets that are stable under the orbifold action, i.e. the multiplets whose U(1) charge is divisible by $k$ \cite{Liu:2016dau}.\footnote{Here we choose the periodic spin structure on $S^7/\mathbb{Z}_k$ in order to impose the same constraint on the bosonic and fermionic modes, see \cite{Bobev:2021wir} for details.} In addition, we only keep the multiplets for which $n$ is even, see~\cite{Bobev:2021wir}. As a result, the  supermultiplets that make up the $\mathcal{N}=6$ KK spectrum are those of the form
\begin{equation}
	B_1[0]_{h,-2r}^{(0,h-r,h+r)} \qquad \text{with} \quad h-|r|\geq0 \quad \text{and}\quad k\,|\,2r \,,\label{N=6:spectrum}
\end{equation}
where we have defined $h = \frac{n}{2} + 1 \in \mathbb{N}$ and $r = h - i - 1$ with $0\leq i \leq n$. The field content of these supermultiplets can be read off from \cite{Liu:2016dau}. Note that to obtain the correct field content for the special cases $0 \leq h - |r| \leq 3$, one must use the Racah-Speiser algorithm (reviewed e.g. in Appendix A.3 of \cite{Cordova:2016emh}). We will omit the details here and proceed to compute the heat kernel coefficients $(a_E,c,b_1)$ based on the KK spectrum~\eqref{N=6:spectrum}.

The SO(6) Weyl dimension formula for a representation with Dynkin label $(a,b,c)$ is
\begin{equation}
	\dim(a,b,c)=\fft{1}{12}(1+a)(1+b)(1+c)(2+a+b)(2+a+c)(3+a+b+c)\,.\label{Weyl:dim:so(6)}
\end{equation}
Using this with the field content of the KK spectrum just discussed, we can proceed in the same way as in Appendix~\ref{App:KK:S7}. For the $(c,b_1)$ heat kernel coefficients, we find a perfect cancellation at any allowed given value of the quantum numbers $(h,r)$,
\begin{equation}
c(h,r) = b_1(h,r) = 0 \, .
\end{equation}
For the $a_E$ coefficient, we obtain a non-vanishing result:
\begin{equation}
	a_E(h,r) = \fft{1}{24}(1+2h)\Big(h(1+h)(-4+5h+5h^2)+(7-10h-10h^2)r^2+5r^4\Big) \, .
\end{equation}
We discuss the sum over the spectrum in more details in Section~\ref{sec:KK-S7-k}.

\subsection{KK supergravity for mABJM}
\label{App:KK:mABJM}

Here we provide some details on the calculation of the SdW coefficient for the KK modes of the AdS$_4$ solution of 11d supergravity holographically dual to the mABJM SCFT. Since this background preserves $\mathcal{N}=2$ superconformal symmetry, we first present a short summary of the relevant superconformal multiplets. To facilitate the comparison with the literature we provide a map between superconformal multiplets in the supergravity conventions of Appendix A in \cite{Klebanov:2008vq} and the SCFT conventions of Table 4 in \cite{Cordova:2016emh}:
\begin{equation}
	\begin{split}
 		A_1\bar{A}_1[2]_{\Delta=2}^{(r=0)}\quad&\leftrightarrow\quad \text{MGRAV}\,,\\
 		L\bar{A}_1[2]_{r+2}^{(r>0)}\quad&\leftrightarrow\quad \text{SGRAV with the upper sign}\,,\\
 		A_1\bar{L}[2]_{-r+2}^{(r<0)}\quad&\leftrightarrow\quad \text{SGRAV with the lower sign}\,,\\
 		L\bar{L}[2]_{\Delta>|r|+2}^{(r)}\quad&\leftrightarrow\quad \text{LGRAV}\,,\\
 		L\bar{A}_1[1]^{(r>0)}_{r+\fft32}\quad&\leftrightarrow\quad \text{SGINO with the upper sign}\,,\\
 		A_1\bar{L}[1]^{(r<0)}_{-r+\fft32}\quad&\leftrightarrow\quad \text{SGINO with the lower sign}\,,\\
 		L\bar{L}[1]_{\Delta>|r|+\fft32}^{(r)}\quad&\leftrightarrow\quad \text{LGINO}\,,\\
 		A_2\bar{A}_2[0]_{\Delta=1}^{(r=0)}\quad&\leftrightarrow\quad \text{MVEC}\,,\\
 		L\bar{A}_2[0]_{r+1}^{(r>0)}\quad&\leftrightarrow\quad \text{SVEC with the upper sign}\,,\\
 		A_2\bar{L}[0]_{-r+1}^{(r<0)}\quad&\leftrightarrow\quad \text{SVEC with the lower sign}\,,\\
 		L\bar{L}[0]_{\Delta>|r|+1}^{(r)}\quad&\leftrightarrow\quad \text{LVEC}\,,\\
 		L\bar{B}_1[0]^{(r>\fft12)}_{\Delta=r}~\&~A_2\bar{B}_1[0]^{(r=\fft12)}_{\Delta=r}\quad&\leftrightarrow\quad \text{HYP with the upper sign}\,,\\
 		B_1\bar{L}[0]^{(r<-\fft12)}_{\Delta=-r}~\&~B_1\bar{A}_2[0]^{(r=-\fft12)}_{\Delta=-r}\quad&\leftrightarrow\quad \text{HYP with the lower sign}\,.
	 	\end{split}
\end{equation}
Note that the multiplets in \cite{Klebanov:2008vq} are labeled by their energy $E_0$, R-symmetry charge $y_0$, and the half-integer Lorentz spin $s_0$. The map to the corresponding quantities in \cite{Cordova:2016emh} is given by 
\begin{equation}
(E_0,y_0,s_0)~~ \longleftrightarrow~~ (\Delta,r,j/2)\,.
\end{equation}
We can use the explicit content of each of the multiplets above together with the information in Table~\ref{tab:coeffs} to systematically calculate the heat kernel coefficients for each superconformal multiplet. The result of this analysis is summarized in Table~\ref{HeatKernelCoefficientsMultiplets}. We hasten to note that the long superconformal multiplets have the interesting feature that while the individual states in the multiplet have heat kernel coefficients that depend on the conformal dimension (or mass in AdS) the total coefficients after summing over all fields in the multiplet are independent of this continuous parameter. The same is true for the $a_E$ and $c$ coefficients of short multiplets. The $b_1$ coefficient of short multiplets appears to depend on the conformal dimension, however one should bear in mind that for such multiplets the conformal dimension is uniquely determined by the spin and R-charge of the state.
    \begin{table}[h]
	\centering
	\renewcommand{\arraystretch}{1.5}
        \begin{tabular}{|c|c|c|c|} \hline
             & $a_E$ & $c$ & $b_1$  \\ \hline
            LGRAV & $\frac{5}{4}$ & $\frac{3}{4}$ & $\frac{1}{8}$\\ \hline
            SGRAV & $\frac{71-10\delta_{y_0,0}}{48}$ & $\frac{35-4\delta_{y_0,0}}{24}$ & $- \frac{1}{144} [3(E_0^2+2 E_0-11) + 2(E_0^3-3 E_0^2-9 E_0+25)\delta_{y_0,0}]$ \\ \hline
            MGRAV & $\frac{41}{24}$ & $\frac{13}{6}$ & 0\\ \hline
            LGINO & 0 & $-\frac{1}{2}$ & $-\frac{1}{12}$\\ \hline
            SGINO & $- \frac{11 + \delta_{y_0, 0}}{48}$ & $ - \frac{17 + \delta_{y_0, 0}}{24}$ & $ \frac{1}{2304}[ 48E_0(E_0+1)-276 - (2E_0 - 7)^2 (4E_0(E_0-1)-7)\delta_{y_0,0}] $\\ \hline
            LVEC & $\frac{1}{4}$ & $\frac{1}{4}$ & $\frac{1}{24}$ \\ \hline
            SVEC & $\frac{165 - 2 \delta_{y_0, 0}}{720}$ & $\frac{25-\delta_{y_0,0}}{120}$ & $ - \frac{1}{288}[6(E_0^2 -2) + (3 E_0^4-6 E_0^3-9 E_0^2+12 E_0+4) \delta_{y_0,0}] $\\ \hline
            MVEC & $\frac{5}{24}$ & $\frac{1}{6}$ & 0\\ \hline
            HYP & $\frac{1}{48}$ & $\frac{1}{24}$ & $ \frac{1}{48}(E_0 - 1)^2 $\\ \hline
         \end{tabular}
        \caption{The heat kernel coefficients $(a_E,c,b_1)$ for different $\cN = 2$ superconformal multiplets. The value of $E_0$ for each short multiplet is uniquely determined by the Lorentz spin and R-charge of its primary state. Note that some of the short multiplets do not include certain $y_0 = 0$ states which explains the appearance of the Kronecker in some of the expressions. The usual massless vector and hypermultiplets of matter-coupled 4d $\mathcal{N}=2$ gauged supergravity are MVEC and HYP, respectively.\label{HeatKernelCoefficientsMultiplets}}
    \end{table}

After this general discussion we are ready to focus our attention on the specific AdS$_4$ $\mathcal{N}=2$ vacuum of interest. This solution was found long ago by Warner as a critical point of the potential of the 4d $\mathcal{N}=8$ ${\rm SO}(8)$ gauged supergravity \cite{Warner:1983vz} and then uplifted to a solution of 11d supergravity in \cite{Corrado:2001nv} and we will thus refer to this background as the CPW solution. A notable feature of this solution is that it has an ${\rm SU}(3) \times {\rm U}(1)$ symmetry in the internal space where the ${\rm U}(1)$ corresponds to the R-symmetry in the dual mABJM SCFT while ${\rm SU}(3)$ corresponds to its flavor symmetry. The KK spectrum of the short superconformal multiplets for this background was described in \cite{Klebanov:2008vq}, the spectrum of spin-2 KK modes was computed in \cite{Klebanov:2009kp}, while the full KK spectrum was analyzed in \cite{Malek:2020yue}.

Although there is no clear a priori notion of a ``KK level'' for the spectrum of 11d supergravity AdS$_4$ solutions one can use the fact that the CPW solution is related to AdS$_4 \times S^7$ by a supersymmetric domain wall and use the integer $k$ labelling the $\mathcal{N}=8$ KK spectrum in Table~\ref{tab:S7-m} as giving some notion of a ``KK level''. To avoid confusion and make it clear we refer to the KK spectrum of the CPW solution we will call this integer $n$ below. With this at hand one can compactly summarize the KK spectrum using the results of \cite{Malek:2020yue} as follows
 \begin{equation}
 \label{E0U3}
        \begin{split}
            {\rm GRAV:}\quad E_0 =&\; \frac{1}{2} + \sqrt{\frac{9}{4} + \frac{1}{2}n(n+6) - \frac{4}{3}C_{p,q} + \frac{1}{2}\Bigl(r + \frac{2}{3}(q-p)\Bigr)^2 } \, ,\\[1mm]
            {\rm GINO:}\quad E_0 =&\; \frac{1}{2} + \sqrt{\frac{7}{2} + \frac{1}{2}n(n+6) - \frac{4}{3}C_{p,q} + \frac{1}{2} r^2} \, ,\\[1mm]
            {\rm VEC/HYP:}\quad E_0 =&\; \frac{1}{2} + \sqrt{\frac{17}{4} + \frac{1}{2}n(n+6) - \frac{4}{3}C_{p,q} + \frac{1}{2}r^2} \, ,
        \end{split}
    \end{equation}
where
\begin{equation}
 C_{p,q} = \frac{1}{3}(p^2 + q^2 + pq) + p + q \, .
\end{equation}
Here the integers $[p,q]$ specify the Dynkin labels of the ${\rm SU}(3)$ representation of the given multiplet and $C_{p,q}$ is its quadratic Casimir.\\

The data about the KK spectrum not captured by the information presented above is which ${\rm SU}(3)$ representation precisely are contained in the spectrum. For the short multiplets this was systematically studied in \cite{Klebanov:2008vq,Malek:2020yue}, see in particular Table 6 in \cite{Klebanov:2008vq} and Equation (5.99) in \cite{Malek:2020yue}.\footnote{Note that the relevant case for our discussion was dubbed Scenario I in \cite{Klebanov:2008vq}.} For each non-negative integer $n$ the list of short multiplets is summarized in Table~\ref{ShortMultiplets} where we also present their ${\rm SU}(3)$ Dynkin labels.
    \begin{table}[h]
    \renewcommand{\arraystretch}{1.5}
        \centering
        \begin{tabular}{|c|c|} \hline
           SGRAV/MGRAV & $[0,0]_{\pm n}$ \\ \hline
           SGINO & $[n+1,0]_{\frac{n+1}{3} },\ [0, n+1]_{-\frac{n+1}{3}}$ \\ \hline
           SVEC/MVEC & $[n+1,1]_{\frac{n}{3} },\ [1, n+1]_{-\frac{n}{3}}$ \\ \hline
           HYP & $[n+2,0]_{\frac{n+2}{3} },\ [0, n+2]_{-\frac{n+2}{3}}$ \\ \hline
        \end{tabular}
        \caption{The list of short multiplets in the KK spectrum of the CPW solution along with their ${\rm SU}(3)$ Dynkin labels. Note that $n\geq 0$ and when $n = 0$, the SGRAV and SVEC multiplets reduce to a single MGRAV and MVEC multiplet, respectively.\label{ShortMultiplets}}
    \end{table}
Using this information together with the SU(3) Weyl dimension formula
\begin{equation}
	\dim(p,q)=\fft12(p+1)(q+1)(p+q+2)\,,\label{Weyl:SU(3)}
\end{equation}
and the heat kernel coefficients in Table~\ref{HeatKernelCoefficientsMultiplets}, we find the following list of $(a_E,c,b_1)$ coefficients for all short multiplets at fixed KK level $n$, 
    \begin{align}
\label{eq:HK-mABJM-short}
            {\rm S/MGRAV:} \quad &\;a_E(n) = \frac{71}{24} - \frac{5}{4}\delta_{n,0} \, , \quad c(n) = \frac{35}{12} - \frac{3}{4}\delta_{n,0} \, , \nonumber\\[1mm] 
            &\; b_1(n) =- \frac{ n(n+6) - 3 }{24} - \frac{1}{8}\delta_{n,0} \, , \nonumber\\[1mm]
            {\rm SGINO:} \quad &\;a_E(n) = -\frac{11}{48}(n+2)(n+3)\, , \quad c(n) = -\frac{17}{24}(n+2)(n+3) \, ,\nonumber\\[1mm]
            &\;b_1(n) = \frac{1}{432}(n+2)(n+3)[n(n+14) - 5] \, , \\[1mm]
            {\rm S/MVEC:} \quad &\;a_E(n) = \frac{11}{24}(n+2)(n+4) - 2\delta_{n,0}\, , \quad c(n) = \frac{5}{12}(n+2)(n+4) - 2\delta_{n,0} \, ,\nonumber\\[1mm]
            &\;b_1(n) = - \frac{1}{216}(n+2)(n+4)[n(n+6) - 9] -  \frac{1}{3}\delta_{n,0}\, , \nonumber\\[1mm]
            {\rm HYP:} \quad &\;a_E(n) = \frac{1}{48}(n+3)(n+4) \, , \quad c(n) = \frac{1}{24}(n+3)(n+4) \, ,\nonumber\\[1mm]
            &\;b_1(n) = \frac{1}{432}(n+3)(n+4)(n-1)^2 \, .\nonumber
        \end{align}

We now move on to discuss the long multiplets. Here the situation is more involved and it is harder to obtain compact expressions for the total number of long multiplets, i.e. to sum over the corresponding ${\rm SU}(3)$ representation labels. To make progress we can use the strategy employed in \cite{Nicolai:1985hs,Klebanov:2008vq,Malek:2020yue,Cesaro:2020piw}. Namely we can rely on the fact that the CPW solution is connected by a holographic RG flow to the AdS$_4\times S^7$ background and thus we can use the symmetry breaking pattern from the $\mathcal{N}=8$ supersymmetry with ${\rm SO}(8)$ R-symmetry in the UV to $\mathcal{N}=2$ supersymmetry with ${\rm U}(1)$ R-symmetry and ${\rm SU}(3)$ flavor symmetry in the IR to organize the multiplets.\footnote{We have independently confirmed the calculations in \cite{Nicolai:1985hs,Klebanov:2008vq,Malek:2020yue,Cesaro:2020piw} using results from \cite{Slansky:1981yr} as well as the LieART package \cite{Feger:2019tvk}. We note that there are probably some typos in \cite{Klebanov:2008vq}, namely in Table 18, there should be two LGINO 0 in the $[1,1]$ representation and in Table 19 there should be two LVEC 0 in the $[0,0]$ representation.} Below we present explicit expressions for the total number of long multiplets at KK level $n$ and a short summary on how they are obtained. Since the heat kernel coefficients  $(a_E,c,b_1)$ are constant for all long multiplets, see Table~\ref{HeatKernelCoefficientsMultiplets}, this counting is sufficient for our purposes.\\
  
    {\bf LGRAV:} It was observed in \cite{Malek:2020yue} that at level $n$, the graviton multiplets transform under SU(3)$\times$U(1)$_R$ as follows:
    \begin{equation} \label{mABJMLGRAV}
        [p, q]_{ \frac{p-q}{3} + (a-b) }\, ,\quad \forall\;p,q,a,b\in \mathbb{N} \,, \;\;\text{with}\;\; p + q + a + b = n \, .
    \end{equation}
    Thus we only need to count the ordered partition $(p,q,a,b)$ of the integer $n$; which is $n+1-p-q$ for fixed $p,q$.  Na\"ively, the total number of LGRAV multiplets is then obtained by summing over $p$ and $q$ to find
    \begin{equation}
        \sum_{\substack{p,q = 0 \\ p+q\leq n}} (n+1-p-q) \dim(p,q) = \frac{1}{360} (n+1) (n+2) (n+3)^2 (n+4)(n+5) \, .
    \end{equation}
This result is true except for $p = q = 0$ and $(a,b) = (n,0)$ or $(0,n)$, where the multiplets are short(or massless when $n=0$), thus the corrected weighted sum that gives the total number of LGRAV multiplets at KK level $n$ is:
    \begin{equation}
         \Omega^{\rm{LGRAV}}(n) =  \frac{1}{360} (n+1) (n+2) (n+3)^2 (n+4)(n+5) - (2 - \delta_{n,0}) \, .
    \end{equation}

{\bf LGINO:} The analysis for the LGINO multiplet is more complicated since there is no compact formula analogous to \eqref{mABJMLGRAV}. As explained above we circumvent this difficulty by using the fact that the field content of the mABJM theory is inherited from the $\mathcal{N} = 8$ multiplets of the ABJM theory. This implies that for a given level $n$, the number of particles with the same spin is identical. The number of LGINO multiplets is thus equal to the number of spin-3/2 fields, minus those in L/SGRAV and SGINO multiplets. The total number of spin-3/2 fields is dictated by the following ${\rm SO}(8)$ representations:
    \begin{equation}
    \begin{aligned}
         \dim(n,0,0,1) &= \frac{1}{90} (n+1) (n+2) (n+3) (n+4) (n+5) (n+6) \, ,\\
         \quad \dim(n-1,0,1,0) &= \frac{1}{90} n (n+1) (n+2) (n+3) (n+4) (n+5) \, .
    \end{aligned}
    \end{equation}
The number of spin-3/2 particles in the graviton multiplets is:
    \begin{equation}
        4\Bigl[\frac{1}{360} (n+1) (n+2) (n+3)^2 (n+4)(n+5) - (2 - \delta_{n,0})\Bigr] + 6(1-\delta_{n,0}) + 2\delta_{n,0} \, .
    \end{equation}
     The number of spin-3/2 particles in the two SGINO multiplets is $2 \times \dim(n+1,0) = (n+2)(n+3)$. With this at hand we finally find the total number of LGINO multiplets:
    \begin{equation}
    \Omega^{\rm LGINO}(n) =  \frac{1}{90}n(n+1)(n+4)(n^3+13 n^2+61 n+123) \, .
    \end{equation}

{\bf LVEC:} To find the total number of LVEC multiplets we follow the same approach as above. First, we find the total number of spin-1 modes from the ${\rm SO}(8)$ representations:
    \begin{equation}
    \begin{aligned}
         &\dim(n,1,0,0) + \dim(n-1, 0, 1,1)_{n \ge 1} + \dim(n-2,1,0,0)_{n\ge 2} \\ 
         &\quad = \frac{3}{40}(n+1)(n+2)(n+3)^2(n+4)(n+5) + \delta_{n,0} \, .
    \end{aligned}
    \end{equation}
 The number of spin-1 modes contained in LGRAV and SGRAV/MGRAV is
    \begin{equation}
        6\Bigl[\frac{1}{360} (n+1) (n+2) (n+3)^2 (n+4)(n+5) - (2 - \delta_{n,0})\Bigr] + 6(1-\delta_{n,0}) + \delta_{n,0} \, ,
    \end{equation}
 while the number contained in LGINO and SGINO is
    \begin{equation}
         4\Bigl[\frac{1}{90}n(n+1)(n+4)(n^3+13 n^2+61 n+123)\Bigr] + 6\dim(n+1,0) \, .
    \end{equation}
Finally the number of spin-1 modes in SVEC/MVEC multiplets is
    \begin{equation}
       2 (1-\delta_{n,0}) \dim(n+1,1) + \delta_{n,0} \dim(1,1) \, . 
    \end{equation}
Combining all this we find that the total number of LVEC multiplets at KK level $n$ is
    \begin{equation}
       \Omega^{\rm LVEC}(n) =  \frac{1}{72}(n+3)(n+4) (n^4+11 n^3+41 n^2+61 n-42) + 8 \delta_{n,0} \, .
    \end{equation}

We now have the total number of each type of long and short multiplets in the KK spectrum of the CPW background. We can combine this with the information in Table~\ref{HeatKernelCoefficientsMultiplets} to find the total heat kernel coefficients. A short calculation shows that, at each KK level $n$, we find a perfect cancellation for the $c$ coefficient:
\begin{equation}
c(n) = \frac34\Omega^{\rm{LGRAV}}(n) - \frac12\Omega^{\rm{LGINO}}(n) + \frac14\Omega^{\rm{LVEC}}(n) + c^{\text{short}}(n) = 0 \, ,
\end{equation}
where $c^{\text{short}}$ is the sum of all $c(n)$ in the short sector~\eqref{eq:HK-mABJM-short}. The same cancellation occurs for the $b_1$ coefficient:
\begin{equation}
b_1(n) = \frac18\Omega^{\rm{LGRAV}}(n) - \frac1{12}\Omega^{\rm{LGINO}}(n) + \frac1{24}\Omega^{\rm{LVEC}}(n) + b_1^{\text{short}}(n) = 0 \, .
\end{equation}
Lastly, for the $a_E$ coefficient, we obtain the following compact final result,
\begin{equation}
\begin{split}
        a_E(n) =&\; \frac{5}{4}\Omega^{\rm LGRAV}(n) + \frac{1}{4} \Omega^{\rm LVEC}(n) + a_E^{\text{short}}(n) \\[1mm]
        =&\; \frac{1}{144} (n+1)(n+2)(n+3)^2 (n+4)(n+5) \, .
\end{split}
\end{equation}
Notably, these are exactly the same expressions we found for the heat kernel coefficients of the KK spectrum of the $\mathcal{N}=8$ AdS$_4\times S^7$ background in \eqref{S7:a:bulk}, \eqref{S7:c:bulk}, and \eqref{S7:b:bulk}. The calculation leading to these results for the two distinct 11d supergravity backgrounds is very different and as we discuss in the main text it will be interesting to obtain further insights into why the results end up the same. Regardless, it is clear that the total $c$ and $b_1$ coefficients after summing over $n$ vanish. The total $a_E$ heat kernel coefficient takes the form of a divergent sum which is identical to the one encountered in the AdS$_4\times S^7$ and thus should be regularized in the same way.

\subsection{KK supergravity on $N^{0,1,0}$}
\label{App:KK:N010}

The KK supergravity spectrum of the AdS$_4\times N^{0,1,0}$ 11d supergravity solution was presented in \cite{Fre:1999gok} where it was organized into multiplets of the ${\rm OSp}(3|4)$ superconformal algebra and the ${\rm SU}(3)$ flavor symmetry. We will use both the conventions of \cite{Fre:1999gok} and \cite{Cordova:2016emh} when we discuss our results. The map between the two sets of conventions is given by
\begin{equation}
	(E_0,J_0,s_0)~~ \longleftrightarrow~~ (\Delta,R/2,j/2) \,,
\end{equation}
where $\Delta\in\mathbb Z/2$ is the conformal dimension, $R\in\mathbb Z_{\geq0}$ denotes the ${\rm SO}(3)$ $R$-symmetry representation, and $j\in\mathbb Z_{\geq0}$ is the Lorentz spin. Note that the dimension of the ${\rm SO}(3)$ representation with label $R$ is given by $\text{dim}(R)=R+1$ and analogously the dimension of the Lorentz spin-$j$ representation is $\text{dim}(j)=j+1$.

The complete list of 3d $\mN=3$ multiplets can be found in Table 5 of \cite{Cordova:2016emh} and is summarized in Table~\ref{table:N010}.
\begin{table}[h]
	\centering
	\renewcommand{\arraystretch}{1.5}
	\begin{tabular}{|c|c|c|}
		\hline
		name & primary & unitarity bound\\\hline
		$L$ & $[j]_\Delta^{(R)}$ & $\Delta>\fft12j+\fft12R+1$ \\\hline
		$A_1$ & $[j]_\Delta^{(R)}~~(j\geq1)$ & $\Delta=\fft12j+\fft12R+1$ \\\hline
		$A_2$ & $[0]_\Delta^{(R)}$ & $\Delta=\fft12R+1$ \\\hline
		$B_1$ & $[0]_\Delta^{(R)}$ & $\Delta=\fft12R$ \\\hline
	\end{tabular}\caption{The complete list of $\mN=3$ multiplets\label{table:N010}}
\end{table}
The $\mN=3$ multiplets arising from the $N^{0,1,0}$ KK spectrum discussed in \cite{Fre:1999gok} read
\begin{equation}
\begin{split}
	L[1]_\Delta^{(R)}\quad&\leftrightarrow\quad SD(s_\text{max}=2,E_0>J_0+3/2,J_0|3)\,,\\
	L[0]_\Delta^{(R)}\quad&\leftrightarrow\quad SD(s_\text{max}=3/2,E_0>J_0+1,J_0|3)\,,\\
	A_1[1]_\Delta^{(R)}\quad&\leftrightarrow\quad SD(s_\text{max}=2,E_0=J_0+3/2,J_0|3)\,,\\
	A_2[0]_\Delta^{(R)}\quad&\leftrightarrow\quad SD(s_\text{max}=3/2,E_0=J_0+1,J_0|3)\,,\\
	B_1[0]_\Delta^{(R)}\quad&\leftrightarrow\quad SD(s_\text{max}=1,E_0=J_0,J_0|3)\,,
\end{split}
\end{equation}
where on the right hand side we used the notation of \cite{Fre:1999gok}. The descendants for generic multiplets and conserved current multiplets are presented in Section 4.3 and 5.4.3 of \cite{Cordova:2016emh}, respectively (see also Table 1-5 of \cite{Fre:1999gok}). Note that the $A_2[0]_\Delta^{(0)}$ multiplet is not present in the spectrum since it contains an extra supercurrent multiplet not present for the AdS$_4\times N^{0,1,0}$ solution which is genuinely $\mathcal{N}=3$ supersymmetric.

To obtain Tables 1-5 of \cite{Fre:1999gok} from the generic expressions given in Section 4.3 of \cite{Cordova:2016emh}, one needs to use the Racah-Speiser (RS) algorithm. For the $\mathfrak{so}(3)$ Lie algebra of Lorentz spin, the RS algorithm is simply given as
\begin{equation}
	[j]^{(R)}=\begin{cases}
		0 & (j=-1) \\
		-[-j-2]^{(R)}& (j=-2,-3,\cdots)
	\end{cases}\qquad(R\geq0)\,,
\end{equation}
where the overall minus sign on the right-hand side means that it ``eats'' the representation without the minus sign. Since the $R$-symmetry is also controlled by the same algebra, we can apply the same rule, i.e.
\begin{equation}
	[j]^{(R)}=\begin{cases}
		0 & (R=-1) \\
		[j]^{-(-R-2)}& (R=-2,-3,\cdots)
	\end{cases}\qquad(j\geq0)\,,
\end{equation}
with the same interpretation for the overall minus sign in the superscript of the RHS.

To calculate the heat kernel coefficients of each $\mathcal{N}=3$ superconformal multiplet of interest, we need to add up the coefficients of every field in the multiplet and use the values in Table~\ref{tab:coeffs}. The result for each of the multiplets in Table~\ref{table:N010} is summarized below.\footnote{Note that the rightmost entry of Table 5 of \cite{Fre:1999gok} is missing one of the two spin-$\fft12$ representations.}
\begin{align}
L[1]_\Delta^{(R)} \;\text{with}\;\Delta>\tfrac12 R + \tfrac32: \quad &a_E(R) = \fft{3(1+R)}{2} \, , \quad c(R) = 0 \, , \quad b_1(R) = 0 \, , \nonumber\\[1mm]
L[0]_\Delta^{(R)} \;\text{with}\;\Delta>\tfrac12 R+1: \quad &a_E(R) = \fft{1+R}{2} \, , \quad c(R) = 0 \, , \quad b_1(R) = 0 \, , \nonumber\\[1mm]
A_1[1]_\Delta^{(R)} \;\text{with}\;\Delta=\tfrac12 R+\tfrac32: \quad &a_E(R) = \fft{5(1+R)}{4} \, , \quad c(R) = \frac{5+R}{4} \, , \quad b_1(R) = 0 \, , \nonumber\\[1mm]
A_2[0]_\Delta^{(R)} \;\text{with}\;\Delta=\tfrac12 R+1: \quad &a_E(R) = -\fft{(1-R)}{4} \, , \quad c(R) = -\frac{3+R}{4} \, , \quad b_1(R) = 0 \, , \nonumber\\[1mm]
B_1[0]_\Delta^{(R)} \;\text{with}\;\Delta=\tfrac12 R: \quad &a_E(R) = \begin{cases} -\fft{1-R}{4} & R\geq2 \\ \fft{1+12R}{360} & R=0,1\end{cases} \, , \\[1mm]  
&c(R) =\begin{cases} -\fft{1-R}{4} & R\geq2 \\ \fft{1+7R}{120} & R=0,1 \end{cases} \, , \quad b_1(R) = 0 \, .\nonumber
\end{align}
Then the total heat kernel coefficients are obtained after summing over all multiplets in the KK tower while taking into account their SU(3) representations. This can be done using the dimension formula~\eqref{Weyl:SU(3)}.

Since all $\mathcal{N}=3$ multiplets have vanishing $b_1(R)$, we immediately get that
\begin{equation}
	b_1^\text{tot}=0 \, .
\end{equation}
Next we consider the total heat kernel coefficient $c^\text{tot}$. Since the $L[j]_\Delta^{(R)}$ long multiplets have vanishing $c$ coefficients, we only need to consider the $A_{1,2}$ and $B_{1}$ short multiplets and their flavor representations given in Sections 3.2 and 3.3 of \cite{Fre:1999gok}. For example, the massive short graviton multiplet $A_1[1]_\Delta^{(R)}$ falls in SU(3) representations with Dynkin label $(k,k)$ with $k\geq 1$, and the R-charge is given by $R=2k$. In addition, there is a massless graviton multiplet in the $(0,0)$ representation with $R=0$. Therefore, the short graviton sector contributes to the $c$ coefficient with
\begin{equation}
	c^{\text{short}(2)} = \sum_{k=0}^\infty\dim(k,k)\fft{5+2k}{4} = \fft14\sum_{k=0}^\infty(1+k)^3(5+2k) \, .
\end{equation}
The massive short gravitino multiplet $A_2[0]_\Delta^{(R)}$ falls in the $(k,k+3)$ representation with $k\geq 0$. In addition, we must consider the conjugate representation $(k+3,k)$ to obtain the complete spectrum. The R-charge is fixed to $R = 2(k + 1)$ and the short gravitino sector contributes to the $c$ coefficient with
\begin{equation}
\label{eq:ciiiN010}
\begin{split}
	c^{\text{short}(3/2)} =&\; \sum_{k=0}^\infty[\dim(k,k+3)+\dim(k+3,k)]\fft{-3-2(k+1)}{4} \\
=&\; -\fft14\sum_{k=0}^\infty(1+k)(4+k)(5+2k)^2 \, .
\end{split}
\end{equation}
The massive short vector multiplet $B_1[0]_\Delta^{(R)}$ falls in the $(k,k)$ representation with $k \geq 2$, and with $R = 2k$. Together with the massless vector multiplet in the $(1,1)$ representation with $R=2$, this gives 
\begin{equation}
	c^{\text{short}(1)} = \sum_{k=0}^\infty\dim(k+1,k+1)\fft{-1+2(k+1)}{4} = \fft14\sum_{k=0}^\infty(2+k)^3(1+2k) \, .
\end{equation}
Lastly, the (massless) Betti multiplet is in the $(0,0)$ SU(3) representation and has $R=2$, for a contribution of
\begin{equation}
c^{\text{Betti}} = \dim(0,0)\fft{-1+2}{4}=\fft14 \, .
\end{equation}
We should now regularize all the above infinite series. In Appendix~\ref{app:reg}, we present a regularization method adapted from~\cite{friedman2012special}. However, this method is based on zeta-function regularization and is therefore not stable. This means that we will obtain different results depending on whether we first sum all short sector contributions and regularize the resulting series, or if we regularize each short sector separately before taking the sum. By doing the former, we find that $c^{\text{tot}} \neq 0$, in violation of our bootstrap constraints~\eqref{eq:no-Cloc}. However, we do find a remarkable cancellation if we decide to regulate each series independently. Our regularization prescription based on~\eqref{eq:FP-zeta-reg} attaches the following finite values:
\begin{equation}
\label{eq:c-N010-reg}
c^{\text{short}(2)} = \frac{55}{1024} \, , \qquad c^{\text{short}(3/2)} = -\frac12 \, , \qquad c^{\text{short}(1)} = \frac{201}{1024} \, ,
\end{equation}
and together with the Betti multiplet contribution, we arrive at
\begin{equation}
c^{\text{tot}} = 0 \, .
\end{equation}
Thus, it is possible to adopt a regularization prescription for the sum over the KK spectrum that is compatible with the bootstrap constraints~\eqref{eq:no-Cloc}. 

The calculation of the $a_E^\text{tot}$ coefficient is more subtle. First, we need to take into account the contributions from both long and short multiplets. Moreover, the resulting infinite sums involve multiple summation indices, which complicates the task of finding a suitable regularization. We give each contribution to $a_E^{\text{tot}}$ in turn below.

i) From Table (9) of \cite{Fre:1999gok}, we have
\begin{equation}
\label{eq:aE-N010-long}
\begin{split}
	a_E^\text{i)}&=\sum_{k=0}^\infty\sum_{j\geq2}^{\infty}\text{dim}(k,k+3j)\sum_{J_0=j}^{k+j}\bigg[\fft{3(1+2J_0)}{2}+\fft{1+2J_0}{2}+\fft{1+2J_0}{2}\bigg]\\
	&=\fft54\sum_{k=0}^\infty\sum_{j\geq2}^{\infty}(1+k)^2(1+2j+k)(1+3j+k)(2+3j+2k)\,.
\end{split}
\end{equation}

ii) From Table (10) of \cite{Fre:1999gok}, we have
\begin{equation}
\begin{split}
	a_E^\text{ii)}&=\sum_{k=0}^\infty\text{dim}(k,k+3)\Bigg[\sum_{J_0=1}^{k+1}\bigg[\fft{3(1+2J_0)}{2}+\fft{1+2J_0}{2}\bigg]+\sum_{J_0=1}^{k}\fft{1+2J_0}{2}\Bigg]\\
	&=\frac14\sum_{k=0}^\infty(1+k)(4+k)(5+2k)(12 + 18k + 5k^2) \, .
\end{split}
\end{equation}

iii) From Table (11) of \cite{Fre:1999gok}, we have
\begin{equation}
\begin{split}
	a_E^\text{iii)}&=\sum_{k=0}^\infty\text{dim}(k,k)\Bigg[\sum_{J_0=0}^{k-1}\bigg[\fft{3(1+2J_0)}{2}+\fft{1+2J_0}{2}\bigg]+\sum_{J_0=0}^{k}\fft{1+2J_0}{2}\Bigg]\\
	&= \frac12\sum_{k=0}^{\infty}(1+k)^3(1+2k+5k^2) \, .
\end{split}
\end{equation}

iv) From (12), (13) and (18), (19) of \cite{Fre:1999gok}, we have
\begin{equation}
	a_E^{\text{short}(2)}=\sum_{k=0}^\infty\dim(k,k)\fft{5(1+2k)}{4}=\fft54\sum_{k=0}^\infty(1+k)^3(1+2k)\,.
\end{equation}

v) From (14), (15) of \cite{Fre:1999gok}, we have
\begin{equation}
	a_E^{\text{short}(3/2)}=\sum_{k=0}^\infty\dim(k,k+3)\fft{-1+2(k+1)}{4}=\fft18\sum_{k=0}^\infty(1+k)(4+k)(1+2k)(5+2k)\,.
\end{equation}

vi) From (16), (17) and (20), (22) of \cite{Fre:1999gok}, we have
\begin{equation}
	a_E^{\text{short}(1)}=\sum_{k=0}^\infty\dim(k+1,k+1)\fft{-1+2(k+1)}{4}=\fft14\sum_{k=0}^\infty(2+k)^3(1+2k)\,.
\end{equation}

vii) From (21), (22) of \cite{Fre:1999gok}, we have
\begin{equation}
	a_E^{\text{Betti}}=\dim(0,0)\fft{-1+2}{4}=\fft14\, .
\end{equation}

Since we do not have bootstrap constraints involving $a_E$, it is difficult to come up with the right prescription for regularizing the above sums. In addition, the method presented in Appendix~\ref{app:reg} does not apply to the long sector series~\eqref{eq:aE-N010-long} which involve both $k$ and $j$ quantum numbers.

\section{Regularizing sums over KK spectra}
\label{app:reg}

Throughout the paper, we encounter a number of infinite sums that require regularization. In this appendix, we review a couple of methods that can in some cases attach a finite value to a divergent series. We hasten to note that these methods, being essentially based on zeta-function regularization, are neither linear nor stable (see e.g.~\cite{Monin:2016bwf}). As such, we do not claim that they can be used to regularize all infinite sums we encounter in an unambiguous way. But we are inclined to trust these methods whenever they yield results that are compatible with AdS/CFT expectations, as discussed in the main text.\\

The first method is a spectral regularization, which we apply to the sum
\begin{equation}
\label{eq:sum-proto}
\mathcal{S} = \sum_{n\geq0}(n+1)(n+2)(n+3)^2(n+4)(n+5) \, .
\end{equation}
We follow \cite{Gibbons:1984dg}. First, we introduce the following zeta function:
\begin{equation}
	\label{eq:spectral-zeta}
	\zeta_\Delta(z) = 1 + \sum_{n\geq1}D(n)\bigl(n(n+6)\bigr)^{-z} \, ,
\end{equation}
where
\begin{equation}
D(n) = \frac{1}{360}(n+1)(n+2)(n+3)^2(n+4)(n+5) \, .
\end{equation}
The function $\zeta_\Delta$ is associated to the scalar Laplacian $\Delta$ on the seven-sphere. Indeed, this differential operator has eigenvalues $n(n+6)$ with $n\geq 0$ and their multiplicities are given by $D(n)$. This spectral zeta function has an integral representation,
\begin{equation}
	\zeta_\Delta(z) = \frac{1}{\Gamma(z)}\,\int_0^\infty\,t^{z-1}\,\mathrm{Tr}\bigl[e^{-t\Delta}\bigr]\,dt \, ,
\end{equation}
and the value of $\zeta_\Delta(0)$ can be obtained from a heat kernel computation as~\cite{gilkey2018invariance}
\begin{equation}
	\zeta_\Delta(0) = a_{d}(\Delta)\,\text{res}_{z=0}\Gamma(z) \, ,
\end{equation}
where $d$ is the dimension of the space on which the scalar Laplacian acts. Crucially, the coefficient $a_d(\Delta)$ vanishes when $d$ is odd~\cite{gilkey2018invariance}, leading to $\zeta_\Delta(0) = 0$ for the seven-sphere. Thus, this method attaches the finite value
\begin{equation}
\mathcal{S} = 0 \, ,
\end{equation} 
to the infinite sum~\eqref{eq:sum-proto}. \\

Another regularization method can be adapted from~\cite{friedman2012special}, and is more general than the spectral zeta-function method. Namely, consider a divergent series of the form
\begin{equation}
\label{eq:sum-proto-gen}
\sum_{n\geq 0} f(n) \, .
\end{equation}
It will be important in what follows that $f$ satisfies the following two properties: (i) $f(n) \neq 0$ for $n \in [0,\infty)$ and, (ii) $f_{\text{top}}(n) = 0$ only for $n=0$. Here the subscript ``top'' indicates the term of highest homogeneity degree in $f$. The way to attach a finite value to~\eqref{eq:sum-proto-gen} is to introduce the zeta function
\begin{equation}
\zeta(s;f) = \sum_{n\geq 0} f(n)^{-s} \, . 
\end{equation}
Then, under the assumptions (i) and (ii) above required for convergence,~\cite{friedman2012special} showed that 
\begin{equation}
\label{eq:FP-zeta-reg}
\zeta(-1;f) = \sum_{L\geq0} c_L(f)\,B_L \, ,
\end{equation}
where $B_L$ is the $L$-th Bernoulli number and $c_L(f)$ is read off from the expansion of
\begin{equation}
\frac{f_a^{[\text{deg}(f)]}}{\text{deg}(f)}\sum_{\ell = 2}^{\text{deg}(f) + 1}\frac{(-1)^{\ell - 1}}{\ell(\ell - 1)}\,C_\ell(f_a) = \sum_{L\geq0} c_L(f)\,a^L \, .
\end{equation}
Here we have defined $f_a(n) = f(n+a)$, 
\begin{equation}
f^{[m]} = \frac{1}{m!} \frac{d^m f(n)}{dn^m}\Big\vert_{n=0} \, , 
\end{equation}
and $C_\ell(f)$ is the coefficient of $y^{1+\text{deg}(f)}$ in
\begin{equation}
\Bigg(\sum_{i=1}^{\text{deg}(f)}\frac{f^{[\text{deg}(f) - i]}}{f^{[\text{deg}(f)]}}\,y^i\Bigg)^\ell \, .
\end{equation}

Let us apply this regularization to the series $\mathcal{S}$ in~\eqref{eq:sum-proto}. We find that the coefficients $c_L$ are given by
\begin{equation}
	c_L(\mathcal{S}) = \Bigl\{-\frac{738}{7},\,-360,\,-471,\,-\frac{949}{3},\,-120,\,-26,\,-3,\,-\frac{1}{7}\Bigr\} \, ,
\end{equation}
for $L=0 \ldots 7$. Remarkably, using this specific linear combination of Bernoulli numbers in~\eqref{eq:FP-zeta-reg} shows that
\begin{equation}
\mathcal{S} = 0 \, ,
\end{equation}
in agreement with the spectral regularization method. We can also use this prescription to attach a finite values to other sums. In Appendix~\ref{App:KK:N010}, we encountered the divergent series 
\begin{equation}
\begin{split}
\mathcal{S}_1 =&\; \frac14\sum_{n\geq 0}(1+n)^3(5 + 2n) \, , \\
\mathcal{S}_2 =&\; -\frac14\sum_{n\geq 0}(1+n)(4+n)(5+2n)^2 \, , \\
\mathcal{S}_3 =&\; \frac14\sum_{n\geq 0}(2+n)^3(1+2n) \, .
\end{split}
\end{equation}
The relevant $c_L$ coefficients are
\begin{equation}
\begin{split}
c_L(\mathcal{S}_1) =&\; \left\{-\frac{1229}{5120},-\frac{5}{4},-\frac{17}{8},-\frac{7}{4},-\frac{11}{16},-\frac{1}{10}\right\}_{L=0\ldots 5} \, , \\[1mm]
c_L(\mathcal{S}_2) =&\; \left\{\frac{125}{16},25,\frac{205}{8},\frac{47}{4},\frac{5}{2},\frac{1}{5}\right\}_{L=0\ldots 5} \, , \\[1mm]
c_L(\mathcal{S}_3) =&\; \left\{-\frac{1267}{5120},-2,-\frac{7}{2},-\frac{5}{2},-\frac{13}{16},-\frac{1}{10}\right\}_{L=0\ldots 5} \, ,
\end{split}
\end{equation}
which lead to the finite values quoted in~\eqref{eq:c-N010-reg}.

\newpage

\bibliography{AdS_Logs}
\bibliographystyle{JHEP}

\end{document}